\documentclass[superscriptaddress,prx,aps,longbibliography,twocolumn,floatfix]{revtex4-2}
\pdfoutput=1

\usepackage{graphicx}
\usepackage{dcolumn}
\usepackage{bm}

\usepackage{amsmath,amssymb,amsfonts,amsthm}
\usepackage{array}
\usepackage{dsfont}
\usepackage{xcolor}

\usepackage{tikz}
\usetikzlibrary{shapes,arrows,positioning,automata,backgrounds,calc,er,patterns}

\usepackage{todonotes}
\usepackage[final]{changes} 
\newcolumntype{P}[1]{>{\centering\arraybackslash}p{#1}}

\usepackage[colorlinks=true,urlcolor=blue,citecolor=blue,allcolors=teal]{hyperref}
\usepackage{cleveref}
\usepackage{multirow}

\newcommand{\Tr}{\mathrm{Tr}}

\newcommand{\ket}[1]{\left|#1\right\rangle}
\newcommand{\bra}[1]{\left\langle#1\right|}
\newcommand{\braket}[2]{\left\langle#1\mid#2\right\rangle}

\newcommand{\sz}{\hat{\sigma}^z}
\DeclareMathOperator{\Var}{Var}
\begin{document}

\title{The ebbs and flows of quantum learning and sensing}
\newcommand{\Aalto}{Department of Applied Physics, Aalto University, FI-00076 Aalto, Espoo, Finland}
\newcommand{\xfiles}{Department of Paranormal Activities, Machine City, Planet E1234}
\newcommand{\ICMM}{Interdisciplinary Centre for Mathematical Modelling and Department of Mathematical Sciences,\\ Loughborough University, Loughborough, Leicestershire LE11 3TU, United Kingdom}
\newcommand{\LUPhys}{Department of Physics, Loughborough University, Loughborough, LE11 3TU, United Kingdom}
\newcommand{\Tampere}{Computational Physics Laboratory, Physics Unit, Faculty of Engineering and Natural Sciences, Tampere University, P.O. Box 692, FI-33014 Tampere, Finland}
\newcommand{\Helsinki}{Helsinki Institute of Physics P.O. Box 64, FI-00014, Finland}

\author{Matias Karjula}
\address{\Aalto}
\author{Teemu Ojanen}
\address{\Tampere}
\address{\Helsinki}
\author{Tapio Ala-Nissila}
\email[\vspace{-3pt}]{tapio.ala-nissila@aalto.fi}
\address{\Aalto}
\address{\ICMM}
\author{Moein N. Ivaki}
\email[\vspace{-3pt}]{moein.najafiivaki@aalto.fi}
\address{\Aalto}


\begin{abstract}
What is the relation between subsystem quantum complexity and the emergence of computationally useful structure? We address this by studying a family of minimally tunable postvariational quantum circuits, and show how spectral nonflatness and metrological response directly control the ensemble-typical information processing power. This unveils an intermediate ``learning phase’’ that precedes the onset of quantum chaos, characterized by pronounced nonflatness and sensitivity of readout states. The optimal information processing capacity improves with system size, while deep scrambling suppresses observable response. The results reveal how such features of random quantum dynamics can be viewed as computational resources for scalable nonlinear computation.
\end{abstract}

\maketitle
\section{Introduction} 

\subsection{Motivation}
Understanding how quantum systems process, learn, and store information is a central problem at the interface of quantum information science, manybody physics, thermodynamics, and computation~\cite{landauer1961irreversibility,hopfield1982neural,bennett1982thermodynamics,deutsch1985quantum,mandelstam1991uncertainty,margolus1998maximum,lloyd2002computational,PhysRevLett.109.120604, lloyd2000ultimate,PhysRevLett.96.010401,peruzzo2014variational,farhi2014quantum}. A principal aspect of this problem is determining how globally generated complexity is reflected in structures accessible to subsystems. This distinction becomes essential when quantum complexity is treated not as an abstract property of a wavefunction or quantum channel, but as a computational resource whose operational usefulness is tied to its accessibility to the degrees of freedom used for encoding, evolution, and decoding~\cite{RevModPhys.91.025001,PhysRevD.97.086015,PRXQuantum.2.030316,hayden2007black,yoshida2017efficient,chapman2022quantum,PhysRevA.67.052301}.
\begin{figure}[t]
    \centering
    \includegraphics[width=1.\columnwidth]{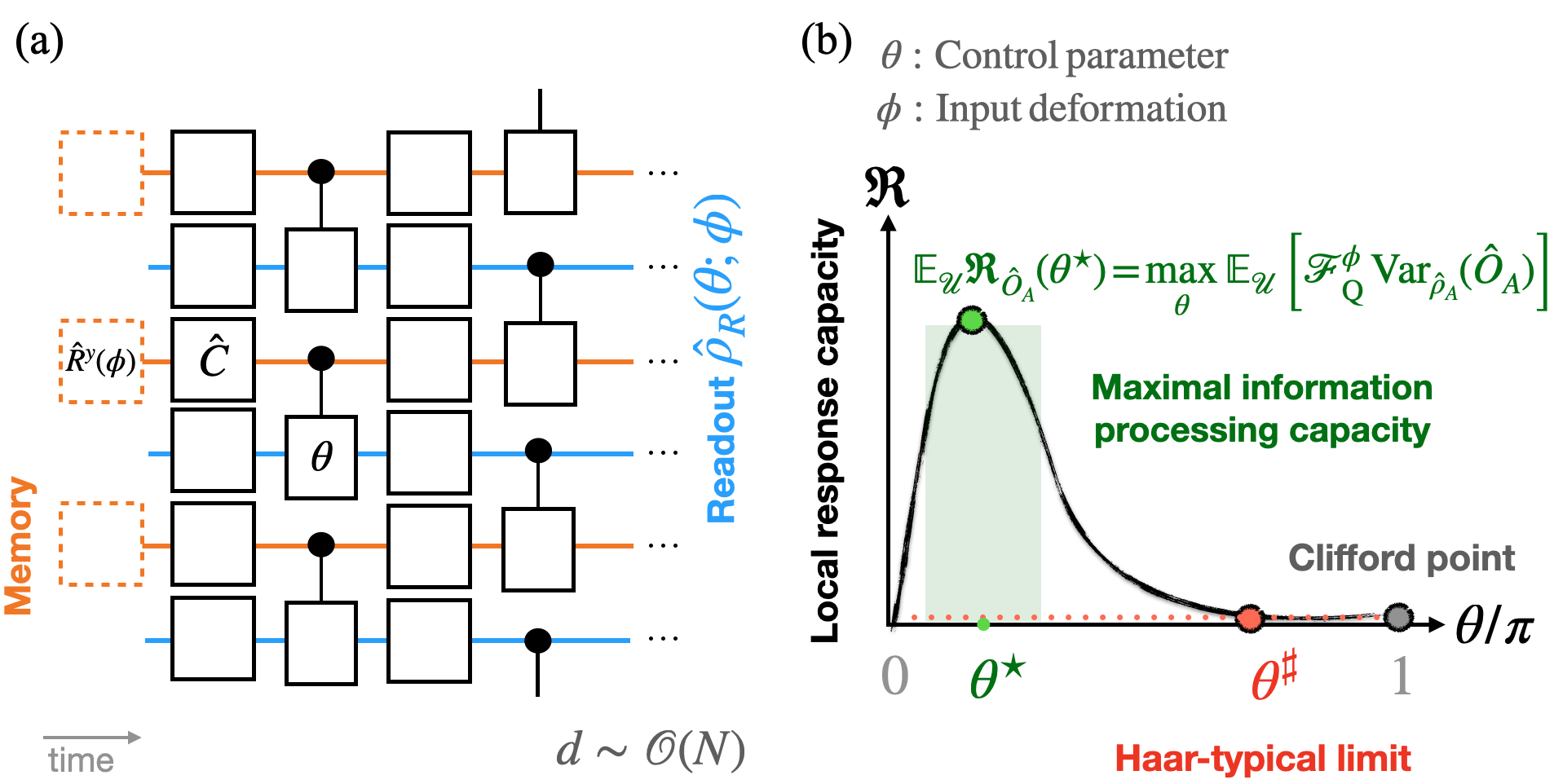}
    \caption{\textbf{The rise and fall of subsystem response in a tunable random quantum circuit}. \textbf{(a)} The one-dimensional model considered here shown for a single depth unit, \(d\!=\!1\). The two-qubit gate denoted by $\theta$ is \(\hat{P}(\theta)\!=\!\mathrm{diag}(1,1,1,e^{i\theta})\), and \(\hat C\) denotes a random single-qubit Clifford. The initial deformation \(\phi\) is imprinted on the memory \(\bar R\) subset via \(\hat R^{y}(\phi)\!=\!\exp[-i(\phi/2)\hat Y]\). \textbf{(b)} At fixed depth \(d\!\sim\!\mathcal{O}(N)\), by tuning \(\theta\) the model interpolates from a weakly entangling and magical regime to the quantum chaotic limit. At \(\theta\!=\!\pi\), the unencoded unitary circuit is maximally entangling but nonmagical, i.e., Clifford-only. We refer to \(\mathfrak R_{\hat O_R}\), {\color{brown}Eq.}~\eqref{eq:complex_cap}, as the response capacity. A maximal response around \(\theta\!\approx\!\theta^{\star}\!<\!\theta^{\sharp}\) indicates the coexistence of a hierarchical and thermodynamically nondegenerate entanglement spectrum with maximal distinguishability under encoding perturbations. This points to the  \textit{formation of a computationally optimal structure before the onset of quantum chaos}. In the Haar-typical limit, \(\pi\!>\!\theta\!>\!\theta^{\sharp}\), the local response is exponentially suppressed.}
    \label{fig:scheme}
\end{figure}

Interacting chaotic quantum dynamics can generate inequivalent forms of complexity and correlations, including entanglement~\cite{PhysRevLett.71.1291,RevModPhys.81.865,RevModPhys.80.517} and nonstabilizerness (aka magic)~\cite{gottesman1998heisenberg,veitch2014resource,PhysRevLett.128.050402,haug2025probing}, the latter being essential for universal quantum computation beyond classically simulable regimes~\cite{PhysRevA.70.052328,leone2021quantum}. Sufficiently expressive noiseless random circuits can further produce approximately Haar-typical unitary-design ensembles with universal spectral features~\cite{PhysRevA.80.012304,harrow2009random,roberts2017chaos,cotler2017chaos,baiguera2026quantum,mi2021information}. In this regime, subsystems become increasingly \textit{flat}, and their spectra approach the corresponding random-matrix distributions. For common quantum-learning architectures, local cost functions, kernels, and other experimentally accessible observables then concentrate exponentially around their ensemble-typical values~\cite{mcclean2018barren,larocca2025barren,thanasilp2024exponential,xiong2025role,tnfv-lzfx}. Spectral flattening and observable concentration therefore signal the onset of subsystem thermalization, where trainability, generalizability, memory retention, and parameter dependent sensitivity may become strongly suppressed. These considerations are particularly relevant to near term architectures, where restricted measurement access and noise further limit which features can be estimated with sufficient accuracy~\cite{RevModPhys.94.015004}.

Between the weakly interacting and fully scrambled limits, however, a dynamical quantum map may remain expressive while retaining operationally accessible features. Such an intermediate regime is commonly associated with the long-standing concept of \textit{edge-of-chaos} computing, pointing to the broader principles that information processing and sensitivity may be enhanced near critical dynamical regimes~\cite{langton1990computation,mitchell1993revisiting,krakauer2011darwinian,mora2011biological,PhysRevLett.113.068102,aaronson2014quantifying}. Despite the recent attention to this regime as a quantum ``Goldilocks'' zone~\cite{j2qj-vwcl,vcindrak2026memory,PhysRevLett.127.100502,tnfv-lzfx,xia2022reservoir,gq9r-d5q8,PRXQuantum.5.040325}, a general framework is still lacking for identifying which subsystem resources make this intermediate regime computationally useful. Taking the perspective of \textit{quantum speed limits}~\cite{deffner2017quantum}, in this work we establish that measures of \textit{spectral nonflatness} and \textit{metrological susceptibility} jointly provide a unifying diagnostic of computationally accessible and usable subsystem complexity, potentially relevant to a broad family of quantum learning and sensing algorithms. Speed limits and thermodynamic uncertainty relations connect achievable precision in parameter estimation and computation to the thermodynamic costs and quantum resources required in both quantum and classical stochastic dynamics~\cite{PhysRevLett.133.247101,PhysRevLett.126.010602,escher2011general,nicholson2020time}.

\subsection{Central results}
The central operational question is therefore whether an accessible subsystem retains both nontrivial spectral structure and a measurable response to an encoded perturbation. We formalize this through an observable speed limit that separates state distinguishability from the fluctuation scale of the measured observable. Let \(\ket{\psi}\!\in\!\mathbb{C}^{2^{N_R}}\!\otimes\!\mathbb{C}^{2^{N_{\bar R}}}\) be a pure state on a complex bipartite Hilbert space \(R\cup\bar R\). The reduced density matrix on subsystem \(R\) is \(\hat\rho_R\!:=\!\Tr_{\bar R}\bigl(\ket{\psi}\bra{\psi}\bigr)\). Here, \(\hat\rho_R\!\equiv\!\hat\rho_R(\theta;\phi)\), where \(\theta\) is a \textit{control parameter} of the evolution and \(\phi\) is a \textit{deformation} of interest for parameter-dependent computation ({\color{brown}Fig.}~\ref{fig:scheme}(a)). Generalized Cramér--Rao-type bounds assert that~\cite{PhysRevX.12.011038,sidhu2020geometric}, under arbitrary valid quantum dynamics and for a \(\phi\)-independent Hermitian subsystem observable \(\hat O_R\), the state-induced infinitesimal response satisfies
\begin{align}
\left|\partial_\phi\langle\hat O_R\rangle\right|^2
\leq
\mathcal F_{\rm Q}^{\phi}\,
\mathrm{Var}_{\hat\rho_R}(\hat O_R).
\label{eq:gen_cramer_Rao}
\end{align}
Here, \(\partial_\phi\!\equiv\!\partial/\partial\phi\) and \(\langle\hat O_R\rangle\!=\!\Tr[\hat\rho_R\hat O_R]\); see Appendix~\ref{app:Response_bound} for a derivation. The speed of an observable change is therefore limited jointly by two fundamentally \textit{distinct}, although in some cases closely related, quantities. The first is the geometric distinguishability generated in the reduced state, captured by the \textit{quantum Fisher information} \(\mathcal F_{\rm Q}^{\phi}\!\equiv\!\mathcal F_{\rm Q}^{\phi}[\hat\rho_R(\theta;\phi)]\)~\cite{montenegro2025quantum,PhysRevLett.127.200402,PhysRevLett.72.3439}, which quantifies the infinitesimal Bures metric, \(ds_{\rm B}^2\!\propto\!\mathcal F_{\rm Q}^{\phi}d\phi^2\), for mixed states~\cite{paris2009quantum}. The second is the in-state fluctuation scale of the measured observable, \(\mathrm{Var}_{\hat\rho_R}(\hat O_R)\!=\!\langle\hat O_R^2\rangle-\langle\hat O_R\rangle^2\), which can be directly related to spectral nonflatness. In particular, the \textit{capacity of entanglement}~\cite{PhysRevD.99.066012,okuyama2021capacity} and \textit{antiflatness}~\cite{jasser2026journey} quantify how far the eigenvalue distributions of reduced states, or of propagated operators in an appropriate operator-space representation, remain from flat or nearly degenerate limits. Such measures connect aspects of the resource theories of entanglement and nonstabilizerness~\cite{tirrito2024quantifying}, with implications for classical simulability through probes of nonlocal magic~\cite{z3vr-w5c5,robin2025anti,grieninger2026nonlocal}. To connect these spectral diagnostics to an operational learning setting, we consider input-dependent expectation values, which form the measured features of a broad class of quantum-learning models~\cite{PhysRevApplied.8.024030,mujal2021opportunities,nakajima2020physical,cerezo2021variational,schuld2021effect}. Since {\color{brown}Eq.}~\eqref{eq:gen_cramer_Rao} bounds the squared input susceptibility of each such feature by the product of reduced-state distinguishability and the observable fluctuation scale, this product provides a natural task-agnostic upper scale for accessible input response. We therefore define the \textit{response capacity}
\begin{align}
\mathfrak R_{\hat O_R}:=
\mathcal F_{\rm Q}^{\phi}\,
\mathrm{Var}_{\hat\rho_R}(\hat O_R).
\label{eq:complex_cap}
\end{align}
The underlying response bound holds for each unitary-circuit realization and therefore also constrains averages, \(\mathbb E_{\mathcal U}\), over independent realizations \(\mathcal U\). The quantity \(\mathbb E_{\mathcal U}\{\mathfrak R_{\hat O_R}\)\} thus sets the ensemble-averaged upper response scale of \(\hat O_R\) to the encoded parameter. Given a family of experimentally accessible readout operators, the aim is to identify the control-parameter regime in which this scale is largest and remains thermodynamically finite ({\color{brown}Fig.}~\ref{fig:scheme}(b)). This construction admits a natural interpretation in the context of \textit{quantum reservoir learning}~\cite{PhysRevApplied.8.024030} and related paradigms of postvariational and \textit{physical learning}~\cite{mujal2021opportunities,nakajima2020physical}. In contrast to variational algorithms based on task-driven, end-to-end optimization of a quantum architecture~\cite{cerezo2021variational}, the main computational object is not an explicitly trained quantum circuit, but rather a fixed quantum dynamical substrate whose internal random evolution generates an observable feature map.

We show that spectral nonflatness and metric susceptibility identify two complementary ingredients of computationally useful subsystem complexity, revealing an intermediate regime in which maximal usable subsystem structure emerges while entanglement and magic remain substantial but submaximal. In this regime, subsystem spectra remain strongly nonflat and encoding perturbations remain sufficiently distinguishable through accessible observables. We then demonstrate the operational relevance of this regime in a postvariational learning setting, where temporal nonlinear learnability, memory, and measurement-accessible Fisher information exhibit a similar rise–peak–fall structure, and the total information-processing capacity grows with system size near the optimum. These probes provide a necessary, although not generally sufficient, condition on average learnability, granting physically grounded proxies for optimizing computational capabilities and analyzing thermodynamic scalability. Our results thereby open a route toward constructing learning and sensing machines whose design principles are rooted in the fundamental physics of quantum information processing, quantum thermodynamics, and resource theories of quantum complexity.



\section{Quantum circuit model} 
To make our arguments concrete, in this work we consider a family of one-dimensional brickwork quantum circuits with $N$ qubits placed in a ring geometry (periodic boundary condition). One-step Floquet operator is given by
\begin{equation}
\begin{aligned}
\hat{\mathcal{U}}_{\rm F}(\theta)
&=\left[\prod_{i\in {\mathrm{odd}}}
\hat{P}_{i,i+1}(\theta)\right]
\left[\bigotimes_{j=0}^{N-1}\hat C^{(1)}_j\right]
\\&\times\left[\prod_{i\in  {\mathrm{even}}}
\hat{P}_{i,i+1}(\theta)\right]
\left[\bigotimes_{j=0}^{N-1}\hat C^{(2)}_j\right].
\end{aligned}
\label{eq:floquet_compact}
\end{equation} 
Here, \(\hat C^{(m)}_j\) denotes the random single-qubit Clifford gate acting on site \(j\) in the \(m\)-th Clifford layer. The gate \(\hat{P}_{jl}(\theta)\!=\!\mathrm{diag}(1,1,1,e^{i\theta})\) acts on qubits \(j,l\). The two-qubit interaction admits the factorization \(\hat{P}_{jl}(\theta) \!=\! e^{i\theta/4}\,
(e^{-i\theta\sz_j/4}\!\otimes e^{-i\theta\sz_l/4})\, e^{\,i\theta\sz_j\otimes\sz_l/4}\), where $\sigma$'s are Pauli operators. More generally, any two-qubit unitary is locally equivalent to \(\hat U(\theta_x,\theta_y,\theta_z)\! = \!\exp[ -\frac{i}{2} \sum_{k=x,y,z}
\theta_k\,\hat\sigma^k\!\otimes\!\hat\sigma^k
]\), with \(0\le\! \theta_z\le\! \theta_y\le\! \theta_x\le\! \pi/2\). The parameters $(\theta_x,\theta_y,\theta_z)$ characterize the nonlocal part of the gate up to arbitrary one-qubit unitaries before and after the interaction. 

To place this in the stabilizer framework, recall first the \(N\)-qubit Pauli group
\[\mathcal P_N=\{\pm 1,\pm i\}\cdot \{\hat I,\hat X,\hat Y,\hat Z\}^{\otimes N}.\]
A unitary \(\hat U\) is called Clifford if it normalizes the Pauli group, namely if \(\hat U\,\mathcal P_N\,\hat U^\dagger\!=\!\mathcal P_N.\) Thus Clifford unitaries map Pauli operators to Pauli operators under conjugation. A pure \(N\)-qubit stabilizer state \(\ket{\psi}\) is a common \(+1\) eigenstate of an abelian stabilizer group \(\mathcal S\subset\mathcal P_N\) of size \(2^N\), i.e. \(\hat g\ket{\psi}\!=\!\ket{\psi},\ \forall\, \hat g\in\mathcal S.\) Equivalently, stabilizer states are precisely those obtained from computational-basis product states by Clifford circuits~\cite{gottesman2024surviving}. Clifford dynamics remains within the accurately and efficiently classically simulable stabilizer manifold~\cite{PhysRevA.70.052328}. Non-Clifford resources, such as a \(\hat T\!=\!{\rm diag} (1,e^{i\pi/4})\) gate, induce Pauli branching in operator space~\cite{rudolph2025pauli}, providing a microscopic mechanism for the emergence of universal random circuit features commonly associated with quantum chaotic dynamics~\cite{c7k1-xcwy, leone2021quantum}.

A key intuition motivating the construction of the family of models here is the following. The average entangling power of a bipartite unitary \(\hat U\) can be defined as the Haar average of the output linear entropy generated from product inputs, \(\mathfrak e (\hat U)\!:=\!\mathbb E_{\ket{\psi_{R\bar R}}}[\,1\!-\!\Tr(\hat\rho_{\bar R}^2)\,]\), where \(\hat\rho_{\bar R}\!=\!\Tr_R(\hat U\ket{\psi_{R\bar R}}\!\bra{\psi_{R\bar R}}\hat U^\dagger)\) and \(\ket{\psi_{R\bar R}}\!=\!\ket{\psi_R}\!\otimes\!\ket{\psi_{\bar R}}\)~\cite{PhysRevA.62.030301,PhysRevA.67.042313}. Similarly, the average nonstabilizing (magic) power can be defined as the average linear stabilizer entropy generated from stabilizer inputs, \(\mathfrak m(\hat U)\!:=\!|\mathrm{STAB}|^{-1}\sum_{|\psi\rangle\in\mathrm{STAB}} \mathcal{M}^{\mathrm{lin}}(\hat U|\psi\rangle)\), where \(\mathcal{M}^{\mathrm{lin}}(|\psi\rangle)\!:=\!1\!-\!2^{-N}\sum_{\hat P\in\mathcal P_N}\langle\psi|\hat P|\psi\rangle^4\) is a faithful monotone of magic for pure states~\cite{PhysRevLett.128.050402,PhysRevA.110.L040403}. For \(\hat{P}(\theta)\) one may take \((\theta_x,\theta_y,\theta_z)\!=\!(\theta/2,0,0)\) up to local Clifford equivalence, and the average entangling power and nonstabilizing (magic-generating) power of the elementary two-qubit gate respectively are \(\mathfrak e(\theta)\propto\sin^2\!\frac{\theta}{2}, \mathfrak m (\theta)\propto\sin^2\theta\)~\cite{varikuti2026impact}. The random single-qubit Clifford layers do not change the gatewise values of \(\mathfrak m(\theta)\), since both are invariant under local Clifford dressing, but, importantly, they do strongly affect the dynamical buildup of manybody magic. Repeated Clifford-interlaced application is expected to drive the gate-level power exponentially toward the Haar-typical value~\cite{varikuti2026impact}. Gate-level thermalization of entangling power \(\mathfrak e(\theta)\) obeys a similar relation, where entangling operations interlaced with random local
gates result in exponential equilibration to the typical value~\cite{PhysRevResearch.2.043126, PhysRevA.95.040302}. Overall, in this model and over the interval $\theta/\pi\in[0,1]$, the entangling power of the gates varies monotonically, whereas the nonstabilizing power is nonmonotonic. Both are mirror symmetric about integer values of \(\theta/\pi\), so it is sufficient to restrict the analysis to this interval. The onset of manybody quantum chaotic behavior and subsystem thermalization is therefore controlled by the interplay between these resources in the presence of random single qubit gates. This provides a minimal and powerful tuning mechanism for probing how the evolution of dynamical resources shape memory and nonlinear processing capacity. 
\begin{figure}[t]
    \centering
    \includegraphics[width=0.99\linewidth]{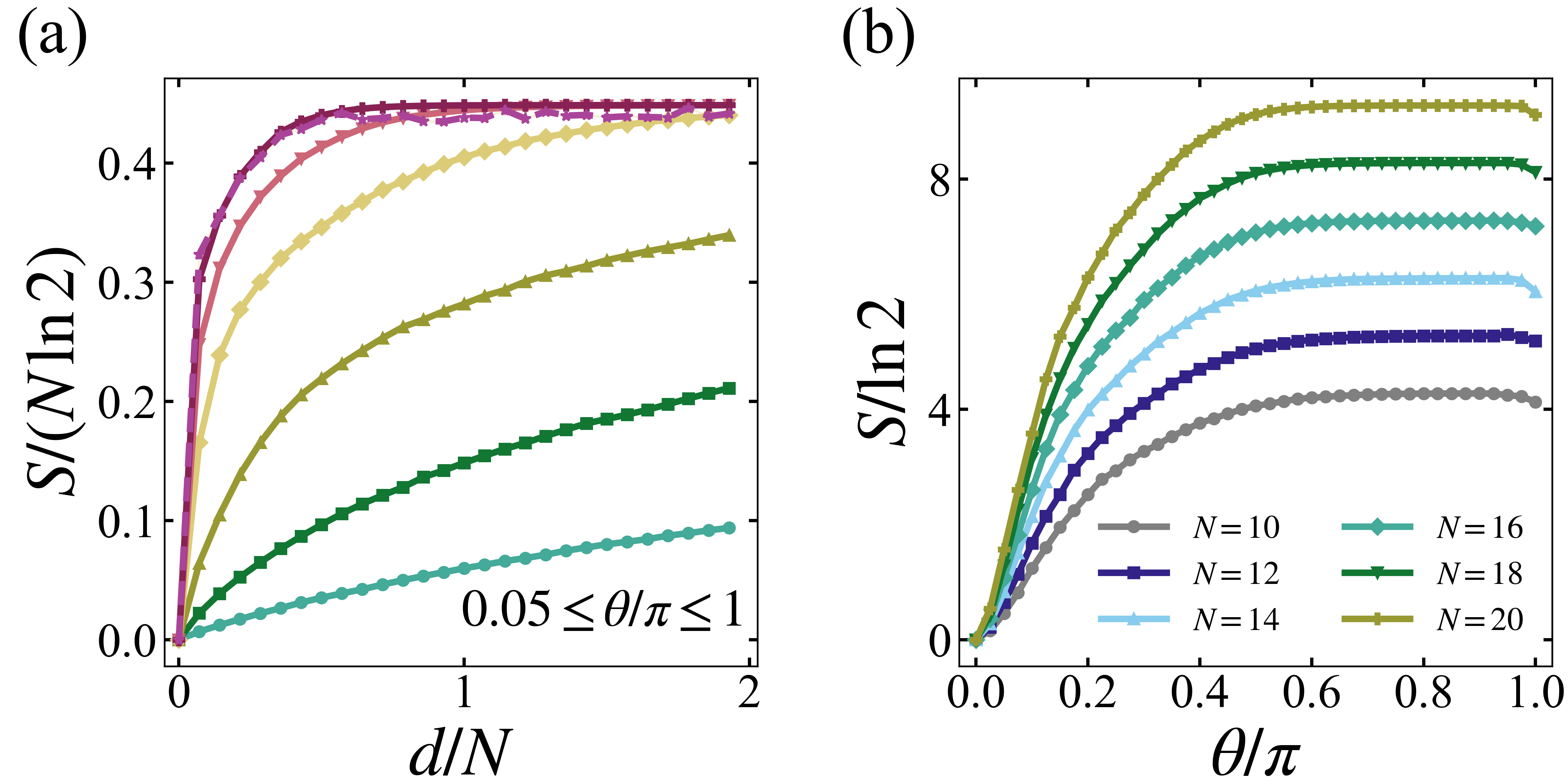}
    \includegraphics[width=0.99\linewidth]{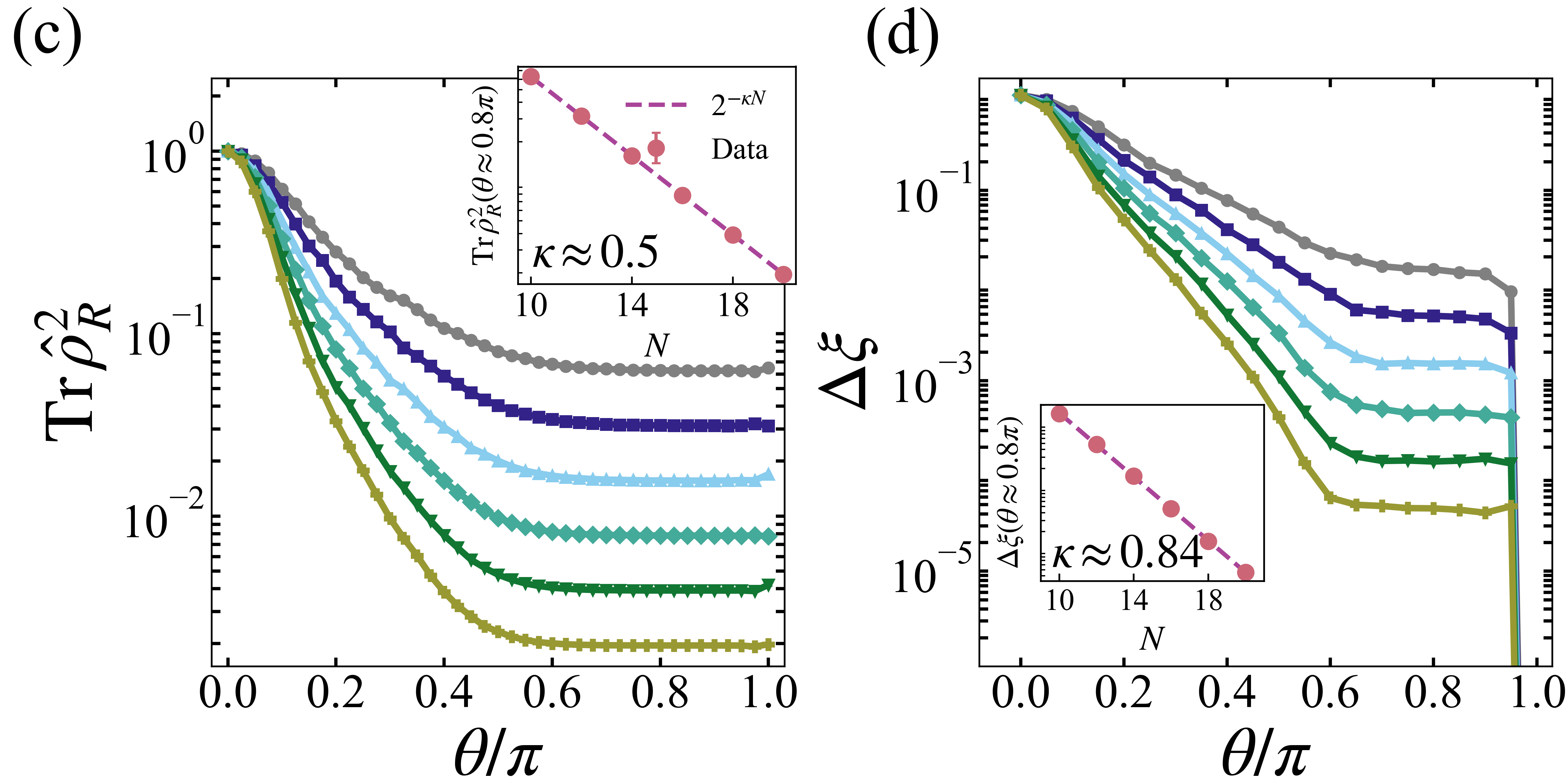}
    \includegraphics[width=0.99\linewidth]{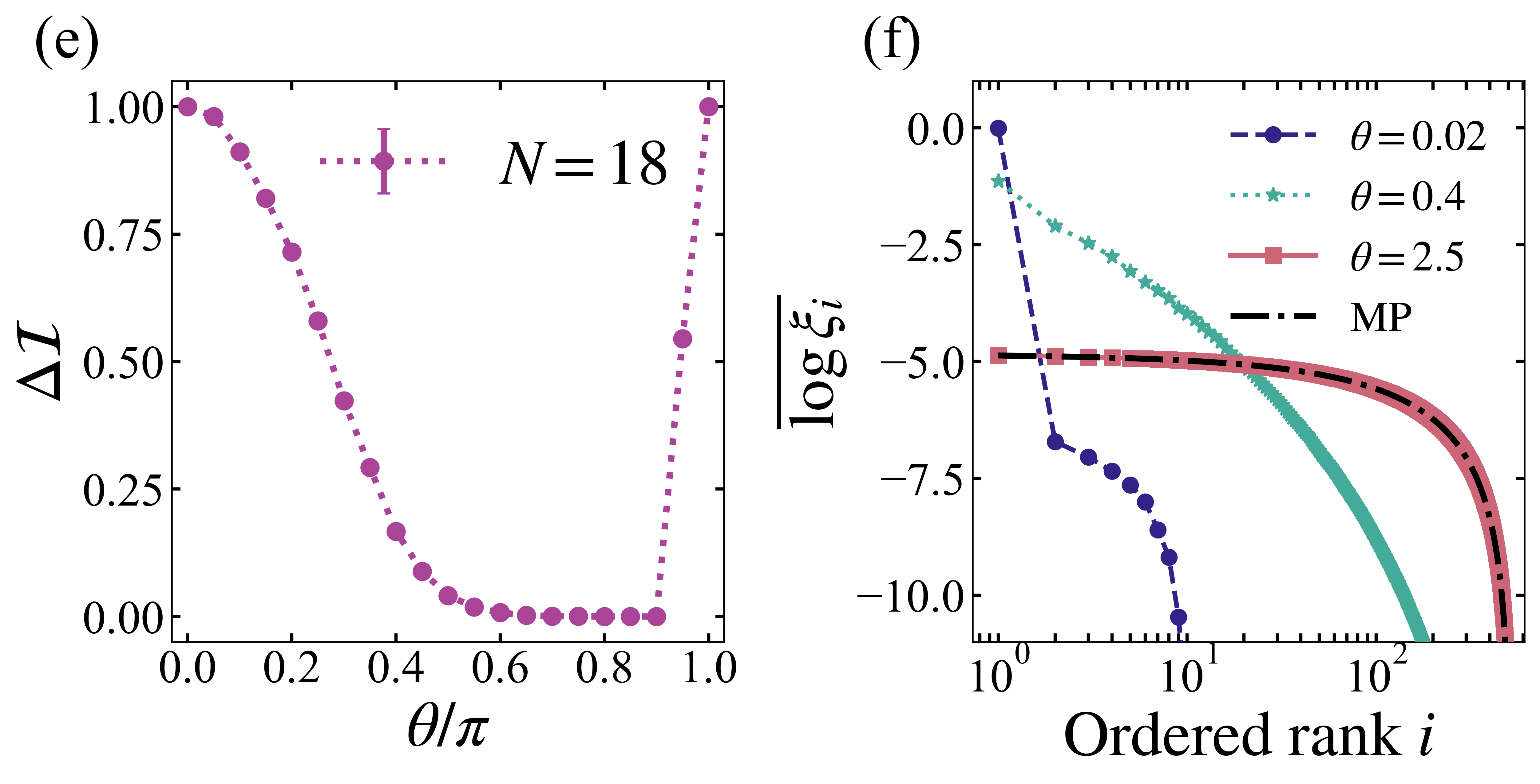}
\caption{\textbf{General properties of the circuit model.}
\textbf{(a)} Entanglement-entropy density \(S_{\rm vN}/N\) as a function of the rescaled circuit depth \(d/N\), for an equal bipartition \(|R|\!=\!|\bar R|\!=\!N/2\) and \(N\!=\!14\), showing how the \(\theta\)-dependent entanglement-growth rate increases with increasing $\theta$. \textbf{(b)} \(S_{\rm vN}\) as a function of \(\theta/\pi\) for \(d/N\!=\!1\), exhibiting the expected volume-law scaling with system size \(N\). \textbf{(c)} Purity and \textbf{(d)} Schmidt gap for several system sizes. Both quantities decay exponentially with \(N\), as \(2^{-\kappa N}\), in the typical regime as shown in the insets. \textbf{(e)} Mutual-magic gap \(\Delta\mathcal I\) as a function of \(\theta/\pi\) for \(d/N\!=\!1\), characterizing the crossover toward the quantum-chaotic regime. \textbf{(f)} Ensemble-averaged entanglement spectra at representative interaction strengths for \((N,d/N)\!=\!(18,1)\). The intermediate regime exhibits a broad, hierarchical distribution of eigenvalues, in sharp contrast to both the product-state and random-state limits, with the latter described by the Marchenko--Pastur (MP) distribution. Data are averages over \(40\)--\(250\) independent circuit realizations and error bars are of the size of the symbols or smaller.}
\label{fig2:general_prop}
\end{figure}

{\color{brown}Figure}~\ref{fig2:general_prop} illustrates the crossover from weakly entangled dynamics to the Haar-typical regime in our model. The entanglement entropy \(S_{\rm vN}\), purity \(\Tr \hat\rho^2_R\), and the Schmidt gap \( \Delta\xi\!:=\!\xi_1-\xi_2,\) where \(\xi_1\) and \(\xi_2\) are the two largest eigenvalues of the reduced density matrix~\cite{PhysRevLett.109.237208}, all track the growth and redistribution of manybody correlations between the memory $\bar R$ and readout $R$ subsystems when increasing \(\theta\) and system size $N$. The interleaved placement of the memory $\bar R$ and readout $R$ qubits is operationally motivated (cf. {\color{brown}Fig.}~\ref{fig:scheme}(a)), as it maximizes their interface and avoids the propagation bottleneck of contiguous partitions, for which the typical encoding-to-readout distance may also grow with \(N\). As one may expect, the entanglement entropy displays a \(\theta\)-dependent growth and, deep in the typical limit, it also exhibits volume-law scaling at fixed depth. At the same time, both the subsystem purity and \(\Delta\xi\) decrease as function of $\theta$ and decay exponentially \(\sim\!2^{-\kappa N}\) in the deeply chaotic regime. Further, for mixed reduced states, the mutual magic is defined in terms of a mixed-state magic measure \(\mathcal M\) as \(\mathcal I_{\mathcal M}(R\!:\!\bar R)\!=\!\mathcal M(\hat\rho_{R\bar R})\!-\!\mathcal M(\hat\rho_R)\!-\! \mathcal M(\hat\rho_{\bar R})\) which quantifies the nonadditive (or long-range) component of the magic resource shared between the two subsystems~\cite{tarabunga2025efficient,tnfv-lzfx}. In {\color{brown}Fig.~\ref{fig2:general_prop}}(e), we plot the relative gap \(\Delta\mathcal I\!=\! |\mathcal I_{\mathcal M}\!-\!\mathcal I_{\mathcal M}^{\rm H}|/{\mathcal I_{\mathcal M}^{\rm H}}\) from the Haar-random reference value, \(\mathcal I_{\mathcal M}^{\rm H}\!\simeq\! N\!-4\), where \({\mathcal M}\) represents the second stabilizer R\'enyi entropy~\cite{PhysRevLett.128.050402}. Similar to other measures, \(\mathcal I_{\mathcal M}\) approaches its Haar-typical value beyond a characteristic interaction strength $\theta^{\sharp}$, \(\Delta\mathcal I_{\mathcal M}(\theta\!>\!\theta^{\sharp})\!\approx\!0\), providing an independent indication of the onset of genuine quantum scrambling. 

Finally, the behavior of averaged entanglement spectrum $\log\xi_i$ as function of the ordered rank $i$ is shown in {\color{brown}Fig.~\ref{fig2:general_prop}}(f). This reveals an intermediate regime with a broad hierarchical distribution of eigenvalues, absent in both the product-state limit, which is dominated by a single large eigenvalue $\xi_1\approx1$, and the fully scrambled, which follows a universal Marchenko-Pastur distribution~\cite{marvcenko1967distribution,yang2015two}. The intermediate phase corresponds to the regime in which the entanglement spectrum has its largest variance, which, as we show next, can also remain extensive with size in the thermodynamic limit. This signals a strongly nondegenerate distribution of Schmidt weights~\cite{serbyn2016power}, before the typicality-induced spectral flattening sets in. 

We next present a detailed analysis of the spectral and metrological diagnostics used to quantify accessible subsystem response, and study their behavior across the circuit's dynamical regimes. We further present an application of our results in a postvariational learning setting.


\section{Nonflatness, metrology, and learning}

\subsection{Spectral nonflatness}
Response functions are natural objects to study in quantum systems since, in close analogy with heat capacities and susceptibilities in equilibrium statistical mechanics, they quantify sensitivity to perturbations and are often controlled by underlying fluctuations~\cite{marconi2008fluctuation}. In equilibrium settings, they may also be interpreted as curvatures of an appropriate thermodynamic potential~\cite{PhysRevD.99.066012, okuyama2021capacity,marconi2008fluctuation, jasser2026journey}. As such, they may distinguish states with similar entropy or average energy but very different spectral organization, thereby revealing intermediate regimes of structure that are invisible to coarse measures alone. 

In the present case, the entanglement spectrum~\cite{PhysRevLett.101.010504, yang2015two,PhysRevLett.105.080501,PhysRevB.83.115322,geraedts2016many,shaffer2014irreversibility}, a fundamental quantity in characterizing complexity and chaos, admits a natural thermodynamic interpretation in terms of the entanglement Hamiltonian \(\hat{\mathcal{H}}_R\!=\!-\log\hat\rho_R, \) whose partition function \(\mathcal Z_{\beta}\!=\!\Tr(\hat\rho_R^\beta)\) generates both R\'enyi entropies \(S_{\beta}\!=\!(1\!-\!\beta)^{-1}\log \mathcal Z_{\beta}\) and a family of modular response functions~\cite{baez2022renyi, andrzejewski2023evolution, PhysRevD.99.066012, rangamani2017holographic, nandy2021capacity, dong2016gravity, okuyama2021capacity, kawabata2021probing}. Here $\beta\!>\!0$ plays the role of an inverse modular temperature, and varying \(\beta\) amounts to probing different regions of the entanglement spectrum, where larger \(\beta\) emphasizes the dominant Schmidt coefficients associated with the bipartition. A natural Gibbs-like ensemble associated with \(\hat{\mathcal{H}}_R\) is the escort state $\hat\omega_{\beta}\!:=\! e^{-\beta\hat{\mathcal{H}}_R}/{\mathcal Z_{\beta}}
\!=\!\hat\rho_R^\beta/{\Tr(\hat\rho_R^\beta)}.$ The corresponding modular internal energy is \(I_{\beta} := \Tr(\hat\omega_{\beta}\hat{\mathcal{H}}_R)
\!=\!-\partial_\beta \log \mathcal Z_{\beta},\) and its derivative is controlled by fluctuations of the modular Hamiltonian, \(\partial_\beta I_{\beta}\!=\!-\Var_{\hat\omega_{\beta}}(\hat{\mathcal{H}}_R).\) The entropy of the modular ensemble itself is $ \mathbb S_\beta
\!:=\!-\Tr\!\left(\hat\omega_{\beta}\log\hat\omega_{\beta}\right)
\!=\!\log \mathcal Z_{\beta}-\beta\,\partial_\beta \log\mathcal Z_{\beta},
\label{eq:modular_entropy}
$ which is the standard thermodynamic identity $\mathbb S_{\beta}\!=\!\beta(I_{\beta}\!-\!F_{\beta})$, with $F_{\beta}\!:=\!-\beta^{-1}\log \mathcal Z_{\beta}$ the ``free energy" of the reduced state. Since \(F_{1}\!=\!0\), we get \(\mathbb S_1\!=\!I_{1}\!=\!S_{\rm vN}\!=\!S_1\). These also suggest the \textit{spectral complexity capacity} (aka the modular heat capacity) \cite{PhysRevD.99.066012, andrzejewski2023evolution}, 
\begin{equation}
\mathcal{C}_{\beta}
:=\beta^2 \Var_{\hat\omega_{\beta}}(\hat{\mathcal{H}}_R)
=\beta^2 \partial_\beta^2 \log \mathcal Z_{\beta}\ge 0,
\label{eq:Cmod_def}
\end{equation}
which measures the response of the entanglement spectrum to changes in the modular temperature $\beta$. At \(\beta\!=\!1\), this becomes the familiar capacity of entanglement \cite{PhysRevD.99.066012, okuyama2021capacity},
\begin{equation}
\mathcal{C}_{\rm E}:= \sum_i \xi_i \left(\log \xi_i\right)^2 - \left( \sum_i \xi_i \log \xi_i \right)^2,
\label{eq:CE_def}
\end{equation}
where \(\{\xi_i\}\) are the eigenvalues of \(\hat\rho_R\). In other words, \(\mathcal{C}_{\rm E}\) is the variance of the entanglement energies \(\epsilon_i\!:=\!-\log\xi_i\). It therefore vanishes for a perfectly flat spectrum on its support, as occurs for stabilizer, pure, and product states, and grows as the spectrum becomes more nonuniform and nondegenerate. In quantum field theory and holography, the capacity of entanglement has been used to probe universal scaling and renormalization-group flows, nonequilibrium entanglement dynamics, and fluctuations or phase transitions in dual gravitational descriptions~\cite{PhysRevD.99.066012}. Importantly, \(\mathcal{C}_{\rm E}\) (and more generally spectral capacity at a temperature $\beta$) is a faithful spectral witness of nonflatness, i.e., \(\mathcal{C}_{\rm E}(\hat\rho_R)\ge 0\) and \(\mathcal{C}_{\rm E}(\hat \rho_R)\!=\!0\) iff the reduced state \(\hat\rho_R\) is proportional to a projector (flat on its support). It is also additive under tensor product \(\mathcal{C}_{\rm E}(\hat\rho_R{\otimes \hat\rho'_R})\!=\!\mathcal{C}_{\rm E}(\hat\rho_R)+\mathcal{C}_{\rm E}(\hat\rho'_R)\) and uniformly continuous in finite dimension~\cite{PhysRevA.106.042419}. A full resource-theoretic interpretation requires additional care, since the appropriate class of free operations and associated state-conversion preorder must be specified; the nonflatness measures considered here are generally neither Schur-convex nor -concave under ordinary majorization, which does not consistently order spectral nonflatness~\cite{jasser2026journey}. In {\color{brown}Appendix}~\ref{app:capacity_power_cp} we also calculate the average nonflattening power of $\hat{P}(\theta)$ gates over several candidate input ensembles, revealing a nonmonotonic dependence on the interaction angle \(\theta\).
\begin{figure}[t]
    \centering
    \includegraphics[width=0.99\linewidth]{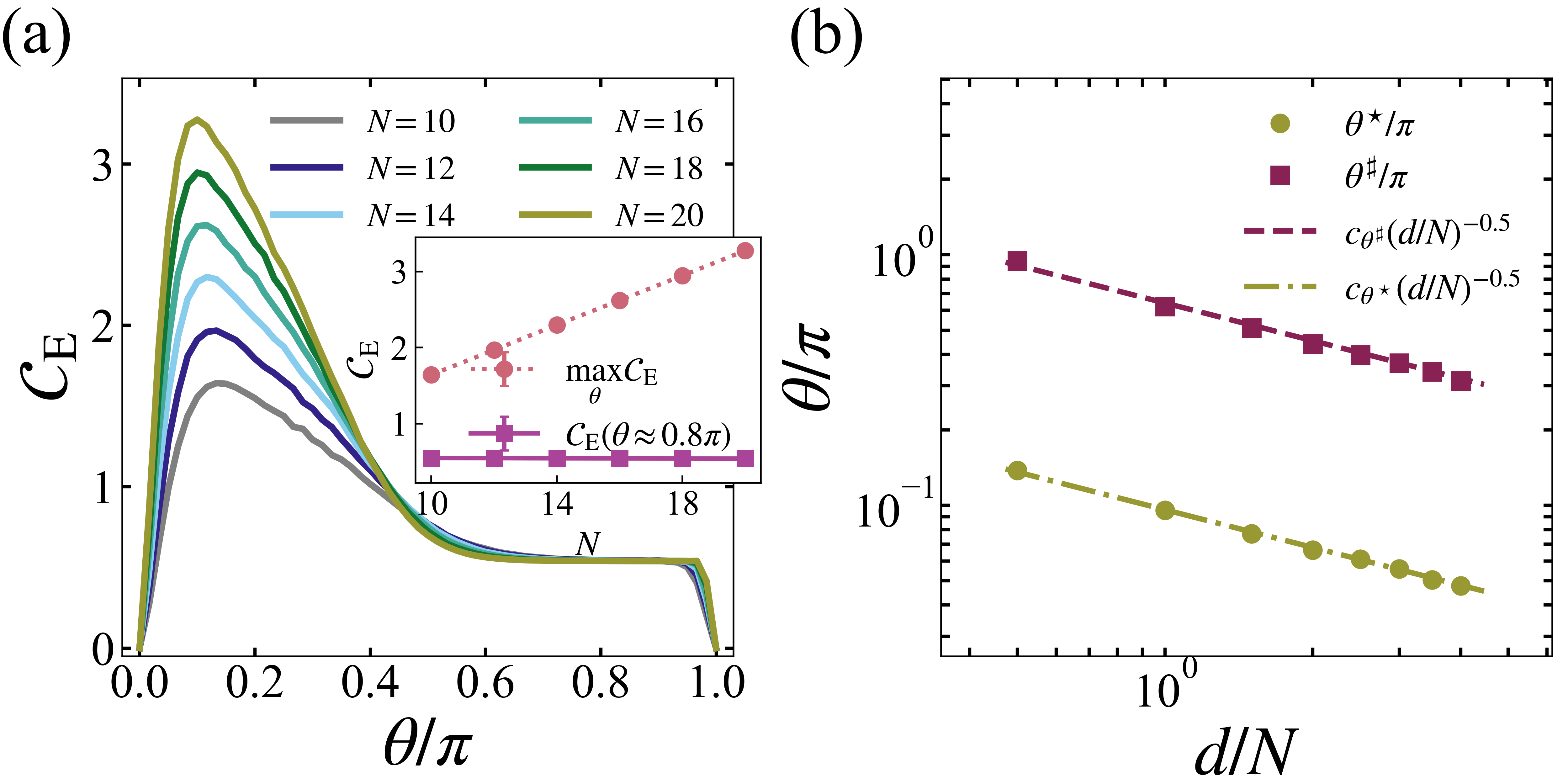}
    \includegraphics[width=0.99\linewidth]{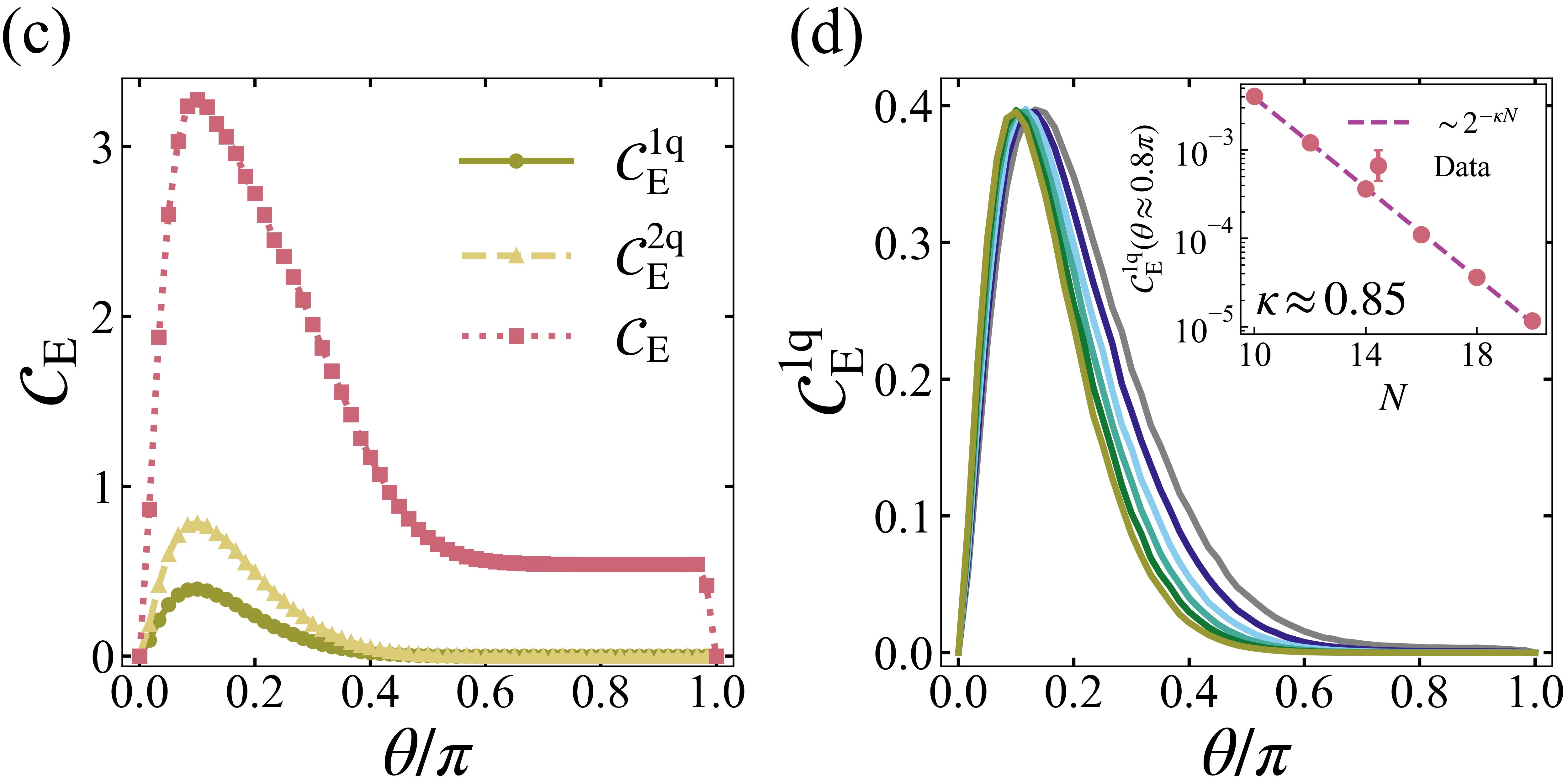}
    \includegraphics[width=0.99\linewidth]{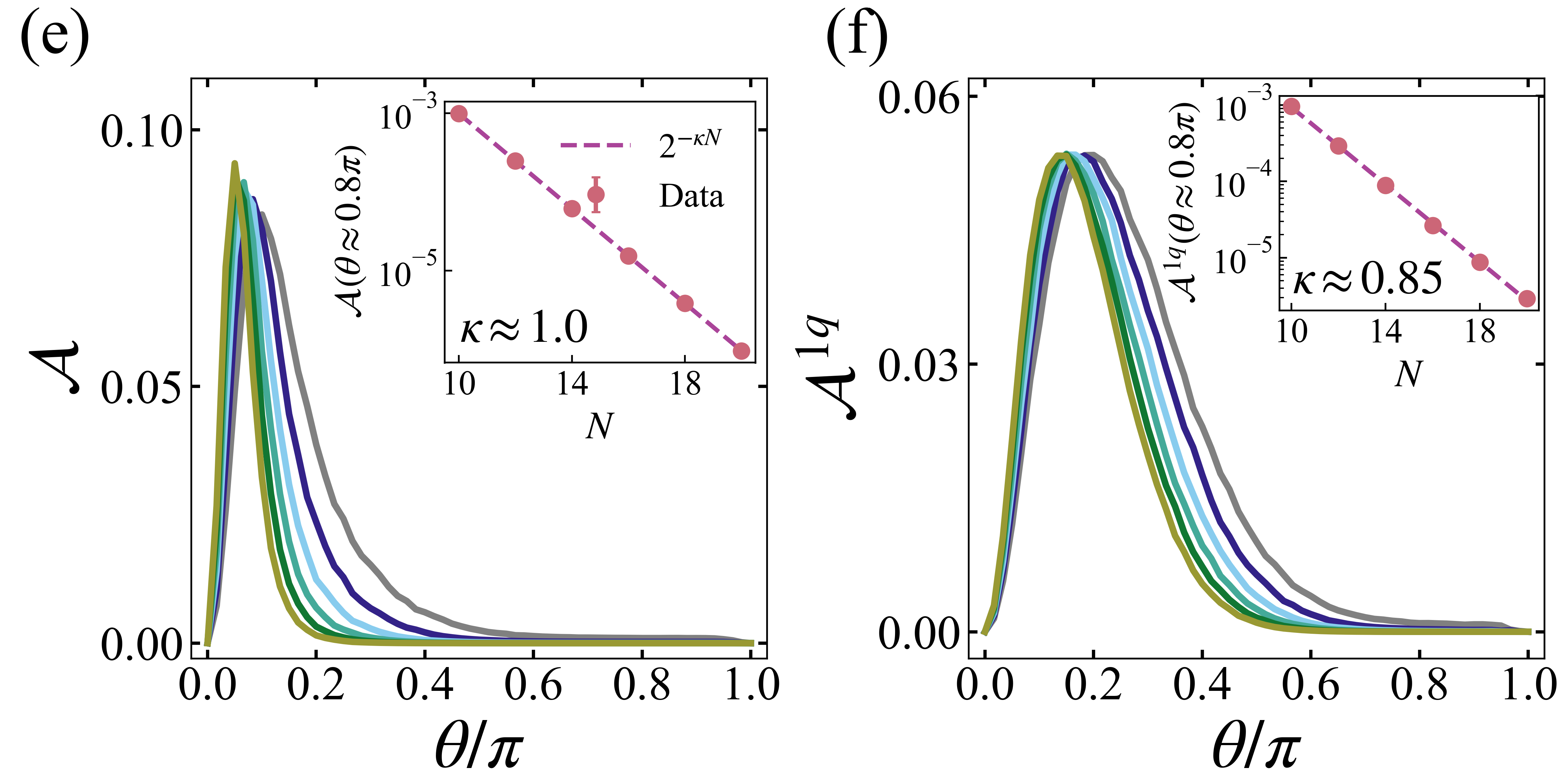}
\caption{\textbf{Spectral nonflatness.} \textbf{(a)} Growth and decay of the entanglement capacity \(\mathcal{C}_{\rm E}\), {\color{brown}Eq}.~\eqref{eq:CE_def}, for \(d/N\!=\!1\). For an equal bipartition, \(\mathcal{C}_{\rm E}\) reaches its maximum near \(\theta^{\star}/\pi\!\approx\!0.15\), with \( \mathcal{C}_{\rm E}^{\rm max} \!=\!  \max_{\theta}\mathbb{E}_{\mathcal U} \{ \mathcal{C}_{\rm E} \}\!\approx\!0.16N,\) as shown in the inset. For \(\theta>\theta^{\sharp}\), it saturates to the universal value \(\mathcal{C}_{\rm E}\!\approx\!0.539\). At the stabilizer point, \(\theta/\pi\!=\!1\), one finds \(\mathcal{C}_{\rm E}\!=\!0\). \textbf{(b)} The characteristic points \(\theta^{\star}\) and \(\theta^{\sharp}\) both exhibit an approximate power-law decay, \(\theta^{\star,\sharp}\!\propto\!(d/N)^{-1/2}\), with a constant ratio \(c_{\theta^{\sharp}}/c_{\theta^{\star}}\!\approx\!6.7\). Results are shown for \(N\!=\!20\). \textbf{(c)} Comparison between small and large subsystems for \((N,d/N)\!=\!(20,1)\). \textbf{(d)} For vanishingly small subsystems, \(\mathcal{C}_{\rm E}\) remains \(\mathcal{O}(1)\) throughout an intermediate regime before converging to its typical value. Inset shows that its saturation value in the chaotic regime is exponentially suppressed as \(2^{-\kappa N}\). \textbf{(e)} Antiflatness \(\mathcal{A}\) for an equal bipartition and \textbf{(f)} for a one-qubit subsystem. In both cases, the \(\mathcal{A}\) exhibits a finite intermediate-regime maximum followed by exponential suppression with \(N\) in the scrambling regime. Data are averages over 350--800 random circuit realizations and error bars are of the size of the symbols or smaller.}
\label{fig3:capacity_antiflatness}
\end{figure}

In the Haar-random state limit, \(\hat\rho_R\) is described by a normalized Wishart random matrix, whose eigenvalue density approaches the Marchenko--Pastur distribution~\cite{PhysRevD.99.066012}. In the limit \(d_R,d_{\bar R}\!\to\!\infty\) at fixed subsystem-size ratio \(\alpha=d_R/d_{\bar R}\), the Haar-typical capacity depends only on \(\alpha\), with an analytical expression given in Refs.~\cite{okuyama2021capacity,kawabata2021probing}. It is maximized for an equal bipartition \(\alpha\!=\!1\), where \(\mathcal C_{\rm E}^{\rm H}\approx0.539\). For comparison, the antiflatness \(\mathcal A:=\mathrm{Var}_{\hat\rho_R}(\hat\rho_R)=\mathrm{Tr}(\hat\rho_R^3)
\!-\!\mathrm{Tr}(\hat\rho_R^2)^2\) also probes spectral nonuniformity but in the typical limit it decays exponentially as \(2^{-N}\)~\cite{tnfv-lzfx}. It can be expressed in terms of the modular partition function as \(\mathcal A \!=\! \mathcal Z_{3}\!-\!\mathcal Z_{2}^2\), or equivalently, \(\mathcal A\!=\! e^{-2S_3}\!-\!e^{-2S_2}\). Note that \(\mathcal A \) is neither additive nor multiplicative. For a one-qubit subsystem, \(\mathcal A^{\max}_{\rm 1q}\!=\!1/16\), but this maximum as a function of Bloch sphere length occurs at a lower value than that of \(\mathcal{C}_{\rm E, 1q}^{\rm max}\!\approx\!0.439\). The former is controlled by low-order spectral moments and dominated by the largest Schmidt coefficients, whereas the latter is sensitive to the spread of the entanglement energies. Both nonflatness measures vanish for (subsystem) spectra that are flat on their support, \(\hat\rho_R\!=\!\hat\Pi_R/r_R\), where \(\hat\Pi_R\) is a rank-\(r_R\) projector and all nonzero eigenvalues equal \(1/r_R\), and \(\mathcal C_{\rm E},\mathcal A\!=\!0\) for product and stabilizer states~\cite{jasser2025stabilizer, y9r6-dx7p, viscardi2026interplay}. A hierarchy of discrete (logarithmic) antiflatness measures \(\mathcal A_{\beta\beta'}\) can also be constructed from the differences between different (or neighboring) R\'enyi entropies. While several conventional measures of spectral nonflatness are maximized by highly degenerate two-valued, or "jump", spectra, which represents an extreme spectral configuration, no single spectrum universally maximizes all notions of nonflatness. Moreover, for certain nonflatness measures, rigorous entropy-dependent upper bounds are available that vanish at maximal subsystem entropy, as required for maximally entangled states~\cite{PRXQuantum.3.010325,jasser2026journey,PhysRevD.99.066012}.

{\color{brown}Figure}~\ref{fig3:capacity_antiflatness} characterizes the behavior of the entanglement capacity and antiflatness in our model at a fixed evolution time $d\!\sim\!\mathcal{O}(N)$. $\mathcal{C}_{\rm E}$ is negligible near the low-rank product and stabilizer limits, but develops a pronounced intermediate peak where the spectrum is highly nonuniform and nondegenerate. For an equal bipartition, the height of this peak at $\theta^\star\!=\!\arg\max_{\theta} \mathbb{E}_{\mathcal U}\{ \mathcal C_{\rm E}(\theta) \}$ grows linearly with size, $\mathcal{C}_{\rm E}(\theta^{\star})\!\propto\!N$. This maximum appears near the point where the reduced density matrix is close to full rank while retaining a broad and highly structured distribution of entanglement energies~\cite{serbyn2016power}. Beyond this peak, $\mathcal{C}_{\rm E}$ decreases and approaches the typical value for $\pi\!>\!\theta\!>\!\theta^{\sharp}$, coinciding with the saturation of the entanglement entropy and the onset of the strongly scrambled regime. The crossover scale \(\theta^{\sharp}\) is determined numerically as the smallest value of \(\theta\) for which the chosen diagnostic lies within a fixed relative tolerance \(\varepsilon\) of its saturation value, and we set \(\varepsilon\!=\!0.02\). 

Further, as shown in {\color{brown}Figs.}~\ref{fig3:capacity_antiflatness} (c)(d), for vanishingly small subsystems $\mathcal{C}_{\rm E}$ remains $\mathcal{O}(1)$ in the intermediate regime, as expected from the finite Hilbert-space dimension of a local reduced density matrix. However, in this case, in the strongly scrambled regime it decreases exponentially with total system size \(2^{-\kappa N}\), consistent with the fact that local subsystems of Haar-typical states become exponentially close to maximally mixed; a feature also observed for other quantum resources~\cite{aditya2025growth}. Antiflatness \(\mathcal{A}\) also exhibits similar behaviors; although, as pointed out earlier, it is not additive under tensor product, and its maximum need not in general coincide with that of the \(\mathcal{C}_{\rm E}\). 

Before moving on, it is instructive to note a relation between nonlocal magic, \(\mathcal{M}^{\rm NL}(\ket{\psi})\!:=\!\min_{\hat U_R\otimes\hat U_{\bar R}}\mathcal{M}(\hat U_R\otimes\hat U_{\bar R}\ket{\psi})\), and nonflatness of the entanglement spectrum~\cite{z3vr-w5c5,ahmad2025experimental,torre2026non}. For Haar-random states, \(\mathcal{M}^{\rm NL}\) is controlled primarily by fluctuations about a flat spectrum and therefore remains \(\mathcal O(1)\) for a symmetric bipartition, while it is suppressed toward zero for strongly asymmetric cuts. Similar to nonflatness measures, volume-law entanglement alone is not sufficient to guarantee extensive nonlocal magic. Our preliminary numerics (not shown) indicate the same behavior in the present circuit model, closely paralleling the rise-and-fall behavior. This is consistent with \(\mathcal{M}^{\rm NL}\) capturing the nonstabilizerness that is irreducibly stored across a bipartition, and being controlled by the organization and correlations of the Schmidt spectrum~\cite{huang2026intrinsic}. 

In a nutshell, despite the large global complexity of the state, the stationary regime marks the point at which ``spectral complexity" is no longer locally consumable for computation. The most favorable regime is instead expected near the preceding maximum of spectral nonflatness, where the reduced state retains the richest structure before relaxing toward the typical form. Remarkably, as we show next, in the local limit the subsystem quantum Fisher information and its components can exhibit closely related qualitative behavior, strongly suggesting that the same intermediate regime can simultaneously maximize spectral structure and parameter sensitivity.

\subsection{Metrological sensitivity}
\subsubsection{Quantum Fisher information and its decomposition}

The spectral nonflatness discussed above quantifies how much structure is present in the eigenvalue distribution of a reduced state. An important complementary question is how fast can such a reduced state change under an infinitesimal deformation of some dynamical parameters~\cite{RevModPhys.89.035002, v7mf-yh8n, PhysRevLett.110.050403, PhysRevLett.110.050402}. We denote this deformation parameter by \(\phi\), and in our study it is a temporal input deformation. In general, the parameter $\theta$ is an independent control parameter, yet in certain cases one may take it as the probed deformation also. The resulting state-space displacement is naturally quantified through the squared Uhlmann fidelity \(\mathcal X(\hat\rho,\hat\rho') \!=\! [ \Tr\sqrt{\sqrt{\hat\rho}\hat\rho'\sqrt{\hat\rho}}
]^2\), \( \mathcal X(\hat\rho_R(\phi) \hat\rho_R(\phi\!+\!d\phi)) \!=\! 1\!-\frac{1}{4}\mathcal F_{\rm Q}^{\phi}d\phi^2\!+\!\mathcal O(d\phi^3), \) up to convention-dependent factors. Thus, \(\mathcal F_{\rm Q}^{\phi}\) is proportional to the fidelity susceptibility \(\partial^2_{\phi}\mathcal{X}\) of the reduced state with respect to a chosen deformation~\cite{gu2010fidelity}. Formally, it is defined through the symmetric logarithmic derivative \(\hat{\mathcal L}_\phi\) as \(\partial_\phi\hat\rho_R\!=\!\frac{1}{2}\{\hat\rho_R,\hat{\mathcal L}_\phi\}\) and \(\mathcal F_{\rm Q}^{\phi}\!:=\!\Tr[\hat\rho_R\hat{\mathcal L}_\phi^2]\), which sets the maximum classical Fisher information obtainable from measurements on subsystem \(R\). For pure states, the fidelity susceptibility coincides with the quantum metric, given by the real part of the quantum geometric tensor~\cite{ProvostVallee1980, liu2020quantum}. It therefore determines the ultimate precision with which \(\phi\) can, in principle, be estimated from that subsystem. In the eigenbasis \(\hat\rho_R\!=\!\sum_n \xi_n\ket{n}\bra{n}\), one can write
\begin{align}
\mathcal F^{\phi}_{\rm Q} = 2\sum_{m,n}
\frac{ |\bra{m}\partial_\phi \hat\rho_R\ket{n}|^2}{ \xi_m+\xi_n },\quad \xi_m+\xi_n>0.\end{align}

The metrological response contains both incoherent and coherent contributions \(\mathcal F^{\phi}_{\rm Q}\!=\!\mathcal F_{\rm inc}\!+\!\mathcal F_{\rm coh}\). The diagonal part, 
\begin{align}
\mathcal F_{\rm inc}^{\phi}\!=\!\sum_n \frac{(\partial_\phi \xi_n)^2}{{\xi_n}},
\label{eq:fisher_inc}
\end{align}
quantifies changes in the eigenvalues (population) of the reduced state, and is exactly the classical Fisher information obtained by measuring in the instantaneous eigenbasis of \(\hat{\rho}_R\). The off-diagonal part,
\begin{align}
\mathcal F_{\rm coh}^{\phi} = 4\sum_{m<n}
\frac{(\xi_m-\xi_n)^2}{\xi_m+\xi_n}
|\braket{m}{\partial_\phi n}|^2,
\label{eq:fisher_coh}
\end{align}
quantifies rotations of the eigenbasis and is therefore the coherent, basis-changing contribution to distinguishability. It captures distinguishable motion associated with quantum coherences between instantaneous eigenvectors, which can be visible to suitable observables even when entropy-like spectral quantities remain nearly unchanged~\cite{tang2025estimating}. This decomposition is useful in the present setting as a subsystem may remain highly sensitive through coherent eigenbasis motion while exhibiting little first-order spectral response, as can occur for flat-on-support stabilizer reduced states or Haar-typical extensive subsystems. Conversely, a subsystem may retain substantial spectral structure while responding only weakly to a temporal deformation, for example when information about earlier encoded inputs has largely decayed from the readout state. The coherent-incoherent separation therefore differentiates the existence of spectral structure from its sensitivity to the encoded parameter. As mentioned before, the operational relevance of \(\mathcal{F}^{\phi}_{\rm Q}\) follows from generalized Cram\'er--Rao-type response bounds, whose tightness depends on how strongly the chosen observable couples to the coherent and incoherent components of the state motion~\cite{PhysRevX.12.011038}. 
\begin{figure*}[ht]
    \centering
    \includegraphics[width=0.99\linewidth]{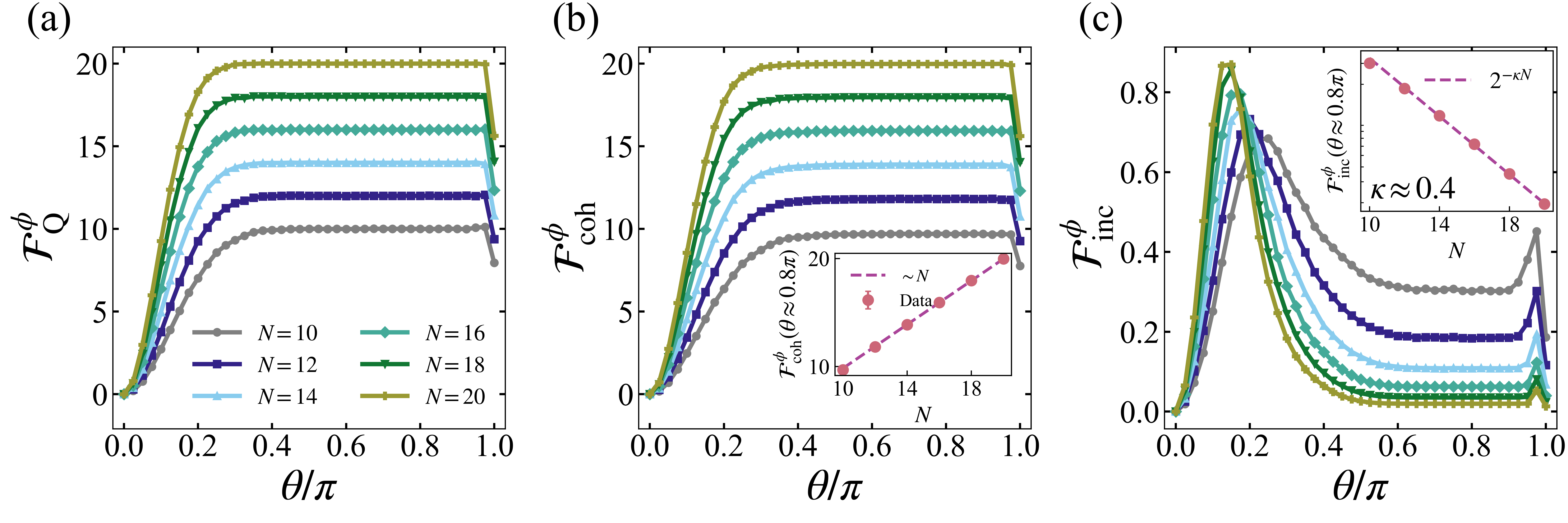}
\includegraphics[width=0.99\linewidth]{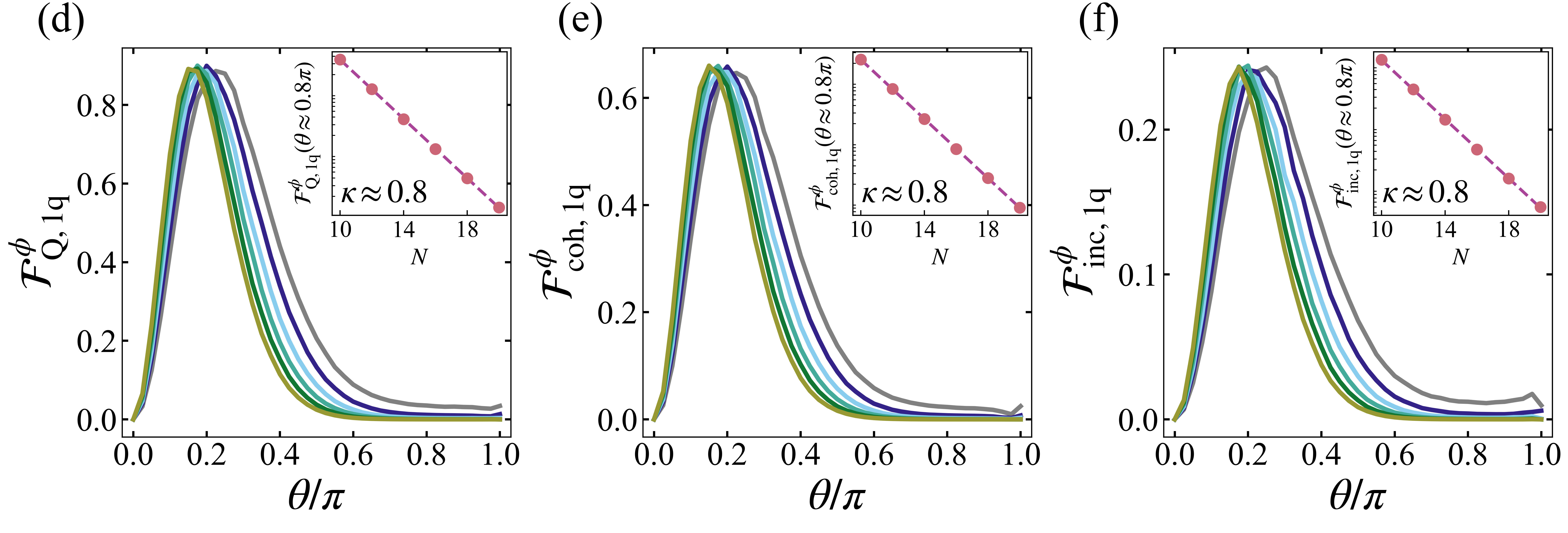}
\caption{\textbf{Metrological response.} \textbf{(a)} Quantum Fisher information, and its \textbf{(b)} coherent, and \textbf{(c)} incoherent parts for equal bipartitioning and \(d/N\!=\!1\). In the volume-law regime, the readout subsystem can retain a finite fraction of the input sensitivity initially encoded in the memory qubits, leading to a quantum Fisher information that grows linearly with system size \(\propto N\). Deep in the chaotic phase, however, this sensitivity is carried almost entirely by coherent eigenbasis rotations, while the incoherent contribution associated with spectral deformation is exponentially suppressed as \(2^{-\kappa N}\). \textbf{(d)} For small readout subsystems, however, all components of the subsystem quantum Fisher information, shown in \textbf{(e)} and \textbf{(f)}, remain at most \(\mathcal{O}(1)\) and exhibit a rise-peak-fall behavior, peaking in the intermediate regime before becoming exponentially suppressed under deep scrambling. The input derivatives are evaluated numerically by a finite-difference scheme around \(\phi=0\), using a perturbation amplitude \(\delta\phi=10^{-3}\). Data are an average over 2000 independent realizations. For local probes, we additionally average over subsystems to suppress site-specific fluctuations.}
\label{fig:fisherinfo}
\end{figure*}

We encode the inputs into an initially pure product-state $\ket{0}^{\otimes{N}}$ by applying \(\hat R^y(\phi)\) rotations on \(|\bar R|\!=\!N/2\) memory qubits, with the encoding generator \(\mathcal{\hat {G}}_\phi\!=\!(1/2)\sum_{j\in \bar R}\hat Y_j\). The pure state initial Fisher information is therefore \(\mathcal F_{\rm Q}^{\rm glob}\!\propto\!\mathrm{Var}({\mathcal{\hat {G}}}_\phi)\) and the globally encoded sensitivity scales as \(\mathcal F_{\rm Q}^\phi\!\propto\! N\). Importantly, in our model this also fixes the ceiling for the scaling that can be recovered by the complementary subsystem after \(\theta\)-dependent scrambling. Sufficiently large complementary subsystem can retain an \(\mathcal O(N)\) information, whereas a vanishingly small subsystem can access only an exponentially suppressed fraction in the Haar-scrambled limit~\cite{wysocki2026volume, tang2025estimating}. An extensive growth of recoverable input sensitivity therefore generally requires the signal to be encoded across an extensive number of qubits, rather than on a fixed local subset. {\color{brown} Figure}~\ref{fig:fisherinfo} summarizes such behaviors. For an equal bipartition and in the deeply chaotic regime, the coherent contribution remains extensive and linear in $N$, while the incoherent contribution becomes exponentially small. The total \(\mathcal F_{\rm Q}^{\phi}\) in this limit is therefore dominated by eigenbasis rotations, while spectral quantities such as the entanglement entropy, purity, and associated nonflatness measures have already approached their stationary random-state values. This highlights that strong subsystem distinguishability can persist entirely through coherent motion even after spectrally useful structure is exhausted. For vanishingly small subsystems, by contrast, \(\mathcal F_{\rm Q}^{\phi}\) develops a finite \(\mathcal O(1)\) maximum at intermediate interaction strength $\theta$, with both contributions becoming relevant and displaying a qualitatively similar rise-peak-fall behavior. These closely parallel the behavior of the spectral nonflatness measures in small subsystems. 

\subsubsection{Entanglement response}
\begin{figure}[t]
    \centering
    \includegraphics[width=0.99\linewidth]{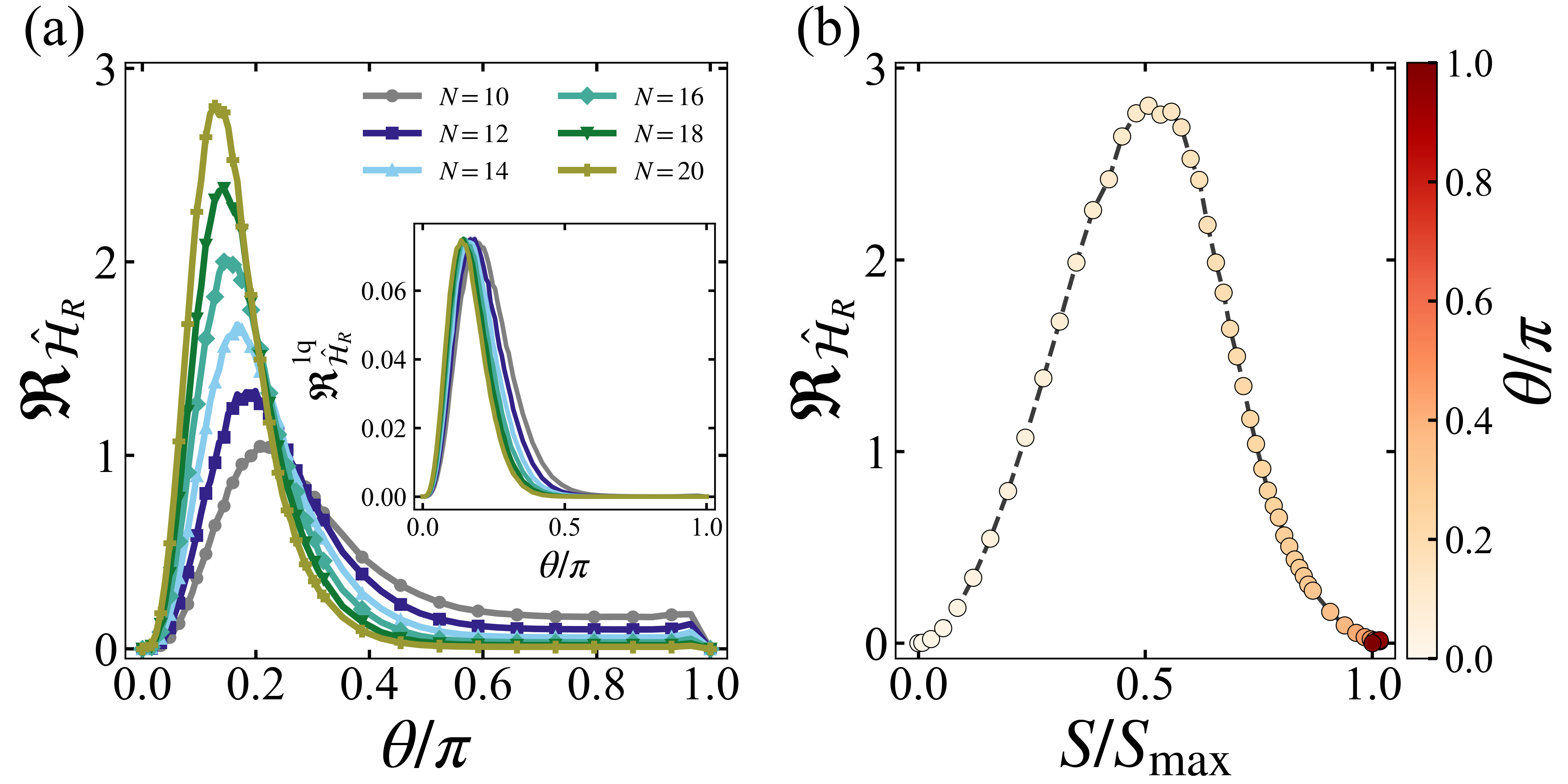}
    \caption{\textbf{Entanglement response.} \textbf{(a)} $\mathfrak R_{\hat{\mathcal{H}}_R}\!=\!\mathcal F_{\rm {inc}}^{\phi}\,\mathcal C_{\rm E}$ as a function of $\theta$, for \(d/N\!=\!1\) and equal bipartitioning. Inset shows the response for one qubit subsystems. \textbf{(b)} \(\mathfrak R_{\hat{\mathcal{H}}_R}\) as a function of rescaled entanglement entropy \(S/S_{\max}\), for $(N,d/N)=(20,1)$. Markedly, the entanglement response to input deformation is maximal when the entanglement itself is submaximal. Data are an average over \(2800\) independent circuit realizations. Ensemble averaging is performed on the realization-wise product \(\mathcal F_{\rm inc}^{\phi}\mathcal C_{\rm E}\).}
    \label{fig:ent_resp}
\end{figure}

To complete our analysis, let us consider the response of the state-dependent spectral functional $\hat{\mathcal H}_R\!=\!-\log\hat\rho_R$. Noting \(\langle\partial_{\phi}\hat{\mathcal{H}}_{R}\rangle\!=\!0\), {\color{brown}{Eq.}}~\eqref{eq:gen_cramer_Rao} gives \(\left|\partial_\phi S_{\rm vN}\right|^2\!\leq\! \mathfrak R_{\hat{\mathcal{H}}_R}\), where $\mathfrak R_{\hat{\mathcal{H}}_R}\!=\!\mathcal F_{\rm {inc}}^{\phi}\,\mathcal C_{\rm E}$. Since $\hat{\mathcal H}_R$ is diagonal in the instantaneous eigenbasis, only the incoherent channel contributes. The first-order response of the von Neumann entanglement entropy is therefore tightly constrained by the product of entanglement capacity and the incoherent part of metrological sensitivity. Plotted in {\color{brown}Fig.}~\ref{fig:ent_resp}(a), one can see that the response product displays the familiar rise–peak–fall structure as a function of \(\theta\). This behavior identifies the regime in which the available spectral resources are maximally susceptible to changes in the inputs \(\phi\). Across an extensive bipartition, in the intermediate regime the response capacity also grows extensively with \(N\). Remarkably, as can be seen from {\color{brown}Fig.}~\ref{fig:ent_resp}(b), this peak occurs at around half the maximum entanglement entropy, providing a more precise characterization of the intermediate regime as the point where entanglement response is strongest, before deep scrambling drives the reduced state toward a typical structure. This regime, for which local distinguishability and spectral structure remain simultaneously finite and maximal, is therefore expected to coincide with a dynamical regime of enhanced nonlinear information processing capacity accompanied by relatively long memory. We establish this next.

\subsection{Quantum reservoir learning}
We now present an application of our results in a postvariational quantum learning setting. Quantum reservoir computing~\cite{PhysRevApplied.8.024030} provides a natural playground in which to study the relation between quantum resources and the intrinsic learnability and memory of quantum systems. In this framework, the main interest lies in a fixed quantum dynamical substrate whose evolution maps input signals into a high-dimensional feature space, while only a simple readout layer is trained~\cite{mujal2021opportunities, ding2026thermodynamics, ricci2026quantum, r8ww-qw7j}. This closely parallels the broader logic of reservoir and neuromorphic computing, where useful computation emerges from the internal dynamics of a complex physical system rather than from deep optimization of all microscopic parameters. From this perspective, quantum reservoirs may also be viewed as brain-inspired processing devices, in which memory, nonlinearity, and rich internal dynamics play roles analogous to those of recurrent activity in biological or neuromorphic architectures~\cite{markovic2020quantum}. A key question is therefore which physical resources of the underlying quantum dynamics, rather than explicit parameter optimization, govern memory retention, nonlinear feature generation, and ultimately computational performance. This makes the framework particularly well suited for isolating the resources that are genuinely useful for various information processing tasks.

The construction of the reservoir from the introduced circuit model is detailed in {\color{brown}Appendix}~\ref{app:reservoir_protocol}. In a suitable dynamical regime, the reservoir enjoys a strictly contractive map and convergence toward an input-driven trajectory, thanks to measure and reset operations. The joint state evolves as
\begin{align}
\hat\rho^{\rm}_{R\bar R}(t)=\hat{\mathcal U}(\theta)
\Bigl[\ket{0}\!\bra{0}_{R}\otimes\hat\varrho_{\bar R}(t)
\Bigr]\hat{\mathcal U}^{\dagger}(\theta),
\end{align}
where \(\hat\varrho_{\bar R}(t)\!=\!\hat{\mathcal V}_{\bar R}(\phi_t)\hat\rho_{\bar R}(t)\hat{\mathcal V}_{\bar R}^\dagger(\phi_t)\), and \({\hat{\mathcal{V}}}_{\bar R}(\phi_t)\!=\! \otimes_{j\in\bar R}\hat{R}_j^y(\phi_t)\), and \(\hat\rho_{\bar R}(0)\!=\!\ket{0}\!\bra{0}_{\bar R}\). Here, \(\hat{\mathcal U}(\theta)\!=\!\prod_i^d \hat{\mathcal{U}}_{\rm F} (\theta)\) represents the circuit unitary with \(d\!\sim\!\mathcal{O}(N)\). The updated memory is \(\hat\rho_{\bar R}(t+1)\!=\!\Tr_{R}\hat\rho_{R\bar R}(t)\). In this way, it acts as a (dissipative) driven manybody medium whose recurrent internal dynamical evolution accurately stores a finite number of past inputs, mixes them, and maps them into a high-dimensional feature space. As opposed to the more conventional approaches~\cite{PhysRevApplied.8.024030}, here the reset operation is input independent. The unitary quantum evolution is linear in the density operator, yet the observable features are generally nonlinear function of the classical encoding angle. Encoding the inputs through rotations on an extensive number of qubits allows the eigenvalue-difference spectrum of the encoding generator, and hence the available instantaneous Fourier bandwidth, to grow extensively with system size, rather than remaining bounded by a fixed local encoding dimension~\cite{schuld2021effect, schutte2025expressivity, mujal2021analytical,xia2026quantum}. Numerically, we simulate the reservoir protocol exactly via the Kraus representation of a quantum channel, and evaluate the expectation values of all $\hat Z$-type Pauli strings supported on the readout subsystem directly. These constitute the feature space and operational objects of learning. Higher-order \(\hat Z\)-strings are experimentally accessible within the same measurement basis. Although the total number of collected features grows exponentially, \(2^{N/2}-1\), the additional contributions from basis-restricted higher-order strings is expected to become marginal deeper in the chaotic regime. There fewbody and high-weight correlators alike concentrate toward their typical values, and scrambling generates volume-law operator entanglement and distributes operator weight over exponentially many Pauli string components~\cite{c7k1-xcwy,PhysRevLett.131.180403}, leaving an exponentially small overlap with the prescribed strings. The resulting standardized and time-dependent feature vectors are then used for Ridge linear regression to determine the optimal output weights, which are subsequently tested on an unseen data stream. In practice no direct knowledge of the full quantum state is required; the protocol only relies on experimentally accessible expectation values. The computational power of the reservoir is therefore expressed operationally in the measured observables rather than in an explicit reconstruction of a manybody state. 

\subsubsection{Classical Fisher information}
A reservoir processes a history of inputs, so the relevant state manifold is multiparametric and indexed by present and past input signal values. The corresponding classical Fisher-information matrices quantify which temporal directions remain distinguishable in the final readout state and measurements distribution. A natural extension is therefore to resolve the metrological response across a finite input history, which is also a measure of (linear) memory and temporal distance.
\begin{figure}[t]
    \centering
    \includegraphics[width=0.99\linewidth]{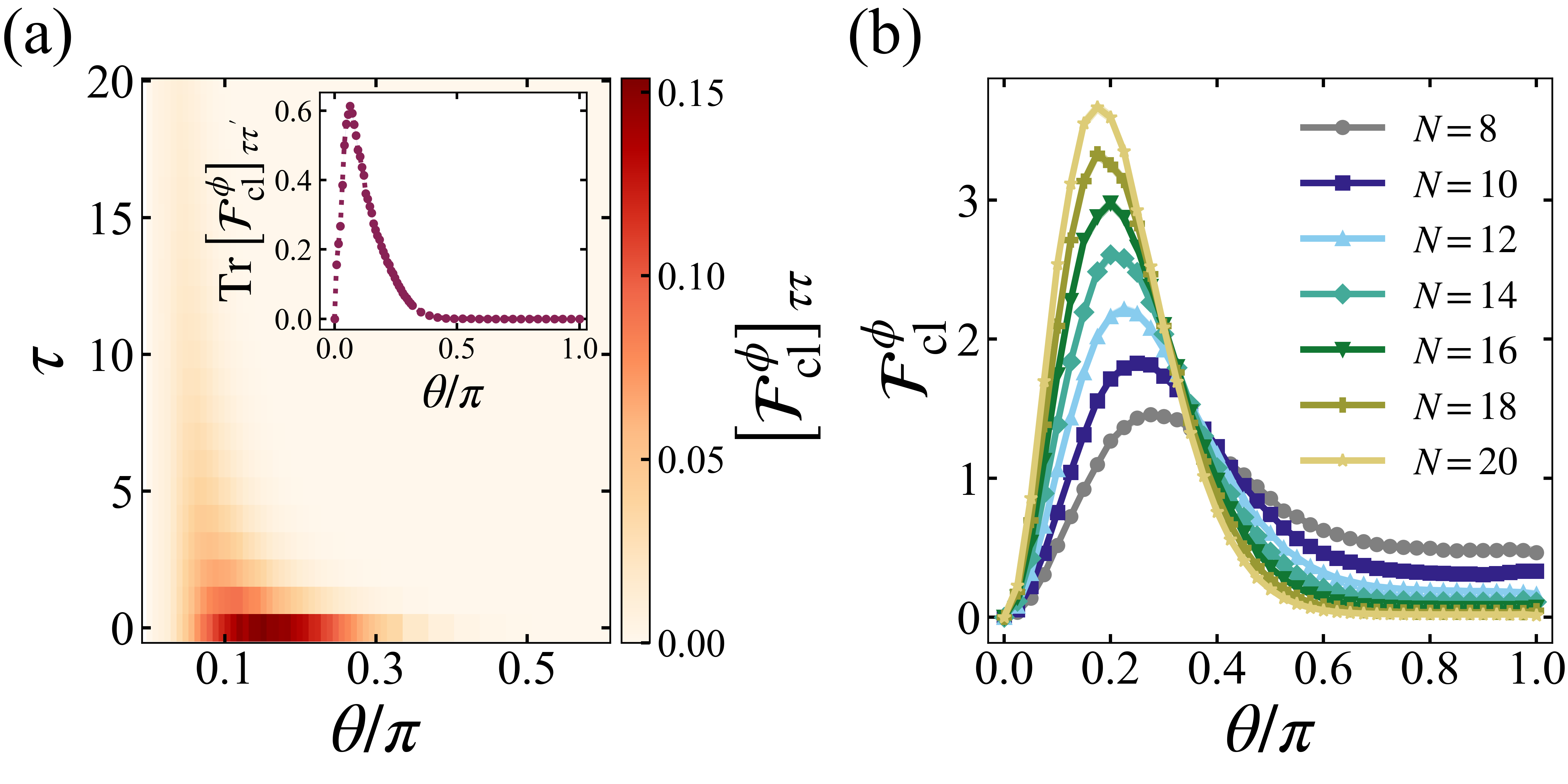}
    \caption{\textbf{Classical Fisher information.} \textbf{(a)} History-resolved temporal Fisher information, given by the diagonal elements of the Fisher-information matrix in {\color{brown}Eq}.~\eqref{eq:clsscl_fish}, which serves as a proxy for linear memory. Its decay with the delay \(\tau\) quantifies the gradual loss of sensitivity to past inputs, a central requirement for effective reservoir computing; larger \(\tau\) corresponds to inputs further in the past. The inset shows \(\Tr\mathcal{F}^{\phi}_{\rm cl}\!=\!\sum_{\tau}[\mathcal{F}_{\rm cl}^{\phi}]_{\tau\tau}\), which provides a measure of the total temporally accessible memory. The Fisher information is evaluated after a washout period to remove dependence on the initial state. Results are shown for \((N,d/N)\!=\!(14,1)\). \textbf{(b)} Static classical Fisher information associated with computational-basis measurements for \(d/N\!=\!1\). It is maximal in the intermediate regime, near \(\theta^{\star}/\pi\approx 0.15\!-\!0.25\), where the output probability distribution is most sensitive to variations of the encoded parameter. The peak grows approximately as \(\max_{\theta}\mathcal{F}^{\phi}_{\rm cl}\!\propto\! N,\) indicating enhanced measurement-accessible distinguishability for larger \(N\). At larger \(\theta\), the Fisher information decreases toward exponentially small values, and the measurement distribution becomes increasingly insensitive to input variations. Data are averages over \(2800\) independent circuit realizations.}
    \label{fig:temp_fish}
\end{figure}
The readout state defines a parametrized manifold of density operators
\(\hat\rho_R(\theta;\boldsymbol{\phi})\), where
\(\boldsymbol{\phi}\!=\!(\phi_1,\ldots,\phi_t)\) denotes the input history vector for \(t\) iterations of the reservoir. Infinitesimal perturbations of the input history induce displacements on this quantum-state manifold. With the Bures metric \(ds_B^2\!=\!\frac{1}{4}\sum_{\tau\tau'}[\mathcal{F}^{\boldsymbol{\phi}}_{\rm Q}]_{\tau\tau'}d\phi_\tau d\phi_{\tau'}\), the matrix \([\mathcal{F}^{\boldsymbol{\phi}}_{\rm Q}]_{\tau\tau'}\) quantifies the multiparameter temporal distinguishability of reservoir states generated by perturbing the input-history components of \(\boldsymbol{\phi}\). Here the indices \(\tau\) and \(\tau'\) label input times within the history vector. A measurement maps the quantum-state manifold to a classical probability manifold. For projective measurements in the computational basis of subsystem \(R\), with the positive operator-valued measure \(\hat{\mathcal{K}}_{\mathbf z}\!=\!|\mathbf z\rangle\langle\mathbf z|\), \(\mathbf z\!\in\!\{0,1\}^{|R|}\), Born's rule gives
\(p_{\mathbf z}^{\boldsymbol{\phi}}\!=\!\operatorname{Tr}[\hat\rho_R\hat{\mathcal{K}}_{\mathbf z}]\), and the associated classical Fisher information matrix is~\cite{meyer2021fisher}
\begin{equation}
    \left[\mathcal{F}_{\rm cl}^{\boldsymbol{\phi}}\right]_{\tau\tau'}
    =
    \sum_{{\mathbf z}:p_{\mathbf z}>0}
    \frac{
        \left(\partial_{\phi_{\tau}} p_{\mathbf z}^{\boldsymbol{\phi}}\right)
        \left(\partial_{\phi_{\tau'}} p_{\mathbf z}^{\boldsymbol{\phi}}\right)
    }{
        p_{\mathbf z}^{\boldsymbol{\phi}}
    }.
    \label{eq:clsscl_fish}
\end{equation}
The classical Fisher information from \(\hat Z\)-basis measurements naturally receives contributions from both eigenvalue changes and eigenbasis rotations. The diagonal elements of the temporal Fisher matrix relate to the intrinsic local distinguishability of past inputs before specifying a particular classical memory estimator or regression task, with its trace a proxy for the qualitative behavior of total memory. Its off-diagonal elements quantify overlap or redundancy between the effects of different input times.

As mentioned, the influence of earlier inputs is progressively contracted by the quantum map (here via measure and reset), inducing fading memory and the echo-state behavior~\cite{PhysRevApplied.8.024030}. The finiteness of memory and a unique input-driven stationary trajectory are crucial to the success of reservoir computing. Under stationary fading-memory dynamics, one expects \([\mathcal F_{\rm cl}^{\boldsymbol{\phi}}]_{\tau\tau}\lesssim [\mathcal F_{\rm cl}^{\boldsymbol{\phi}}]_{\tau'\tau'}\) for \(\tau\!>\!\tau'\), so that older input deformations become progressively less distinguishable at the final readout. As shown in {\color{brown}Fig.}~\ref{fig:temp_fish}(a), temporal sensitivity is maximized at intermediate values of \(\theta\), followed by a fast decay. The fall of the Fisher information associated with an earlier input increases the minimum attainable error in reconstructing that input; long memory and linear information processing capacity therefore directly connect to the persistence of nonvanishing temporal directions at large delays. Further, shown in {\color{brown}Fig.}~\ref{fig:temp_fish}(b), the scaling of the static \(\mathcal F^{\phi}_{\rm cl}\) also reveals whether measurement-accessible input sensitivity is enhanced or suppressed as the reservoir size increases. Larger reservoirs receive a larger input-information budget and a larger classical readout model. The scaling trends are central to assessing whether larger systems may yield genuine improvements in parameter estimation and learnability, or whether those finite-size gains are ultimately lost under deep scrambling. For a Pauli measurement with binary outputs, the alignment factor \(\cos^{2}\gamma_{{\hat P}_R}\!=\!\mathcal{F}^{\phi}_{\rm cl}/\mathcal{F}_{\rm Q}^{\phi}\) has a direct measurement-theoretic interpretation and is precisely the fraction of subsystem quantum Fisher information resolved by a chosen Pauli measurement (see {\color{brown}Appendix}~\ref{app:single_site_Response}). Together, these quantities offer a direct and operationally motivated diagnostic of the expected reservoir functionality, identifying the regime in which the subsystem complexity structure, characterized earlier, may become both computationally useful and operationally accessible. Although these are key ingredients of maximal learnability, they alone do not necessarily guarantee a high nonlinear computational performance. Nevertheless, when the circuit design, including the input encoding arrangement, allows the available spectral and metrological structure to be expressed through the measured features, the learning performance is expected to follow the same qualitative dependence on the control parameter. In the next section we provide direct evidence for this correspondence.

\subsubsection{Information processing capacity} To characterize the intrinsic computational power of the reservoir independently of any particular benchmark, we adopt the information-processing capacity framework introduced for general dynamical systems~\cite{dambre2012information, martinez2023information, vcindrak2026memory,84f3-63mz}. The central idea is to assess how accurately the measured reservoir features can reconstruct a complete family of mutually orthogonal functions of the input history with controlled nonlinear order. Since the trained readout is linear in the measured features, nonlinear dependence on the classical input history originates from the input-dependent encoding channel and from its repeated, generally noncommuting composition with the fixed memory channel. The detailed construction of the orthogonal targets, finite-data estimator, regression protocol, and shuffled-target significance threshold is given in {\color{brown}Appendix}~\ref{app:ipc}. In short, the first order target quantifies linear memory, whereas higher-order targets systematically probe nonlinear transformations and combinations of past inputs. The capacity associated with each target is determined by the squared correlation between the target and its prediction from the trained linear readout, and the total information-processing capacity is bounded by the effective rank of the accessible feature space.
\begin{figure}[t]
    \centering
    \includegraphics[width=0.99\linewidth]{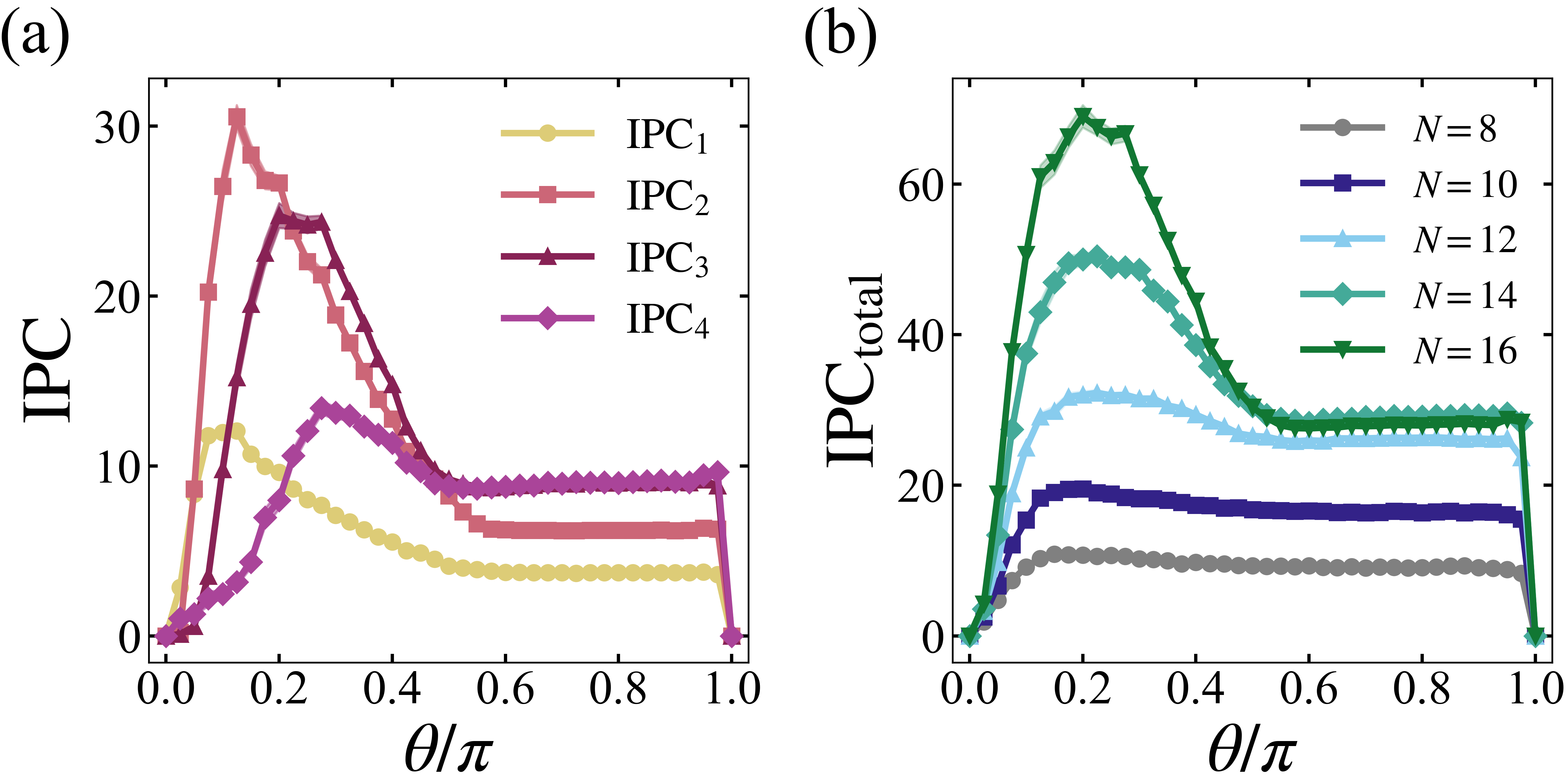}
    \caption{\textbf{Information processing capacity.} \textbf{(a)} Higher-order capacities quantify increasingly nonlinear transformations of the input history at a fixed maximum delay, \(\tau_{\max}=30\). As the nonlinear order increases, the capacity maximum shifts toward the more strongly interacting, thermalizing side of the intermediate regime. \textbf{(b)} The total capacity is obtained by summing over all the calculated nonlinear orders, and may be interpreted as the effective number of mutually orthogonal functions of the input history that are linearly accessible from the measured reservoir features. It is enhanced in the intermediate regime, where it consistently improves as the system size and the number of trainable features increase. We discard the first 500 input steps as a washout period to suppress dependence on the initial state, and divide the remaining data into training and test sets using a \(0.6/0.4\) split. The full input sequence contains 1500 randomly generated inputs \(\phi\!\in\![-1,1]\times10^{-1}\). Data are an average over 20 independent realizations, and standard error bars are negligible.}
    \label{fig:ipc}
\end{figure}

{\color{brown}Figure}~\ref{fig:ipc} shows that the different capacity orders exhibit the same qualitative rise-and-fall structure as the spectral and metrological diagnostics. The optimum depends on nonlinear order. Linear memory is generally favored closer to the weakly interacting regime, where past inputs remain directly recoverable, whereas higher-order capacities peak deeper in the intermediate regime, where stronger manybody mixing and a higher dimensional feature space produces useful nonlinear combinations of the input history~\cite{vcindrak2026memory}, reflecting a trade-off between memory retention and nonlinear processing. Remarkably, in the optimal regime the total information-processing capacity consistently grows with the system size, indicating that larger reservoirs may support an increasing number of independently reconstructible temporal functions. This strongly suggests that our reservoir construction supports a scalable learning gain in the intermediate regime.

\subsubsection{Benchmarks}
We now evaluate the performance of the reservoir using the common short-term memory and nonlinear autoregressive moving-average (NARMA) benchmarks, which probe complementary aspects of temporal information processing. In the linear short-term memory task, the target is a delayed input, \(y_t^{\tau}\!=\!\phi_{t-\tau}\!\in\![0,1]\). The performance at delay \(\tau\) is quantified by the squared Pearson correlation coefficient \(\mathfrak{c}(\tau)\!=\!\texttt{cov}^{2}(\phi_{t-\tau},\tilde y_t^{\tau})/[{\texttt{var}(\phi_{t-\tau}) \texttt{var}(\tilde y_t^{\tau})]}\), where \(\tilde y^{\tau}_t\) is the reservoir prediction. For linear (or memory-dominated) tasks, it is useful to rescale the input amplitude to a relatively small value, so that the encoded perturbations remain approximately within the linear-response regime~\cite{PRXQuantum.5.040325, ivaki2025dynamical}. More generally, the input strength can affect both the degree of input--output nonlinearity and the magnitude of the initial perturbation applied to the dynamics. Its optimal value is in general task dependent and should be tuned according to the desired balance between memory retention and nonlinear processing. The order-$\tau$ NARMA task combines long-term memory with nonlinear processing and is defined recursively as\(y_{t+1}\!=\!\alpha y_t\!+\!\beta y_t\sum_{j\!=\!0}^{\tau\!-\!1}y_{t-j}\!+\!\gamma \phi_{t-\tau+1}\phi_t\!+\!\delta,\) where the standard NARMA-$\tau$ choice is often \((\alpha,\beta,\gamma,\delta)=(0.3, 0.05,1.5,0.1)\). Prediction accuracy is quantified using the normalized root-mean-square error \(\mathrm{NRMSE} \!=\!\sqrt{\sum_t\left(y_t-\tilde y_t\right)^2\!/\!\sum_t\left(y_t-\bar y\right)^2},\) where \(\bar y\) is the mean target value. Smaller values of \(\mathrm{NRMSE}\) correspond to better prediction, while values near unity indicate performance comparable to predicting the target mean.

\begin{figure}[t]
    \centering
    \includegraphics[width=0.99\linewidth]{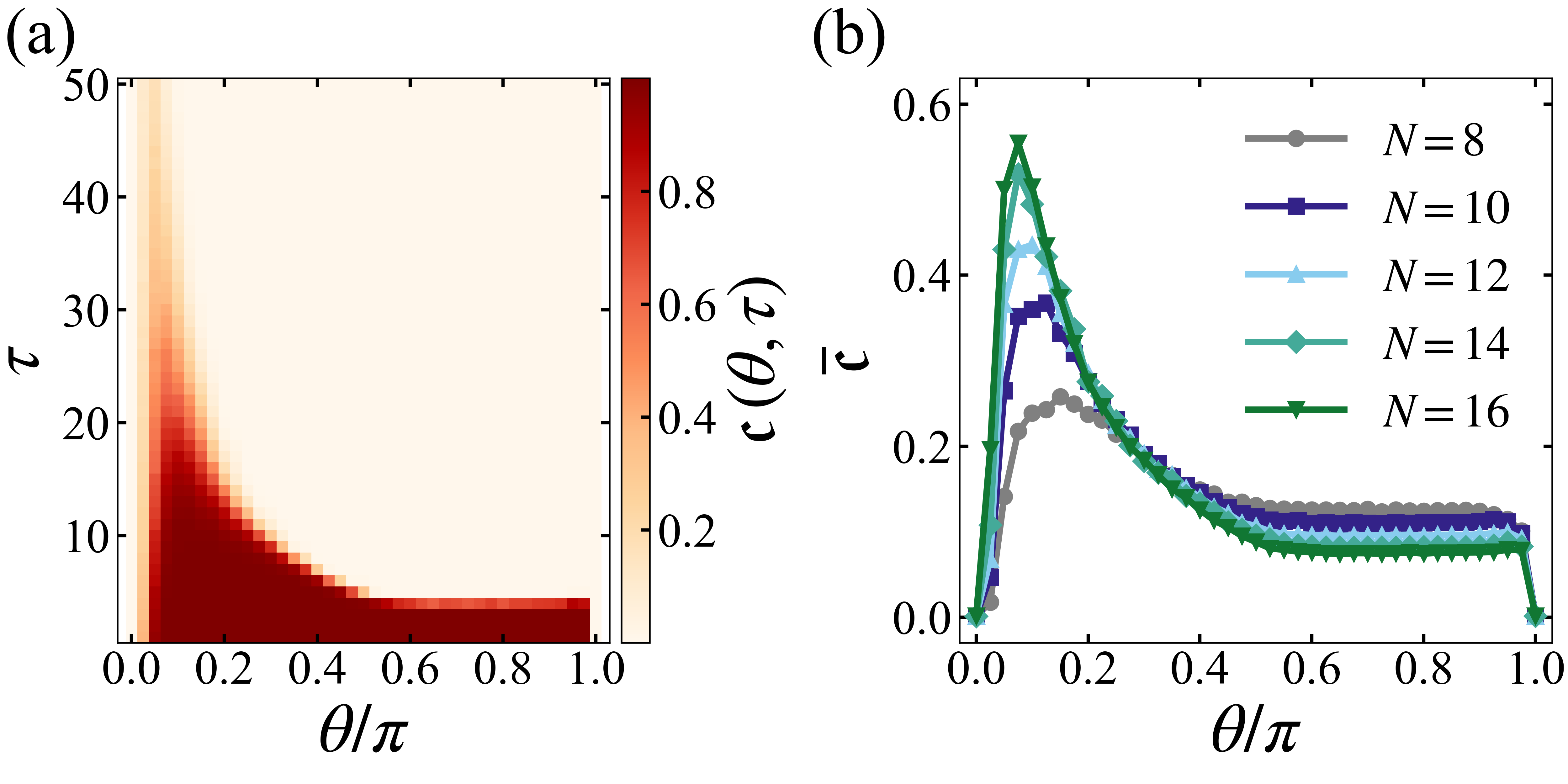}
    \includegraphics[width=0.99\linewidth]{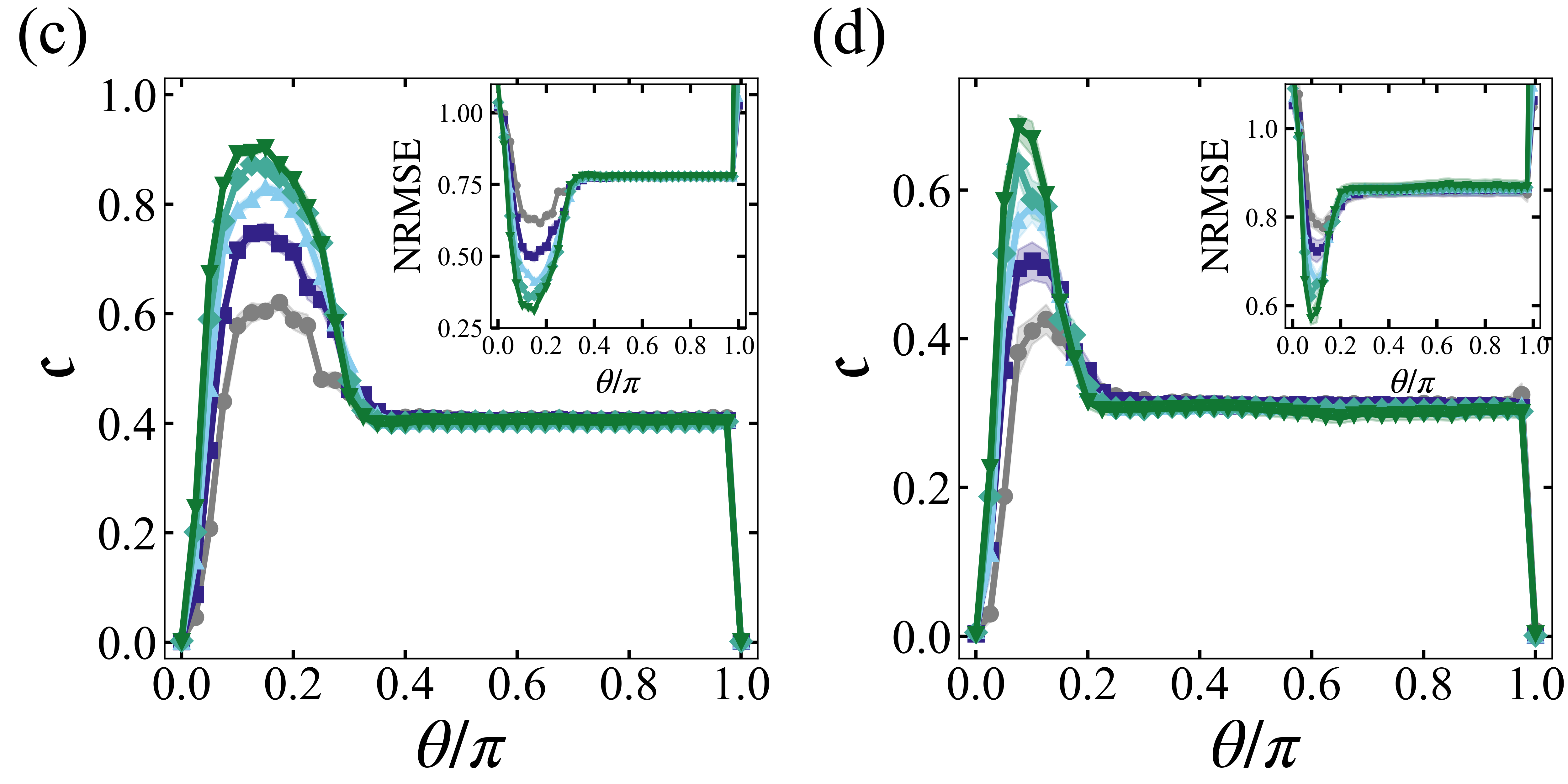}
    \caption{\textbf{Benchmark tasks.} \textbf{(a)} Heatmap of the linear memory metric ${\mathfrak{c}}(\theta,\tau)$ for $(N,d/N)=(16,1)$, and \textbf{(b)} the mean total memory capacity \(\overline{\mathfrak{c}}=(1/\tau_{\max})\sum^{ \tau_{\max}}_{\tau}\mathfrak{c}(\theta,\tau)\) for $\tau_{\rm {\max}}=50$, showing the improvement of the total memory in an intermediate regime as the system size grows. For random inputs with scale set to $\phi \in [0,1]\times10^{-3}$. Overall, memory closely tracks the same qualitative behavior as the spectral and geometric resources, although its peak is typically shifted closer to the product-state limit. \textbf{(c), (d)} Performance for NARMA-10 task with \(\phi\!\in\![0,0.2]\times10^{-1}\), and NARMA-20 with \(\phi\!\in\![0,0.2]\times5\times{10^{-2}}\). The washout is set to 500 steps, and training/testing split is \(0.6/0.4\) for a total of 1500 random inputs. Data are an average over 20 independent realizations.}
    \label{fig:stm_narma}
\end{figure}

As shown in {\color{brown} Fig.~\ref{fig:stm_narma}}, across both benchmarks, the reservoir exhibits very good performance that improves with increasing system size. This provides strong evidence that the useful computational regime may remain thermodynamically scalable. As pointed out before, the precise optimal operating point depends on the task. Memory-dominated reconstruction is shifted toward the lower-resource side of the crossover, where information about past inputs is retained more directly, whereas the NARMA optimum can occur deeper in the intermediate regime, where stronger nonlinear mixing may become beneficial~\cite{j2qj-vwcl}. More generally, the best benchmark performance lies close to the region in which the classical Fisher information and information processing capacity are maximized, strengthening further the interpretation that useful learning requires the coexistence of retained input distinguishability, nonflat subsystem structure, and sufficient observable response. At the Clifford endpoint \(\theta/\pi\!=\!1\), the state for \(\phi\!=\!0\) remains stabilizer and the spectral nonflatness measures vanish. This fact alone does not always imply perfectly vanishing learnability for encoded inputs, since input rotations are non-Clifford for generic random \(\phi\!\neq\!0\). For the computational-basis feature map considered here, however, the performance collapses at this endpoint, indicating that the first order state-space response is poorly aligned with the measured observable algebra.


\section{Summary and Discussion}

\subsection{Outlook}
Since the framework presented in this work follows from established limits on quantum dynamics and parameter estimation, it recasts the success and failure of learnability and sensing directly in terms of thermodynamic and information theoretic quantities, providing a necessary condition for effective information processing across broad classes of parameter-dependent tasks. More generally, this viewpoint offers a layered approach which connects quantum resource theory, entanglement-spectrum structure, quantum sensing, information scrambling, thermodynamic response, and postvariational quantum learning, suggesting that the onset and loss of computational usefulness can be understood as different manifestations of the same underlying redistribution of accessible information. This interpretation places earlier observations of optimal performance near the edge of quantum chaos~\cite{gq9r-d5q8,tnfv-lzfx, j2qj-vwcl} and of a memory–nonlinearity trade-off~\cite{vcindrak2026memory} within a common spectral, geometric, and measurement-dependent framework. 

A broader open question is whether these ingredients can be organized into an operational resource theory of learnability, providing a common framework that can be extended to variational~\cite{cerezo2021variational}, kernel-based~\cite{thanasilp2024exponential}, and other quantum learning architectures. This could also sharpen the meaning of quantum advantage by identifying when optimal information processing coexists with classical intractability of the underlying dynamics, the required observables, or the resulting measurement output distributions. The same logic admits a natural operator-space analogue. After vectorization, Heisenberg-evolved observables define states in a doubled Hilbert space, whose reduced states possess operator-entanglement spectra and corresponding nonflatness measures. Their parameter-dependent motion can likewise be characterized by an operator-space Fisher susceptibility, while their projection onto experimentally accessible operator subspaces determines how much of this structure can contribute to the readout. This suggests a connection between Pauli branching, operator entanglement, classical simulation complexity, and expressivity~\cite{rudolph2025pauli,p7xt-s9nz}.

Several practical concerns follow. The optimal response allowed requires a state-dependent measurement, whereas, practically, the reservoirs use a fixed and restricted family of Pauli strings. The gap between subsystem quantum and classical Fisher information of the implemented measurement therefore quantifies a concrete decoding limitation. Enlarging the observable set or learning an approximate measurement basis may recover part of this inaccessible sensitivity, but at the cost of additional measurements and postprocessing. Further, the shot cost of resolving feature differences may grow rapidly near the Haar-typical regime as expectation values and input derivatives concentrate. This raises whether the intermediate regime can support a genuine quantum advantage and whether the observables needed to extract it are classically intractable. Establishing such an advantage would require a joint analysis of measurement complexity, classical simulation cost, and achievable learning performance. We also note a temporal learning target may be encoded directly into the reservoir's control parameter(s) rather than only through a layer of local rotations, which repeatedly exposes the input to the evolving manybody state~\cite{mccaul2025minimal}. Although this defines an alternative construction which is closer in spirit to quantum neural networks, the same response framework applies. The asymptotic fate of the intermediate regime in higher dimensions, symmetry-constrained systems, and noisy open dynamics remains unresolved and constitutes an important direction for future works~\cite{ivaki2025noise}. The role of spatial geometry in shaping spectral nonflatness and metrological response remains also largely unexplored. The framework very well applies to other random circuit architectures and Hamiltonian dynamics, and is experimentally verifiable on current platforms, including superconducting-qubit devices~\cite{yasuda2023quantum}. This provides a promising route toward the practical deployment beyond purely theoretical demonstrations.

\subsection{Summary}
We have developed a tunable one-dimensional random circuit model that reveals an intermediate regime in which computationally useful subsystem structure is maximized. At weak interaction strength, the dynamics generates insufficient entanglement, spectral hierarchy, and nonlinear mixing. Under strong scrambling, by contrast, globally encoded information is redistributed into Haar-typical and increasingly nonlocal correlations, becoming inaccessible to practical measurements. Between these limits, a finite and scalable ``learning phase" emerges in which reduced states remain strongly nonflat, retain substantial sensitivity to temporal input perturbations, and preserve this sensitivity in simple readout observables. In a postvariational reservoir computing setting, this regime coincides with enhanced memory, nonlinear processing, measurement-accessible Fisher information, and total information processing capacity. This provides a precise sense in which global quantum complexity becomes computationally useful only while it remains locally structured, distinguishable, and convertible into measured features. The underlying response principles apply more generally to arbitrary differentiable quantum dynamical maps.


\textbf{Acknowledgments.}
We gratefully acknowledge the inspiring discussions with Gerard McCaul, Alexander Balanov, Alexandre Zagoskin, Emmanuel Rousseau, Viktor Iv\'ady, Bence Bak\'o, Zolt\'an Kolarovszki, and Kim P\"oyh\"onen. This work was supported by the European Union and the European Innovation Council through the Horizon Europe Project No. QRC-4-ESP (Grant Agreement No. 101129663), and EU Horizon Europe Quest project (Project No. 101156088), and the Academy of Finland through its QTF Center of Excellence program (Project No. 312298). T.O. acknowledges the support by the Finnish Research Council project 362573 and the Finnish quantum flagship program. This work is part of the Finnish Center of Excellence in Quantum Materials (QMAT).

\appendix

\section{Response bound}
\label{app:Response_bound}
Given a \(\phi\)-independent subsystem observable \(\hat{O}_R\), \( \langle\hat{O}_R\rangle\!=\!\Tr(\hat\rho_R\hat{O}_R)\), \( \partial_{\phi}\langle\hat{O}_R\rangle\!=\! \Tr[(\partial_\phi \hat\rho_R)\hat{O}_R ]\), using the definition of the symmetric logarithmic derivative yields \(\partial_\phi\langle\hat{O}_R\rangle\!=\!\frac{1}{2}\Tr[\hat\rho_R\{\mathcal{\hat{L}}_\phi,\hat{O}_R\}].\) Since adding a scalar to \(\hat{O}_R\) does not change the derivative, we can write \(\partial_\phi\langle\hat{O}_R\rangle_\phi\!=\!\frac{1}{2}\Tr[\hat\rho_R\{\hat{\mathcal{L}}_\phi,\delta\hat{O}_R\}]\), where \( \delta\hat{O}_R =\hat{O}_R-\langle\hat{O}_R\rangle\). This is a state-weighted inner product between the quantum-geometric velocity \(\hat{\mathcal{L}}_\phi\) and the observable fluctuation \(\delta\hat{O}_R\). For Hermitian \(\hat{\mathcal{L}}_\phi\) and \(\hat{O}_R\), it holds that 
\begin{equation}
\begin{aligned}  
\left|\partial_\phi\langle \hat{O}_R\rangle\right|^2\!&=\!\left|\mathrm{Re}\Tr\left[\hat\rho_R \hat{\mathcal{L}}_\phi\delta \hat{O}_R\right]
\right|^2\\&\le\left|\Tr\left[\hat\rho_R \hat{\mathcal{L}}_\phi\delta \hat{O}_R\right]\right|^2.
\end{aligned}
\end{equation}
Applying the Cauchy--Schwarz inequality to the weighted Hilbert--Schmidt inner product, \( |\Tr[\hat\rho_R \hat X^\dagger \hat Y]|^2 \leq \Tr[\hat\rho_R \hat{X}^\dagger \hat{X}]\, \Tr[\hat\rho_R \hat{Y}^\dagger \hat{Y}],\) with \(\hat{X}\!=\!\hat{\mathcal{L}}_\phi\) and \(\hat{Y}=\delta\hat{O}_R\), gives 
\begin{equation}|\partial_\phi\langle\hat{O}_R\rangle|^2 \leq \Tr\left[\hat\rho_R \hat{\mathcal{L}}_\phi^2\right]\, \Tr\left[\hat\rho_R(\delta \hat{O}_R)^2\right].
\label{app_eq:bound}
\end{equation}
The first factor on the right hand side of Eq.~\eqref{app_eq:bound} is precisely the quantum Fisher information, while the second factor is the variance of the observable. Therefore \(|\partial_\phi\langle\hat{O}_R\rangle|^2 \leq \mathcal{F}_{\rm Q}^{\phi}\Var_{\hat\rho_R}(\hat{O}_R)\). More generally, if the observable explicitly depends on \(\phi\), then \(|\partial_\phi\langle\hat O_R\rangle-\langle\partial_\phi\hat O_R\rangle|^2\leq\mathcal F_{\rm Q}^\phi\Var_{\hat\rho}(\hat O_R)\). An equivalent derivation can be found in Ref.~\cite{PhysRevX.12.011038}.


\section{Average nonflattening power along the identity-CNOT interpolation}
\label{app:capacity_power_cp}
Here we derive the gate-level average nonflattening power for the controlled-phase family \(\mathrm{diag}(1,1,1,e^{i\theta}),\, 0\le \theta \le \pi\), which lies on the Id--CNOT edge of the two-qubit Weyl chamber~\cite{varikuti2026impact}. In analogy with entangling and nonstabilizing powers, we may define the average nonflattening power of a bipartite gate \(\hat U\) through a faithful measure of spectral nonflatness \(\mathcal{N}\) (such as $\mathcal{C}_{\rm E},\mathcal{A}$)

\begin{equation}
\mathfrak n_{\mathcal N}^{\mathcal E}(\hat U)
:=\mathbb E_{|\psi\rangle\in\mathcal E}
\!\left[\mathcal N\!\left(\Tr_{\bar R}\!\big(\hat U|\psi\rangle\langle\psi|\hat U^\dagger\big)\right)
\right],
\label{eq:avg_capacity_power_def}
\end{equation}
where \(\mathcal E\) denotes an ensemble of spectrally flat input states. We consider different ensembles of pure two-qubit input states: product stabilizer states, general two-qubit stabilizer states, and Haar-random product states. For any pure two-qubit output state \(\ket{\psi}\), the one-qubit reduced density matrix has a binary spectrum, \(\mathrm{spec}(\hat \rho_R)\!=\!\{\xi,1-\xi\}\) with \(0\!\le\!\xi\!\le\! 1\). Let \(\hat \rho\!=\!\mathrm{diag}(\xi,1-\xi)\), with \(\mathcal{Z}_\beta\!=\!\Tr(\hat \rho^\beta)\!=\!\xi^\beta+(1-\xi)^\beta\). The associated escort state is \(\hat \omega_\beta\!=\!\hat \rho^\beta/\mathcal{Z}_\beta\), that is, \(\hat \omega_\beta\!=\!\mathrm{diag}(\xi^\beta/\mathcal{Z}_\beta,(1-\xi)^\beta/\mathcal{Z}_\beta),\) and the entanglement Hamiltonian is \(\mathcal{\hat H}\!=\!-\log\hat\rho\!=\!\mathrm{diag}(-\log \xi,-\log(1-\xi))\). By definition, \(\mathcal{C}_{\beta}\!=\!\beta^2\,\mathrm{Var}_{\hat\omega_\beta}( \mathcal{\hat H}).\) The variance is elementary, \(\mathrm{Var}_{\hat\omega_\beta}(\mathcal{\hat H})=w_{\xi} w_{1-\xi}[\log\frac{\xi}{1-\xi}]^2,\) with \(w_{\xi}\!=\!\xi^\beta/\mathcal{Z}_\beta,
w_{1-\xi}\!=\!(1-\xi)^\beta/\mathcal{Z}_\beta.\) One then obtains
\begin{equation}
\mathcal{C}_{\beta}=\beta^2\,
\frac{[\xi(1-\xi)]^\beta}{\big(\xi^\beta+(1-\xi)^\beta\big)^2}
\left[\log\frac{\xi}{1-\xi}\right]^2.
\label{eq:binary_cmod_final}
\end{equation}
At \(\beta=1\), this reduces to the capacity of entanglement,
\begin{equation}
\mathcal{C}_{\rm E}
=\xi(1-\xi)\left[\log\frac{\xi}{1-\xi}\right]^2.
\label{eq:binary_ce_final}
\end{equation}
Any two-qubit gate average reduces to determining the reduced binary spectrum of the output state and averaging the corresponding one-parameter function. Taking \(2\xi-1\!=\!\tanh x\), gives \(\mathcal{C}_{\rm E}\!=\!x^2\,{\rm sech}^2x\). The nontrivial maximum satisfies \(x\tanh x\!=\!1\), yielding \(\mathcal{C}^{\rm max}_{\rm E,1q}\approx0.4392\) at approximately \((\xi,1-\xi)\approx(0.916, 0.083)\). Similarly, for antiflatness \(\mathcal A\!=\!\xi(1-\xi)(2\xi-1)^2,\) or in terms of concurrence, $C\!=\!2\sqrt{\xi(1-\xi)}$~\cite{wootters2001entanglement}, \(\mathcal A\!=\!(C^2/4)(1-C^2)\). This yields \(\mathcal{A}^{\rm max}_{\rm 1q}=1/16\) for \((\xi,1-\xi)\approx(0.853, 0.146)\). Note \(\mathcal{C}_{\rm E},\mathcal{A}\!=\!0\) when \(\xi\!=\!1/2\), and \(\xi\!\to\!0,1\).

\begin{figure}
    \centering
    \includegraphics[width=0.99\linewidth]{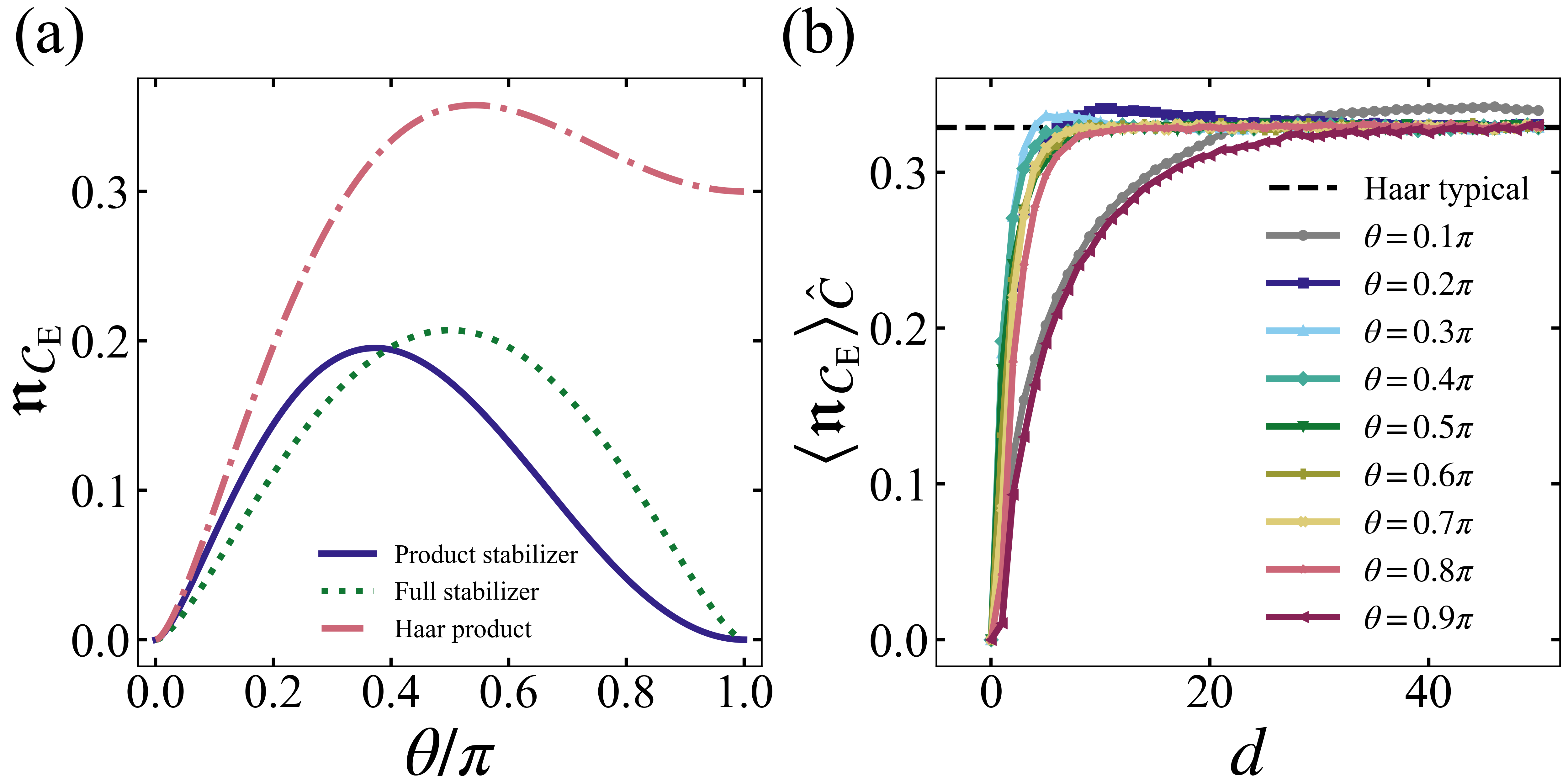}
    \includegraphics[width=0.99\linewidth]{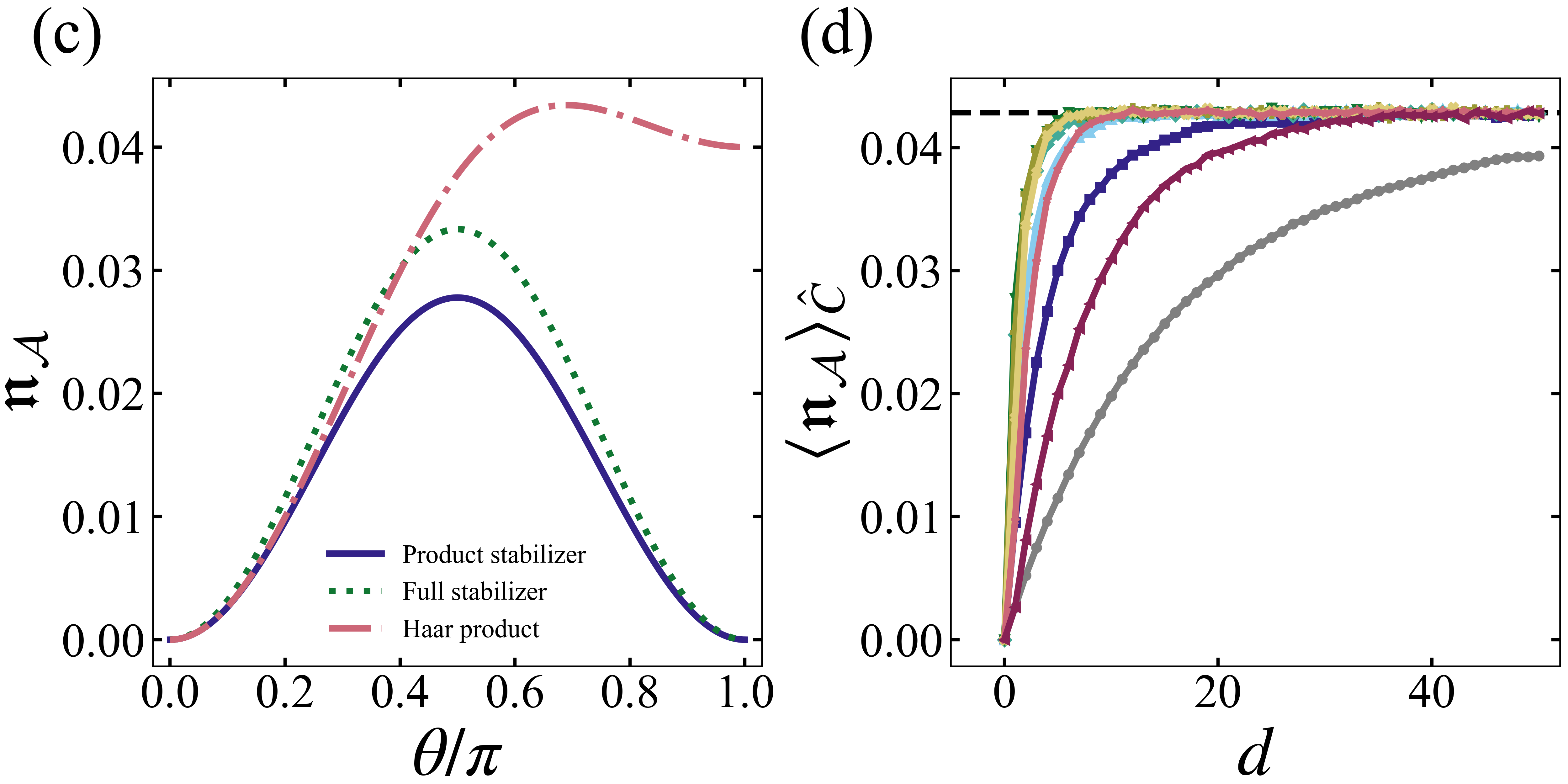}
    \caption{\textbf{Average nonflattening power along the Id--CNOT edge} \textbf{(a)} The product-stabilizer and full-stabilizer curves are given analytically, while the Haar-product curve is evaluated numerically. \textbf{(b)} Exponential approach of the averaged two-qubit nonflattening power \(\langle\mathfrak{n}_{\mathcal{C}_{\rm E}}\rangle_{\hat C}\) to the Haar-typical value under repeated action of single qubit Clifford and \(\hat{P}(\theta)\) gates, shown for the stabilizer product states. \textbf{(c)}, \textbf{(d)} Same as upper panels but for antiflatness. Data are for 12000 samples.}
    \label{fig:flatness_power}
\end{figure}

For two qubits there are \(60\) pure stabilizer states in total, consisting of \(36\) product stabilizer states and \(24\) entangled stabilizer states. We first evaluate the average over the product stabilizer states. A single qubit has six pure stabilizer states, \(\{|0\rangle,|1\rangle,|\pm\rangle,|\pm i\rangle\}\), so there are \(6\times 6=36\) two-qubit product stabilizer inputs. Among these \(36\) inputs, only \(16\) contribute nontrivially. Indeed, if either qubit is a \(\hat Z\)-eigenstate, \(|0\rangle\) or \(|1\rangle\), then the controlled-phase gate only inserts a conditional phase and the output remains a product state. Since the reduced density matrix of a product pure state is pure, any faithful nonflatness vanishes on such inputs. Hence all nontrivial contributions come from the \(4\times 4=16\) states in which both qubits belong to the equatorial stabilizer set \(\{|\pm\rangle,|\pm i\rangle\}\). A general equatorial input may be written as \( |\psi_{\gamma'}\rangle\otimes |\psi_\gamma\rangle \), with \( |\psi_{\gamma}\rangle\!=\!(|0\rangle+e^{i\gamma}|1\rangle)/\sqrt2 \) and \({\gamma'},\gamma\in\{0,\pi,\pi/2,-\pi/2\}\). After applying \(\hat {P}(\theta)\), the output concurrence is \(\sin(\theta/2)\), independently of \(\gamma'\) and \(\gamma\). Therefore the one-qubit reduced spectrum is the same for all \(16\) contributing inputs, namely \((\xi,1-\xi)\!=\!(1\pm \cos(\theta/2))/2\!=\!(\cos^2\frac{\theta}{4},\ \sin^2\frac{\theta}{4})\). The full average therefore reduces to \(\mathfrak n_{\mathcal C_{\beta}}^{\mathrm{ps}}\!=\!\frac{4}{9}\,\mathcal C_{\beta}\!\left(\cos^2\frac{\theta}{4},\,\sin^2\frac{\theta}{4}\right).\) Choosing \(\mathcal C_{\beta}\!=\!\mathcal C_{\rm E}\) gives
\begin{equation}
\mathfrak n_{\mathcal{C}_{\rm E}}^{\mathrm{ps}}\!
=\frac{1}{9}\sin^2\frac{\theta}{2}\left[\log\!\cot^2\frac{\theta}{4}\right]^2,
\label{eq:prod_stab_ce}
\end{equation}
which is plotted in {\color{brown} Fig.}~\ref{fig:flatness_power}. More generally using {\color{brown}Eq.}~\eqref{eq:binary_cmod_final}, one finds
\begin{equation}
\mathfrak n_{\mathcal{C}_{\beta}}^{\mathrm{ps}}\!
=\frac{4}{9}\,\beta^2\,
\frac{\left(\cos^2\frac{\theta}{4}\sin^2\frac{\theta}{4}\right)^\beta}
{\left(\cos^{2\beta}\frac{\theta}{4}+\sin^{2\beta}\frac{\theta}{4}\right)^2}
\left[\log\!\cot^2\frac{\theta}{4}\right]^2,
\label{eq:prod_stab_cmod_beta}
\end{equation}
which has a broad maximum at an intermediate angle, \(\theta\approx1.17\approx0.37\pi\) and vanishes at \(\theta=0,\pi\). Although in this case \(\beta\) can change the shape of the curve as a function of \(\theta\), the peak height is in fact independent of \(\beta\). Starting from {\color{brown}Eq.}~\eqref{eq:prod_stab_cmod_beta}, take \(r:=\tan^2(\theta/4)\), so that \(0\le r\le 1\), \(\cos^2(\theta/4)=1/(1+r)\), \(\sin^2(\theta/4)=r/(1+r)\), and \(\log\!\cot^2(\theta/4)=-\log r\). Equation~\eqref{eq:prod_stab_cmod_beta} then becomes
\begin{equation}
\mathfrak n_{\mathcal{C}_{\beta}}^{\mathrm{ps}}
=\frac{4}{9}\,\beta^2\,\frac{r^\beta}{(1+r^\beta)^2}(\log r)^2.
\label{eq:prodstab_R_form}
\end{equation}
Defining the scaling variable \(x\!:=\!\beta\log r\), one has \(r^\beta\!=\!e^x\) and \((\log r)^2\!=\!x^2/\beta^2\), so
\begin{equation}
\mathfrak n_{\mathcal{C}_{\beta}}^{\mathrm{ps}}
=\frac{4}{9}\,\frac{x^2 e^x}{(1+e^x)^2},\qquad
x=\beta\log\tan^2\frac{\theta}{4}.
\label{eq:universal_scaling_form}
\end{equation}
Thus, the full family collapses onto the universal function \(f(x)\!=\!x^2 e^x/(1+e^x)^2\), and \(\beta\) merely rescales the horizontal axis. The peak height is therefore independent of \(\beta\), while the peak position in \(\theta\) depends on \(\beta\). Differentiating \(f(x)\) gives \(df(x)/dx\!=\!(e^x/(1+e^x)^3)[2x(1+e^x)+x^2(1-e^x)].\) The nontrivial extremum satisfies \(2x(1+e^x)+x^2(1-e^x)\!=\!0\), or equivalently \(x\,\tanh\frac{x}{2}\!=\!2.\) On the physical branch \(0\le \theta\le \pi\), one has \(x\le 0\), so the relevant solution is \(x_\star\approx -2.3993\). The universal peak value is then \(\max_\theta\,\mathfrak n_{\mathcal{C}_{\beta}}^{\mathrm{ps}}\!=\!(4/9)\,x_\star^2 e^{x_\star}/(1+e^{x_\star})^2\approx 0.1952,\) which is independent of \(\beta\). The peak position is obtained by inverting \(x\!=\!\beta\log\tan^2(\theta/4)\), namely \(\theta_\star(\beta)\!=\!4\arctan\!\left(e^{x_\star/(2\beta)}\right).\) Because \(x_\star\!<\!0\), increasing \(\beta\) pushes the peak toward \(\theta\!=\!\pi\), while decreasing \(\beta\) pushes it toward \(\theta\!=\!0\). These agree with direct numerical maximization to machine precision. For antiflatness the expression simply becomes \(\mathfrak{n}_{\mathcal A}^{\mathrm{ps}}\!=(1/36)\sin^{2}{\theta}\!\).

For the additional \(24\) entangled stabilizer states, an explicit classification shows that they split into two subclasses. One subclass contains \(8\) states that remain maximally entangled under \(\hat {P}(\theta)\), and therefore contribute zero to a faithful measure of nonflatness. The second subclass contains \(16\) states that produce a second binary spectrum. A representative of this class is \((\hat I\otimes \hat H)|\Phi^+\rangle\), with $\hat H\!=\!(\hat X+\hat Z)/\sqrt{2}$ the single-qubit Hadamard gate, which is mapped by \(\hat {P}(\theta)\) to a state with one-qubit reduced eigenvalues \((1\pm \sin(\theta/2))/2.\) Since all \(16\) states in this class yield the same spectrum, the complete stabilizer average is

\begin{align}
\mathfrak n_{\mathcal C}^{\mathrm{\mathrm{STAB}}}\!
=&\frac{4}{15}\,
\mathcal {\mathcal{C}_{\beta}}\!\left(\cos^2\frac{\theta}{4},\,\sin^2\frac{\theta}{4}\right)\nonumber\\
+&\frac{4}{15}\,
\mathcal {\mathcal{C}_{\beta}}\!\left(\frac{1+\sin(\theta/2)}{2},\,\frac{1-\sin(\theta/2)}{2}\right).
\label{eq:all_stab_master_formula_simplified}
\end{align}
For \({\mathcal{C}_{\beta}}\!=\!\mathcal C_{\rm E}\), one finds
\begin{align}
\mathfrak n_{\mathcal C_{\rm E}}^{\mathrm{\mathrm{STAB}}}\!
&=
\frac{1}{15}\sin^2\frac{\theta}{2}\left[\log\!\cot^2\frac{\theta}{4}\right]^2 \nonumber\\
\quad
&+\frac{1}{15}\cos^2\frac{\theta}{2}
\left[\log\!\frac{1+\sin(\theta/2)}{1-\sin(\theta/2)}\right]^2,
\label{eq:all_stab_ce}
\end{align}
which is shown in {\color{brown} Fig.}~\ref{fig:flatness_power}. In general one can show that the average over the full stabilizer set takes the form \(\mathfrak n_{\mathcal C_{\rm E}}^{{\mathrm{STAB}}}(\theta)
\!\propto\![\mathfrak n_{\mathcal C_{\rm E}}^{\mathrm{ps}} (\theta)+\mathfrak n_{\mathcal C_{\rm E}}^{\mathrm{ps}} (\theta-\pi)]\). This is maximized at \(\theta^{\star}\!=\!\pi/2\) with \(\mathfrak n_{{\mathcal C}_{\rm E}}(\theta^{\star})\approx0.207\). This expression can be verified by direct brute-force averaging over all \(60\) pure two-qubit stabilizer states to machine precision. For antiflatness one can check that \(\mathfrak{n}_{\mathcal A}^{\mathrm{STAB}}\!=\!(1/30)\sin^{2}{\theta}\!\).

Finally, one may calculate the averaging over Haar-random product inputs,
\begin{equation} 
\begin{aligned}
\mathfrak{n}_{\mathcal N}^{\mathrm{H}}
=\ \int d\psi_R \, d\psi_{\bar R} \;
\mathcal N\!\left(\Tr_{\bar R}\!\left[
\hat{P}(\theta)\ket{\psi_{R} \psi_{\bar R}}\!\bra{\psi_R \psi_{\bar R}}
\hat{P}^\dagger(\theta)\right]\right).
\end{aligned}
\end{equation}
Here $d\psi_R$ and $d\psi_{\bar R}$ denote the Haar measures on the single-qubit pure-state manifolds for subsystems $R$ and ${\bar R}$, respectively. Writing the two input qubit states as $\ket{\psi_R} \!=\! a_0 \ket{0} + a_1 \ket{1}$ and $\ket{\psi_{\bar R}} \!=\! b_0 \ket{0} + b_1 \ket{1}$, with $|a_0|^2 + |a_1|^2 \!=\! 1$ and $|b_0|^2 + |b_1|^2 \!=\! 1$. It is convenient to introduce the populations $x \!:=\! |a_0|^2$ and $y \!:=\! |b_0|^2$, so that $|a_1|^2 \!=\! 1-x$ and $|b_1|^2 \!=\! 1-y$. For a Haar-random single-qubit pure state one may use the standard parametrization $\ket{\psi} \!=\! \cos(\vartheta/2)\ket{0} + e^{i\varphi}\sin(\vartheta/2)\ket{1}$, for which the Haar measure is $d\psi\!=\!(4\pi)^{-1}\sin\vartheta\, d\vartheta\, d\varphi$. Variables $x$ and $y$ are independent uniform random variables on $[0,1]$, and the Haar-product average reduces to an ordinary integral over the unit square. Acting on the product input with $\hat{P}(\theta)$ gives
\begin{align}
\ket{\psi}
&=a_0 b_0 \ket{00}
+ a_0 b_1 \ket{01}\nonumber\\
&+ a_1 b_0 \ket{10}
+ e^{i\theta} a_1 b_1 \ket{11}.
\end{align}
For a pure two-qubit state $\ket{\psi}\!=\!\sum_{ij=0}^1 c_{ij}\ket{ij}$, the concurrence is $C(\ket{\psi})\!=\!2|c_{00}c_{11}-c_{01}c_{10}|$, yielding
\begin{align}
C(\ket{\psi_{\mathrm{}}})
&=2 \left|
(a_0 b_0)(e^{i\theta} a_1 b_1)
-(a_0 b_1)(a_1 b_0)
\right|\nonumber\\
&=2 |a_0 a_1 b_0 b_1|\, |e^{i\theta}-1|\nonumber\\
&=4 |a_0 a_1 b_0 b_1|\, \Bigl|\sin \frac{\theta}{2}\Bigr|.
\label{eq:haar_conc_step}
\end{align}
Squaring and rewriting in terms of $x$ and $y$ gives \(C^2 \!=\! 16\, x(1-x)\, y(1-y)\, \sin^2 \frac{\theta}{2}.\) For a pure bipartite two-qubit state, the spectrum of the reduced density matrix is completely determined by the concurrence, and the Haar-product average of the modified quantity reduces to the two-dimensional integral
\begin{align}
\mathfrak{n}_{\mathcal{C}_{\beta}}^{\mathrm{H}}\!(\theta)
&=\\\nonumber
&\int_0^1 \!\! dx
\int_0^1 \!\! dy\;
\mathcal{C}_{\beta}\!\left(
\frac{1+\sqrt{1-C^2}}{2},
\frac{1-\sqrt{1-C^2}}{2}
\right),
\label{eq:haar_prod_Rvg_final}
\end{align}
The reduced spectrum varies continuously over the Haar-product ensemble, and the integral can be evaluated numerically for \(\mathcal{C}_{\rm E}\). The random-product average behaves qualitatively differently from the stabilizer averages; its peak is shifted to larger angles and its magnitude is substantially larger, reflecting the fact that the Haar ensemble samples a continuous family of product inputs rather than a discrete Clifford orbit. This is maximized at \(\theta^{\star}\!=\!1.701\) with \(\mathfrak n_{\mathcal{C}_{\rm E}}(\theta^{\star})\approx0.357\) and naturally it does not vanish at \(\theta\!=\!\pi\). Both the peak position and the peak height depend nontrivially on $\beta$. For antiflatness the integration is straightforward, yielding the exact expression \(\mathfrak{n}_{\mathcal A}^{\rm H}\!=\!(1/9)\sin^2(\theta/2)-(16/225)\sin^4(\theta/2)\), which peaks at \(\theta^{\star}\!=\!2.166\), with \(\mathfrak n_{\mathcal A}(\theta^{\star})\approx0.043\). The results are plotted in {\color{brown}Fig.}~\ref{fig:flatness_power}.


\section{Construction of the quantum reservoir}
\label{app:reservoir_protocol}
We construct the quantum reservoir from the random circuit introduced in {\color{brown}Eq.}~\eqref{eq:floquet_compact}, using the encoding–evolution–readout protocol illustrated in Fig. {\color{brown}Fig.}~\ref{fig:scheme}. The $N$ qubits are divided into two interleaved sets of equal size, $\bar R\!=\!{0,2,\ldots,N-2}, \quad R\!=\!{1,3,\ldots,N-1}, \quad |\bar R|\!=\!|R|\!=\!N/2,$ which we refer to as the memory and readout subsystems, respectively. The alternating partition is chosen so that every memory qubit is directly coupled to neighboring readout qubits by the nearest-neighbor brickwork dynamics. This maximizes the interface between encoding and readout degrees of freedom, avoids a geometric bottleneck associated with two spatially contiguous halves, and distributes the injected signal uniformly throughout the circuit. Let \(\phi_t\) denote the scalar input presented to the reservoir at discrete time \(t\), injected collectively on the memory qubits through
\begin{equation}
\hat{\mathcal{V}}_{\bar R}(\phi_t)=\bigotimes_{j\in \bar R}\hat R^y_{j}(\phi_t),\quad \hat R^y(\phi_t)=e^{-i\phi_t\hat Y/2}.\end{equation}
The use of identical local rotations provides an extensive but experimentally simple encoding. For the initial product state \(\ket{0}^{\otimes N}\), the globally encoded quantum Fisher information is
\(\mathcal F_Q^\phi\!\propto|\bar R|\), and the input sensitivity of memory subsystem before reservoir evolution is extensive in system size. The subsequent \(\phi\)-independent circuit dynamics does not create additional global Fisher information (with respect to the encoding angle) on the readout subset, but redistributes it among local, nonlocal, coherent, and inaccessible degrees of freedom. At each input step, the encoded memory state is evolved under a fixed realization of the reservoir circuit. Denoting the memory state immediately before encoding by \(\hat\rho_{\bar R}(t)\), we first form \( \hat\varrho_{\bar R}(t)\!=\!\hat{\mathcal V}_{\bar R}(\phi_t)\hat\rho_{\bar R}(t)\hat{\mathcal V}_{\bar R}^\dagger(\phi_t).\) The readout qubits are then initialized in the reference state \(\ket{0}\!\bra{0}_R\), after which the joint system evolves for a prescribed circuit depth,
\begin{equation}   
\hat\rho_{R\bar R}(t) = \hat{\mathcal{U}}(\theta)
\Bigl[\ket{0}\!\bra{0}_R\otimes\hat\varrho_{\bar R}(t)\Bigr]\hat{\mathcal{U}}^{\dagger}(\theta),
\end{equation}
with \(\hat{\mathcal{U}}(\theta)\!=\!\prod_i^d \hat{\mathcal{U}}_{\rm F} (\theta)\), where \( \hat{\mathcal{U}}_{\rm F}(\theta)\) is defined in {\color{brown}Eq.}~\eqref{eq:floquet_compact}. 
The readout state and the updated memory state are
\begin{equation}
\hat\rho_R(t)
=\mathrm{Tr}_{\bar R}\!\left[\hat\rho_{R\bar R}(t)
\right],\quad \hat\rho_{\bar R}(t+1) =\mathrm{Tr}_R\! \left[\hat\rho_{R\bar R}(t)\right].
\end{equation}
Tracing out and reinitializing \(R\) after every input step produces an effective dissipative recurrence on \(\bar R\). The memory subsystem carries information from previous inputs, while the repeated reset prevents the measured degrees of freedom from retaining uncontrolled information between successive steps. The resulting dynamics therefore combines recurrent memory, manybody mixing, and controlled information leakage through the readout channel. Equivalently, the memory evolution can be written as a completely positive trace-preserving map,
\(\hat\rho_{\bar R}(t+1)\!=\!\sum_{\mathbf z}\hat{\mathcal{K}}_{\mathbf z} \hat\varrho_{\bar R}(t)\,\hat{\mathcal{K}}_{\mathbf z}^{\dagger},\) where \(\hat{\mathcal{K}}_{\mathbf z}\!=(\!{}\bra{\mathbf z}_R\otimes {\hat I}_{\bar R})\hat{\mathcal{U}}(\theta)(\ket{\mathbf{0}}_R\otimes {\hat I}_{\bar R})\) are input-independent Kraus operators acting on \(\bar R\), and \({\ket{\mathbf z}_R}\) is the computational basis of the readout subsystem. Since the reservoir realization and the partition are fixed, these operators can be precomputed once. The dependence on the temporal signal enters only through \(\hat{\mathcal V}_{\bar R}(\phi_t)\). For an equal bipartition of an \(N\)-qubit system, the propagated density matrix dimension is reduced from \(2^N\times 2^N\) to \(2^{N/2}\times 2^{N/2}\). Note there are at most \(2^{|R|}\) Kraus operators. 

Information is extracted from the readout state through expectation values of a prescribed observable set \(\{\hat O_\mu\}_{\mu=1}^{F}\), where \(F\!=\!2^{|R|}-1\) is the number of measured features \(x_\mu(t)\!=\!\mathrm{Tr}\left[\hat\rho_R(t)\hat O_\mu\right].\) In the present implementation, the observables are nontrivial diagonal Pauli strings, \(\hat O_\mu=\prod_{j\in S_\mu}\hat Z_j\), with \(S_\mu\subseteq R\) and \(S_\mu\neq\varnothing\). These observables define the instantaneous reservoir feature vector \(\mathbf x(t)\!=\!(1,x_1(t),\ldots,x_F(t))^{\mathsf T}\), and the leading constant provides an intercept for the classical readout. Because all of these observables commute, they can be estimated from the same computational-basis measurement distribution. If \(p_{\mathbf z}(t)\!=\!\bra{\mathbf z}\hat\rho_R(t)\ket{\mathbf z}\), then
\begin{equation}
\left\langle\prod_{j\in S}Z_j\right\rangle_t=\sum_{\mathbf z}
p_{\mathbf z}(t)(-1)^{\sum_{j\in S}z_j}.
\end{equation}
Thus a single measurement basis supplies all diagonal Pauli-string features considered here. State tomography is neither required by the learning protocol nor used by the trained readout; the computational variables are directly measurable expectation values. Although all strings can be estimated from a common measurement basis, estimating the entire exponentially large distribution, or all its moments to controlled simultaneous accuracy, generally requires rapidly increasing numbers of samples. Exact state-vector or density-matrix probabilities remove this cost from the numerics. Before training, an initial washout interval is discarded to suppress dependence on the arbitrary initial memory state. Each remaining feature is standardized using statistics obtained from the training interval, \(\tilde x_\mu(t)=({x_\mu(t)-\overline{x}_\mu^{\,\mathrm{tr}}})/{s_\mu^{\mathrm{tr}}},\), where \(\overline{x}_\mu^{\mathrm{tr}}\) and \(s_\mu^{\mathrm{tr}}\) are the training mean and standard deviation. The same transformation is subsequently applied to validation and test data, without recomputing their statistics. Data standardization is statistically conventional, but it may conceal the physical concentration as it maps a physically tiny feature fluctuation to an order-one numerical variable. When \(\mathcal{F}_{\rm cl}^{\phi}\) is exponentially small, the required shot number is exponentially large even though exact standardized features still produce finite machine-learning performance. Given a target sequence \(y(t)\), the reservoir prediction is taken to be linear in the standardized features, \(\tilde y(t)\!=\!\mathbf w^{\top}\tilde{\mathbf x}(t).\) The weights are obtained by the ridge regression,
\begin{equation}
\mathbf w_\lambda=\underset{\mathbf w}{\operatorname{argmin}}
\left[\sum_t\left|y(t)-\mathbf w^{\top}
\tilde{\mathbf x}(t) \right|^2+\lambda|\mathbf w|_2^2
\right],\end{equation}
or, in matrix form, \(\mathbf w_\lambda\!=\!\left(\tilde X_{\mathrm{tr}}^{\top}\tilde X_{\mathrm{tr}}+\lambda I\right)^{-1}\tilde X_{\mathrm{tr}}^{\top}\mathbf y_{\mathrm{tr}}\). The regularization parameter \(\lambda=10^{-4}-10^{-5}\) is fixed independently of the test set, either globally or through validation. Performance is then evaluated on an unseen temporal interval using the same reservoir realization, observable set, standardization parameters, and trained weights. The quantum circuit itself is therefore never optimized for a particular task. Its role is to transform the input history into a high-dimensional feature vector, while the linear output layer selects the combination of those features relevant to the desired target. Memory arises because the resulting states depend recursively on previous inputs. Although the unitary quantum channel is linear in density operators, the map from the classical input history to the measured features is nonlinear since the input-dependent rotations are repeatedly composed with the fixed, generally noncommuting memory channel; higher-weight observables expose additional nonlinear components of this history dependence. Fading memory also arises from the repeated coupling to and reset of the readout subsystem. The quality of the reservoir is consequently controlled by whether the dynamics preserves input-dependent structure that remains distinguishable through the restricted observable feature map.


\section{Information-processing capacity}
\label{app:ipc}
We quantify the temporal processing capability of the reservoir using information-processing capacities (IPCs). The IPC measures how well information about past inputs, and nonlinear functions of those inputs, can be linearly reconstructed from the reservoir features. For an input sequence $\{\phi_t\}_{t=1}^{T}$ with $\phi_t\in[-1,1]$, the target functions are constructed from Legendre polynomials $P_k(\phi)$, which form an orthogonal basis for uniformly distributed inputs on $[-1,1]$ (this should not be confused with the scaled input used in {\color{brown}Fig.}~\ref{fig:ipc}). 
We use
\begin{equation}
\begin{aligned}
    &P_0(\phi) = 1,\quad
    P_1(\phi) = \phi,\quad
    P_2(\phi) = \frac{1}{2}(3\phi^2-1),\\
    &P_n(\phi)
    = \frac{2n-1}{n}\phi P_{n-1}(\phi) - \frac{n-1}{n}P_{n-2}(\phi)
\end{aligned}
\end{equation}
with the recurrence used for higher orders. The IPC targets are scalar time series constructed from delayed inputs. For a fixed set of Legendre orders $\mathbf{k}\!=\!(k_1,\ldots,k_m)$ and delays 
$\boldsymbol{\tau}\!=\!(\tau_1,\ldots,\tau_m)$, the scalar target value at time $t$ is
\begin{equation}
    y^{\mathbf{k}}_{\boldsymbol{\tau},t}
    =
    \prod_{\ell=1}^{m}
    P_{k_\ell}\!\left(\phi_{t-\tau_\ell}\right),
    \qquad
    D=\sum_{\ell=1}^{m} k_\ell ,
    \label{eq:ipc_target_scalar}
\end{equation}
where $D$ is the total order of the target. 
The delays satisfy $\tau_\ell\in\{1,\ldots,\tau_{\max}\}$, and duplicate targets are avoided by imposing ordering constraints on the delay indices. After discarding the initial washout and the first $\tau_{\max}$ time steps needed for delay alignment, the remaining reservoir feature times are denoted by $t_i\!=\!t_0+i$, with $i\!=\!0,\ldots,M-1$. 
For each choice of $(\mathbf{k},\boldsymbol{\tau})$, Eq.~\eqref{eq:ipc_target_scalar} defines one target vector
\begin{equation}
    \mathbf{y}^{\mathbf{k}}_{\boldsymbol{\tau}}
    =
    \begin{pmatrix}
    y^{\mathbf{k}}_{\boldsymbol{\tau},t_0}\\
    y^{\mathbf{k}}_{\boldsymbol{\tau},t_1}\\
    \vdots\\
    y^{\mathbf{k}}_{\boldsymbol{\tau},t_{M-1}}
    \end{pmatrix}
    \in \mathbb{R}^{M}.
    \label{eq:ipc_target_vector}
\end{equation}
The full target matrix is obtained by placing all such target vectors as columns,
\begin{equation}
    Y=
    \begin{pmatrix}
    | & | &  & |\\
    \mathbf{y}^{\mathbf{k}_{1}}_{\boldsymbol{\tau}_{1}} &
    \mathbf{y}^{\mathbf{k}_{2}}_{\boldsymbol{\tau}_{2}} &
    \cdots &
    \mathbf{y}^{\mathbf{k}_{B}}_{\boldsymbol{\tau}_{B}}\\
    | & | &  & |
    \end{pmatrix}
    \in\mathbb{R}^{M\times B}.
    \label{eq:ipc_target_matrix}
\end{equation}
Thus, each column of $Y$ is one IPC target time series, while each row is aligned with one reservoir feature vector at the same retained time $t_i$.

For first-order IPC, only $P_1$ appears, giving the linear memory targets
\begin{equation}
    y^{(1)}_{\tau,t}
    =
    P_1(\phi_{t-\tau})
    =
    \phi_{t-\tau},
    \qquad
    \tau=1,\ldots,\tau_{\max}.
    \label{eq:ipc1_targets}
\end{equation}
The second-order IPC contains pure second-order targets,
\begin{equation}
    y^{(2)}_{\tau,t}
    =
    P_2(\phi_{t-\tau}),
    \qquad
    \tau=1,\ldots,\tau_{\max},
    \label{eq:ipc2_pure_targets}
\end{equation}
and mixed products of two first-order delayed inputs,
\begin{equation}
\begin{aligned}
    y^{(1,1)}_{(\tau_1,\tau_2),t}
    =
    P_1(\phi_{t-\tau_1})P_1(\phi_{t-\tau_2})
    &=
    \phi_{t-\tau_1}\phi_{t-\tau_2},
    \\
    1\le \tau_1<\tau_2\le \tau_{\max}.
    \label{eq:ipc2_mixed_targets}
\end{aligned}
\end{equation}
Higher-order IPCs are constructed analogously by including all products of delayed Legendre polynomials whose total Legendre degree is $D$. 
For example, third-order targets include 
\begin{equation}
\begin{aligned}
    &y^{(3)}_{\tau,t}=P_3(\phi_{t-\tau}), \quad 
    y^{(1,2)}_{(\tau_1,\tau_2),t}=P_1(\phi_{t-\tau_1})P_2(\phi_{t-\tau_2}), \\\quad&\text{and}\quad 
    y^{(1,1,1)}_{(\tau_1,\tau_2,\tau_3),t}
    =
    P_1(\phi_{t-\tau_1})
    P_1(\phi_{t-\tau_2})
    P_1(\phi_{t-\tau_3}).
\end{aligned}
\end{equation}
In this work, we compute IPCs up to fourth order, $D=1,\ldots,4$, with $\tau_{\max}=30$, giving 
$B_1\!=\!30$, $B_2\!=\!465$, $B_3\!=\!4960$, and $B_4\!=\!40920$ targets for orders $1$, $2$, $3$, and $4$, respectively.

Let $X\in\mathbb{R}^{M\times F}$ denote the reservoir feature matrix aligned with the target matrix $Y$ in Eq.~\eqref{eq:ipc_target_matrix}. 
The rows of $X$ and $Y$ are split into training and test parts. 
Before training, each feature column is standardized using the mean and standard deviation computed from the training set only, and the same transformation is applied to the test set. 
The reservoir readout is trained by multi-output ridge regression,
\begin{equation}
\begin{aligned}
    W =
    \left(\tilde X_{\mathrm{tr}}^{\top}\tilde X_{\mathrm{tr}} + \Lambda\right)^{-1}
    \tilde X_{\mathrm{tr}}^{\top}Y_{\mathrm{tr}},
    \label{eq:ipc_ridge}
\end{aligned}
\end{equation}
where $\tilde X_{\mathrm{tr}}$ includes a bias column and the zero in $\Lambda=\lambda\,\mathrm{diag}(0,1,\ldots,1)$ leaves the bias weight unregularized. 
The test predictions are $\tilde Y_{\mathrm{te}}\!=\!\tilde X_{\mathrm{te}}W$. For each target column $j$, the capacity is defined as the squared Pearson correlation between the true and predicted test targets,
\begin{equation}
    \mathfrak{C}_j =
    \mathrm{corr}^{2}
    \!\left(
        Y_{\mathrm{te},:,j},
        \tilde Y_{\mathrm{te},:,j}
    \right).
    \label{eq:ipc_capacity_single}
\end{equation}
The IPC of order $D$ is obtained by summing the accepted capacities belonging to that order,
\begin{equation}
    \mathrm{IPC}_{D}
    =
    \sum_{j\in\mathcal{T}_{D}}
    \mathfrak{C}_j^{\mathrm{acc}},
    \qquad
    \mathrm{IPC}_{\mathrm{tot}}
    =
    \sum_{D=1}^{4}\mathrm{IPC}_{D}.
\end{equation}

To suppress finite-sample false positives, we use a shuffled-target cutoff. 
For each target $j$, the target values are randomly permuted $S$ times, the ridge readout is retrained for each shuffled target, and the corresponding shuffled capacities 
$\{C_{j,s}^{\mathrm{shuf}}\}_{s=1}^{S}$ are computed. 
The target-specific cutoff is chosen as
\begin{equation}
    \mathfrak{C}_{j,\mathrm{cut}}
    =
    \gamma\,
    Q_{q}
    \left(
        \{\mathfrak{C}_{j,s}^{\mathrm{shuf}}\}_{s=1}^{S}
    \right),
    \label{eq:ipc_shuffle_cutoff}
\end{equation}
where $Q_q$ denotes the $q$-quantile of the shuffled-capacity distribution. In the results reported in the main text, we use a safety factor \(\gamma\!=\!1.2\), \(S\!=\!1000\) shuffled surrogate trials, and the \(q\!=\!0.999\) quantile of the resulting shuffled-capacity distribution.
The accepted capacity is
\begin{equation}
    \mathfrak{C}_j^{\mathrm{acc}}
    =
    \begin{cases}
    \mathfrak{C}_j, & \mathfrak{C}_j > \mathfrak{C}_{j,\mathrm{cut}},\\
    0, & \mathfrak{C}_j \le \mathfrak{C}_{j,\mathrm{cut}}.
    \end{cases}
\end{equation}
Thus, a target contributes to the IPC only if its predictive capacity exceeds the level expected from finite-sample correlations with randomly permuted targets.


\section{Local quantum Fisher information and observable response}
\label{app:single_site_Response}
The decomposition into coherent and incoherent contributions yields refined upper bounds and conditional lower speed limits of the type introduced in Ref.~\cite{PhysRevX.12.011038}. Let \(\hat\rho_R(\phi)\) be a differentiable family of reduced density matrices with instantaneous spectral decomposition \(\hat\rho_R(\phi)\!=\!\sum_n \xi_n(\phi) \ket{n(\phi)}\!\bra{n(\phi)}.\) Its tangent can be separated as \(\partial_\phi \hat\rho_R\!=\!(\partial_\phi \hat\rho_R)_{\rm inc}\!+\!(\partial_\phi \hat\rho_R)_{\rm coh},\) where \((\partial_\phi \hat\rho_R)_{\rm inc}\!=\!\sum_n(\partial_\phi \xi_n)\ket{n}\!\bra{n},\) and \((\partial_\phi \hat\rho_R)_{\rm coh}\!=\!\sum_n \xi_n(\ket{\partial_\phi n}\!\bra{n}+\ket{n}\!\bra{\partial_\phi n}).\) We assume a locally smooth, nondegenerate spectrum. At degeneracies, the decomposition should instead be formulated using spectral projectors onto the degenerate eigenspaces; a decomposition into individual eigenvectors within a degenerate subspace is basis dependent. The incoherent part changes the spectrum, whereas the coherent part rotates the eigenbasis while preserving the eigenvalues to first order. For a Hermitian observable \(\hat{O}_R\), we can define its diagonal and off-diagonal components in the instantaneous eigenbasis of \(\hat\rho_R\), \(\hat O_{\rm inc}\!=\!\sum_n\bra{n}\hat O_R\ket{n}\ket{n}\!\bra{n},\,\hat O_{\rm coh}\!=\!\hat O_{R}-\hat O_{\rm inc}.\) Since the incoherent tangent is diagonal and the coherent tangent is off diagonal in the same basis, \(\partial_\phi\langle \hat O_R\rangle\!=\!v_{\rm inc}+v_{\rm coh},\) with \(v_{\rm inc}\!=\!\Tr[\hat O_{\rm inc}(\partial_\phi\hat\rho_R)_{\rm inc}],\) and \(v_{\rm coh}\!=\!\Tr[\hat O_{\rm coh}(\partial_\phi\hat\rho_R)_{\rm coh}].\) Correspondingly \(\mathcal F_{\rm Q}^\phi\!=\!\mathcal F_{\rm inc}^\phi+\mathcal F_{\rm coh}^\phi\). The coherent--incoherent observable speed limits of Ref.~\cite{PhysRevX.12.011038} give \(|v_{\rm inc}|\leq\Delta \hat O_{\rm inc}\sqrt{\mathcal F_{\rm inc}^\phi}\), and \(|v_{\rm coh}|\!\leq\!\Delta \hat O_{\rm coh}\sqrt{\mathcal F_{\rm coh}^\phi},\) where \(\Delta\hat O_s\!:=\!\sqrt{\Var_{\hat\rho_R}(\hat O_s)}\), with \(s\in\{\mathrm{inc},\mathrm{coh}\}\). Using the ordinary triangle inequality gives the resolved upper bound \(|\partial_\phi\langle\hat O_R\rangle|\!\leq\!\Delta \hat O_{\rm inc}\sqrt{\mathcal F_{\rm inc}^\phi}\!+\!\Delta \hat O_{\rm coh}\sqrt{\mathcal F_{\rm coh}^\phi}.\) Using the reverse triangle inequality gives \(|\partial_\phi\langle\hat O_R\rangle|\!\geq\!||v_{\rm inc}|-|v_{\rm coh}||.\) These yield the conditional lower speed limit \(|\partial_\phi\langle\hat O_R\rangle|\!\geq\! \max\{|v_{\rm coh}|-\Delta \hat O_{\rm inc}\sqrt{\mathcal F_{\rm inc}^\phi},|v_{\rm inc}|-\Delta \hat O_{\rm coh}\sqrt{\mathcal F_{\rm coh}^\phi} ,\;0\}.\) A nontrivial positive lower bound occurs only when a known contribution is larger than the maximum possible complementary contribution from the unknown sector.

Now consider a differentiable family of one-qubit reduced states \(\hat\rho_j(\phi)\!=\!\frac{1}{2}[\hat I+\mathbf r_j(\phi)\cdot\hat{\boldsymbol{\sigma}}]\), where \(\mathbf r_j\!=\!(r_{j,x},r_{j,y},r_{j,z}),r_{j,\alpha}\!=\!\langle\hat\sigma_j^\alpha\rangle,\) and $|\mathbf r_j|\leq 1$. For a mixed one-qubit state, the quantum Fisher information with respect to $\phi$ is~\cite{liu2020quantum}
\begin{equation}
\mathcal F_{{\rm Q},j}^{\phi}
=\left|\partial_\phi\mathbf r_j
\right|^2+\frac{\left(\mathbf r_j\cdot
\partial_\phi\mathbf r_j
\right)^2
}{1-|\mathbf r_j|^2
},\end{equation}
with the pure-state limit understood by continuity. At pure-state or other rank-changing points, the \(\mathcal F_{{\rm Q},j}^{\phi}\) is defined through the symmetric logarithmic derivative or the corresponding limiting spectral expression. The Euclidean term contains both radial and angular Bloch-vector motion, whereas the second term enhances the radial contribution as the state approaches the boundary of the Bloch ball. Using \(\left(\mathbf r_j\cdot\partial_\phi\mathbf r_j\right)^2\leq|\mathbf r_j|^2\left|\partial_\phi\mathbf r_j\right|^2,\) gives~\cite{plodzien2026operator}, \((1-|\mathbf r_j|^2)\mathcal F_{{\rm{Q}},j}^{\phi}\leq|\partial_\phi\mathbf r_j|^2\leq\mathcal F_{{\rm{Q}},j}^{\phi}.\)
Further, we can write \(\hat\rho(\phi)=\frac{1}{2}[\hat I+r(\phi)\,\hat{\mathbf n}(\phi)\cdot\hat{\boldsymbol\sigma}],|\hat{\mathbf n}|=1, \) and consider the most general Hermitian single-qubit observable \(\hat O=o_0\hat I+\mathbf o\cdot\hat{\boldsymbol\sigma}.\) The identity component \(o_0\hat I\) contributes neither to the variance nor to the response. The parts of the observable diagonal and off diagonal in the instantaneous eigenbasis of \(\hat\rho\) are
\begin{align}
\hat O_{\rm inc}&=o_0\hat I+o_\parallel\hat{\mathbf n}\cdot\hat{\boldsymbol\sigma},
\\\hat O_{\rm coh}
&=\mathbf o_\perp\cdot\hat{\boldsymbol\sigma}.
\end{align}
where \(o_\parallel\!=\!\mathbf o\cdot\hat{\mathbf n}\) and \(\mathbf o_\perp\!=\!\mathbf o-o_\parallel\hat{\mathbf n}\). The expectation value is \(\langle\hat O\rangle\!=\!o_0+r\,o_\parallel, \) and its derivative separates as
\begin{equation}
\partial_\phi\langle\hat O\rangle
=\underbrace{o_\parallel\,\partial_\phi r}_{v_{\rm inc}}+\underbrace{r\,\mathbf o\cdot\partial_\phi\hat{\mathbf n}
}_{v_{\rm coh}}.
\end{equation}
The first term changes the eigenvalues, while the second rotates the eigenbasis. The corresponding partial variances are
\begin{align}
(\Delta \hat O_{\rm inc})^2
&=o_\parallel^2(1-r^2),
\\(\Delta \hat O_{\rm coh})^2
&=|\mathbf o_\perp|^2=
|\mathbf o|^2-o_\parallel^2.
\end{align}
In radial--angular variables the quantum Fisher information becomes
\begin{equation}
\mathcal F_{\rm Q}^\phi=\frac{(\partial_\phi r)^2}{1-r^2}
+r^2\left|\partial_\phi\hat{\mathbf n}\right|^2,
\label{eq:qubit_qfi_Radial_Rngular}
\end{equation}
so that \(\mathcal F_{\rm inc}^\phi\!=\!{(\partial_\phi r)^2}/({1-r^2}),
\) and \(\mathcal F_{\rm coh}^\phi\!=\!r^2\left|\partial_\phi\hat{\mathbf n}\right|^2.\) For the incoherent part, the speed limit is saturated identically, \(\Delta \hat O_{\rm inc}\sqrt{\mathcal F_{\rm inc}^\phi}\!=\!|o_\parallel\partial_\phi r|\!=\!|v_{\rm inc}|.\) For the coherent contribution,
\begin{equation}
|v_{\rm coh}|
=r\left|\mathbf o_\perp\cdot\partial_\phi\hat{\mathbf n}
\right|\leq r|\mathbf o_\perp|
\left|\partial_\phi\hat{\mathbf n}
\right|=\Delta \hat O_{\rm coh}
\sqrt{\mathcal F_{\rm coh}^\phi},
\label{eq:coh_bound_qubit}
\end{equation}
with equality when \(\mathbf o_\perp\) is parallel or antiparallel to \(\partial_\phi\hat{\mathbf n}\). The coherent--incoherent upper bound therefore reads
\begin{equation}
\left|\partial_\phi\langle\hat O\rangle
\right|\leq|o_\parallel\partial_\phi r|
+r\sqrt{|\mathbf o|^2-o_\parallel^2}
\left|\partial_\phi\hat{\mathbf n}\right|.
\label{eq:qubit_upper_Resolved}
\end{equation}
The lower bound is \(| \partial_\phi\langle\hat O\rangle|\!\geq\!\max \{ r|\mathbf o\cdot\partial_\phi\hat{\mathbf n}|-|o_\parallel\partial_\phi r|,|o_\parallel\partial_\phi r|-r\sqrt{|\mathbf o|^2-o_\parallel^2}|\partial_\phi\hat{\mathbf n}|,\;0\}.\) A sharper reverse-triangle bound is, \(|\partial_\phi\langle\hat O\rangle|\geq||o_\parallel\partial_\phi r|-r|\mathbf o_\perp\cdot\partial_\phi\hat{\mathbf n}||.\) The full variance of \(\hat O\) is \(|\mathbf o|^2-r^2o_\parallel^2,\) and hence the unresolved response bound is
\begin{equation}
\left|\partial_\phi\langle\hat O\rangle\right|^2
\leq\mathcal F_Q^\phi\left(|\mathbf o|^2-r^2o_\parallel^2\right).
\end{equation}
For a normalized Pauli measurement along a unit direction \(\hat{\mathbf m}\),
\(\hat O\!=\!\hat{\mathbf m}\cdot\hat{\boldsymbol\sigma},\,|\hat{\mathbf m}|\!=\!1\), one sets \(\mathbf o=\hat{\mathbf m}\). The response becomes \(\partial_\phi\langle\hat{\mathbf m}\cdot\hat{\boldsymbol\sigma}\rangle\!=\!(\hat{\mathbf m}\cdot\hat{\mathbf n})\,\partial_\phi r+r\,\hat{\mathbf m}\cdot\partial_\phi\hat{\mathbf n}.\) If \(\hat{\mathbf m}\parallel\hat{\mathbf n}\), the response is purely incoherent. If \(\hat{\mathbf m}\perp\hat{\mathbf n}\), the instantaneous response is purely coherent. For a generic measurement direction, both contributions coexist and may interfere destructively.

For an individual Pauli observable $\hat\sigma_j^\alpha$, one has \(\Var_{\hat\rho_j}(\hat\sigma_j^\alpha)\!=\!1-\langle\hat\sigma_j^\alpha\rangle^2\). The generalized Cram\'er--Rao response inequality therefore becomes \(|\partial_\phi\langle\hat\sigma_j^\alpha\rangle|^2\leq\mathcal F_{{\rm{Q}},j}^{\phi}[1-\langle\hat\sigma_j^\alpha\rangle^2].\) This may also be written in measurement-theoretic form. Measuring $\hat\sigma_j^\alpha$ produces the binary probabilities \(p_{\pm}^{(\alpha)}\!=\!(1\pm\langle\hat\sigma_j^\alpha\rangle)/2\). The corresponding classical Fisher information is
\begin{align}
\mathcal F_{{\rm{cl}},j}^{(\alpha)}
=\sum_{s=\pm}\frac{\left(\partial_\phi p_s^{(\alpha)}
\right)^2}{p_s^{(\alpha)}}=\frac{\left|\partial_\phi\left\langle\hat\sigma_j^\alpha
\right\rangle\right|^2}{1-\left\langle\hat\sigma_j^\alpha\right\rangle^2
}.\end{align}
Consequently, \(\mathcal F_{{\rm cl},j}^{(\alpha)}\leq\mathcal F_{{\rm Q},j}^{\phi}.\) Thus, $\mathcal F_{{\rm Q},j}^{\phi}$ quantifies the maximum parameter sensitivity available from arbitrary measurements on site $j$, whereas $\mathcal F_{{\rm{cl}},j}^{(\alpha)}$ quantifies the sensitivity actually resolved by measuring the Pauli component $\alpha$. More generally, the response may be expressed geometrically as
\begin{equation}
\left|\partial_\phi\left\langle\hat\sigma_j^\alpha
\right\rangle\right|^2=\mathcal F_{{\rm Q},j}^{\phi}
\operatorname{Var}_{\hat\rho_j}
\left(\hat\sigma_j^\alpha\right)\cos^2\gamma_{j,\alpha},
\end{equation}
where $\gamma_{j,\alpha}$ denotes the angle between the symmetric-logarithmic-derivative tangent and the centered observable in the state-weighted operator geometry. Combining gives \(\mathcal F_{\mathrm{cl},j}^{(\alpha)}=\mathcal F_{Q,j}^{\phi}\cos^2\gamma_{j,\alpha},\) whenever \(\operatorname{Var}_{\hat\rho_j}(\hat\sigma_j^\alpha)>0 \),
with deterministic limiting cases understood by continuity. A local feature may therefore become insensitive for three limiting reasons:
\begin{equation}
\mathcal F_{{\rm Q},j}^{\phi}\rightarrow 0,
\qquad\operatorname{Var}_{\hat\rho_j}
\left(\hat\sigma_j^\alpha\right)\rightarrow 0,
\qquad\cos^2\gamma_{j,\alpha}\rightarrow 0.
\end{equation}
These correspond, respectively, to loss of locally available distinguishability, concentration of the state near an eigenstate of the measured observable, and misalignment between the available state motion and the chosen measurement axis. These mechanisms are not, in general, statistically or physically independent.

The pointwise response bound remains valid for each circuit realization $\hat U$. Averaging over the circuit ensemble gives
\begin{equation}
\mathbb E_U\left[\left|\partial_\phi\left\langle
\hat\sigma_j^\alpha\right\rangle_U\right|^2
\right]\leq\mathbb E_U\left[\mathcal F_{{\rm Q},j}^{\phi}(U)\left(
1-\left\langle\hat\sigma_j^\alpha\right\rangle_U^2\right)\right].
\end{equation}
The right-hand side in general cannot be factorized. Since \(0\!\leq\!1-\left\langle\hat\sigma_j^\alpha\right\rangle^2\!\leq\!1,\) one nevertheless obtains
\begin{equation}
\begin{aligned} 
\mathbb E_U
\left[\left|\partial_\phi
\left\langle\hat\sigma_j^\alpha\right\rangle_U \right|^2
\right]
\leq\mathbb E_U\left[\mathcal F_{{\rm Q},j}^{\phi}
\right].
\end{aligned}
\end{equation}
If the ensemble-averaged single-site Fisher information is exponentially suppressed, \(\mathbb E_U[\mathcal F_{{\rm Q},j}^{\phi}]\in\mathcal O(e^{-aN}),\,a\!>\!0,\) then every bounded single-site Pauli response is necessarily exponentially suppressed, \(\mathbb E_U[|\partial_\phi\left\langle\hat\sigma_j^\alpha\right\rangle_U|^2]\in\mathcal O(e^{-aN}).\) The converse does not necessarily follow.

Now regard a single-site reservoir feature as the scalar input-dependent function,
\(C_j^\alpha(\phi,U)\!=\!\left\langle\hat\sigma_j^\alpha\right\rangle_U.\) Its circuit-ensemble gradient variance is
\begin{align}
\operatorname{Var}_U
\left[\partial_\phi C_j^\alpha\right]
&=\mathbb E_U
\left[\left(\partial_\phi C_j^\alpha\right)^2\right]-\mathbb E_U
\left[\partial_\phi C_j^\alpha
\right]^2\nonumber\\
&\leq\mathbb E_U
\left[\left(\partial_\phi C_j^\alpha\right)^2\right],
\end{align}
which gives
\begin{equation}
\operatorname{Var}_U
\left[\partial_\phi C_j^\alpha\right]
\leq\mathbb E_U\left[\mathcal F_{Q,j}^{\phi}
\left(1-\left\langle\hat\sigma_j^\alpha
\right\rangle_U^2\right)\right].
\label{eq:gradient_variance_local_qfi_bound}
\end{equation}
When \(\mathbb E_U\left[\partial_\phi C_j^\alpha\right]\!=\!0,\) the gradient second moment and gradient variance coincide. In this case, the exponential suppression of one is equivalent to the suppression of the other. An exponentially small averaged response scale is then sufficient to imply an exponentially small gradient variance. The converse is again false. Note also this is in general different from conventional variational     barren plateau, but rather an input-response concentration with analogous scaling. 

For a Haar-random pure state on \(N\) qubits, the single-qubit reduced state is exponentially close to maximally mixed. Its Bloch vector satisfies \(\mathbb E_U[r_\alpha]\!=\!0,\,\mathbb E_U[r_\alpha^2] \!=\!{1/(2^N+1)},\,\alpha\!\in\!\{x,y,z\}, \) and therefore \(\mathbb E_U[|\mathbf r|^2] \!=\!{3/(2^N+1)}\). For a traceless single-qubit observable \(\hat O\!=\!\mathbf o\cdot\hat{\boldsymbol\sigma}\), \(\mathbb E_U [\langle\hat O\rangle^2]\!=\!|\mathbf o|^2/(2^N+1),\) while \(\mathbb E_U[\Var_{\hat\rho_U}(\hat O)]\!=\!(2^N|\mathbf o|^2)/(2^N+1)\!\rightarrow\!|\mathbf o|^2.\) Thus the in-state fluctuation of a bounded local observable remains order one, and its expectation value concentrates exponentially close to zero across circuit realizations. When the locally retained information is exponentially suppressed, \(\mathbb E_U[\mathcal F_{{\rm Q},j}^\phi] \in\mathcal O(e^{-aN})\), the averaged response of every bounded single-site observable is correspondingly suppressed \(\mathbb E_U[|\partial_\phi\langle\hat O_j\rangle|^2]\leq\|\hat O_j\|_\infty^2\mathbb E_U[\mathcal F_{{\rm Q},j}^\phi].\) The suppression is therefore controlled by the loss of locally accessible distinguishability or by measurement-axis misalignment. A larger block may retain nonzero quantum Fisher information even though no single-site observable exhibits an appreciable response. Note the Haar typicality of \(\hat\rho_j\) alone does not imply suppression of \(\mathcal F_{Q,j}^{\phi}\), since the latter also depends on the parameter derivative \(\partial_\phi\hat\rho_j\). Such suppression follows for Haar-scrambled parameterized ensembles only because the same random dynamics jointly typicalizes the state and its tangent, leading to an exponentially small reduced tangent norm for fixed local subsystems~\cite{wysocki2026volume}. Suppression of the local quantum Fisher information must therefore be established from the parameter-dependent dynamics or ensemble.

\bibliography{ref.bib}

@article{baez2022renyi,
  title={R{\'e}nyi entropy and free energy},
  author={Baez, John C},
  journal={Entropy},
  volume={24},
  number={5},
  pages={706},
  year={2022},
  publisher={MDPI}
}

@article{andrzejewski2023evolution,
  title={Evolution of capacity of entanglement and modular entropy in harmonic chains and scalar fields},
  author={Andrzejewski, Krzysztof},
  journal={Physical Review D},
  volume={108},
  number={12},
  pages={125013},
  year={2023},
  publisher={APS}
}

@incollection{rangamani2017holographic,
  title={Holographic entanglement entropy},
  author={Rangamani, Mukund and Takayanagi, Tadashi},
  booktitle={Holographic entanglement entropy},
  pages={35--47},
  year={2017},
  publisher={Springer}
}

@article{nandy2021capacity,
  title={Capacity of entanglement in local operators},
  author={Nandy, Pratik},
  journal={Journal of High Energy Physics},
  volume={2021},
  number={7},
  pages={1--23},
  year={2021},
  publisher={Springer}
}

@article{yang2015two,
  title = {Two-Component Structure in the Entanglement Spectrum of Highly Excited States},
  author = {Yang, Zhi-Cheng and Chamon, Claudio and Hamma, Alioscia and Mucciolo, Eduardo R.},
  journal = {Phys. Rev. Lett.},
  volume = {115},
  issue = {26},
  pages = {267206},
  numpages = {5},
  year = {2015},
  month = {Dec},
  publisher = {American Physical Society},
  doi = {10.1103/PhysRevLett.115.267206},
  url = {https://link.aps.org/doi/10.1103/PhysRevLett.115.267206}
}

@article{dong2016gravity,
  title={The gravity dual of R{\'e}nyi entropy},
  author={Dong, Xi},
  journal={Nature communications},
  volume={7},
  number={1},
  pages={12472},
  year={2016},
  publisher={Nature Publishing Group UK London}
}

@article{serbyn2016power,
  title={Power-law entanglement spectrum in many-body localized phases},
  author={Serbyn, Maksym and Michailidis, Alexios A and Abanin, Dmitry A and Papi{\'c}, Zlatko},
  journal={Physical review letters},
  volume={117},
  number={16},
  pages={160601},
  year={2016},
  publisher={APS}
}

@article{aaronson2014quantifying,
  title={Quantifying the rise and fall of complexity in closed systems: the coffee automaton},
  author={Aaronson, Scott and Carroll, Sean M and Ouellette, Lauren},
  journal={arXiv preprint arXiv:1405.6903},
  year={2014}
}

@article{lloyd2002computational,
  title = {Computational Capacity of the Universe},
  author = {Lloyd, Seth},
  journal = {Phys. Rev. Lett.},
  volume = {88},
  issue = {23},
  pages = {237901},
  numpages = {4},
  year = {2002},
  month = {May},
  publisher = {American Physical Society},
  doi = {10.1103/PhysRevLett.88.237901},
  url = {https://link.aps.org/doi/10.1103/PhysRevLett.88.237901}
}

@article{marvcenko1967distribution,
  title={Distribution of eigenvalues for some sets of random matrices},
  author={Mar{\v{c}}enko, Vladimir A and Pastur, Leonid Andreevich},
  journal={Mathematics of the USSR-Sbornik},
  volume={1},
  number={4},
  pages={457},
  year={1967},
  publisher={IOP Publishing}
}

@article{tirrito2024quantifying,
  title={Quantifying nonstabilizerness through entanglement spectrum flatness},
  author={Tirrito, Emanuele and Tarabunga, Poetri Sonya and Lami, Gugliemo and Chanda, Titas and Leone, Lorenzo and Oliviero, Salvatore FE and Dalmonte, Marcello and Collura, Mario and Hamma, Alioscia},
  journal={Physical Review A},
  volume={109},
  number={4},
  pages={L040401},
  year={2024},
  publisher={APS}
}

@article{landauer1961irreversibility,
  title={Irreversibility and heat generation in the computing process},
  author={Landauer, Rolf},
  journal={IBM journal of research and development},
  volume={5},
  number={3},
  pages={183--191},
  year={1961},
  publisher={Ibm}
}

@article{lloyd2000ultimate,
  title={Ultimate physical limits to computation},
  author={Lloyd, Seth},
  journal={Nature},
  volume={406},
  number={6799},
  pages={1047--1054},
  year={2000},
  publisher={Nature Publishing Group UK London}
}

@article{okuyama2021capacity,
  title={Capacity of entanglement in random pure state},
  author={Okuyama, Kazumi},
  journal={Physics Letters B},
  volume={820},
  pages={136600},
  year={2021},
  publisher={Elsevier}
}

@article{varikuti2026impact,
  title={Impact of Clifford operations on non-stabilizing power and quantum chaos},
  author={Varikuti, Naga Dileep and Bandyopadhyay, Soumik and Hauke, Philipp},
  journal={Quantum},
  volume={10},
  pages={2017},
  year={2026},
  publisher={Verein zur F{\"o}rderung des Open Access Publizierens in den Quantenwissenschaften}
}

@article{PhysRevA.67.052301,
  title = {Quantum dynamics as a physical resource},
  author = {Nielsen, Michael A. and Dawson, Christopher M. and Dodd, Jennifer L. and Gilchrist, Alexei and Mortimer, Duncan and Osborne, Tobias J. and Bremner, Michael J. and Harrow, Aram W. and Hines, Andrew},
  journal = {Phys. Rev. A},
  volume = {67},
  issue = {5},
  pages = {052301},
  numpages = {19},
  year = {2003},
  month = {May},
  publisher = {American Physical Society},
  doi = {10.1103/PhysRevA.67.052301},
  url = {https://link.aps.org/doi/10.1103/PhysRevA.67.052301}
}

@article{PhysRevD.97.086015,
  title = {Second law of quantum complexity},
  author = {Brown, Adam R. and Susskind, Leonard},
  journal = {Phys. Rev. D},
  volume = {97},
  issue = {8},
  pages = {086015},
  numpages = {29},
  year = {2018},
  month = {Apr},
  publisher = {American Physical Society},
  doi = {10.1103/PhysRevD.97.086015},
  url = {https://link.aps.org/doi/10.1103/PhysRevD.97.086015}
}

@article{PhysRevA.62.030301,
  title = {Entangling power of quantum evolutions},
  author = {Zanardi, Paolo and Zalka, Christof and Faoro, Lara},
  journal = {Phys. Rev. A},
  volume = {62},
  issue = {3},
  pages = {030301},
  numpages = {4},
  year = {2000},
  month = {Aug},
  publisher = {American Physical Society},
  doi = {10.1103/PhysRevA.62.030301},
  url = {https://link.aps.org/doi/10.1103/PhysRevA.62.030301}
}

@article{PhysRevLett.128.050402,
  title = {Stabilizer R\'enyi Entropy},
  author = {Leone, Lorenzo and Oliviero, Salvatore F. E. and Hamma, Alioscia},
  journal = {Phys. Rev. Lett.},
  volume = {128},
  issue = {5},
  pages = {050402},
  numpages = {5},
  year = {2022},
  month = {Feb},
  publisher = {American Physical Society},
  doi = {10.1103/PhysRevLett.128.050402},
  url = {https://link.aps.org/doi/10.1103/PhysRevLett.128.050402}
}

@article{jasser2025stabilizer,
  title = {Stabilizer entropy and entanglement complexity in the Sachdev-Ye-Kitaev model},
  author = {Jasser, Barbara and Odavi\ifmmode \acute{c}\else \'{c}\fi{}, Jovan and Hamma, Alioscia},
  journal = {Phys. Rev. B},
  volume = {112},
  issue = {17},
  pages = {174204},
  numpages = {16},
  year = {2025},
  month = {Nov},
  publisher = {American Physical Society},
  doi = {10.1103/rz86-47h3},
  url = {https://link.aps.org/doi/10.1103/rz86-47h3}
}

@article{y9r6-dx7p,
  title = {Stabilizer entropy in nonintegrable quantum evolutions},
  author = {Odavi\ifmmode \acute{c}\else \'{c}\fi{}, J. and Viscardi, M. and Hamma, A.},
  journal = {Phys. Rev. B},
  volume = {112},
  issue = {10},
  pages = {104301},
  numpages = {14},
  year = {2025},
  month = {Sep},
  publisher = {American Physical Society},
  doi = {10.1103/y9r6-dx7p},
  url = {https://link.aps.org/doi/10.1103/y9r6-dx7p}
}

@article{PhysRevD.99.066012,
  title = {Aspects of capacity of entanglement},
  author = {de Boer, Jan and J\"arvel\"a, Jarkko and Keski-Vakkuri, Esko},
  journal = {Phys. Rev. D},
  volume = {99},
  issue = {6},
  pages = {066012},
  numpages = {34},
  year = {2019},
  month = {Mar},
  publisher = {American Physical Society},
  doi = {10.1103/PhysRevD.99.066012},
  url = {https://link.aps.org/doi/10.1103/PhysRevD.99.066012}
}

@article{z3vr-w5c5,
  title = {Gravitational Backreaction is Magical},
  author = {Cao, ChunJun and Cheng, Gong and Hamma, Alioscia and Leone, Lorenzo and Munizzi, William and Oliviero, Savatore F.E.},
  journal = {PRX Quantum},
  volume = {6},
  issue = {4},
  pages = {040375},
  numpages = {39},
  year = {2025},
  month = {Dec},
  publisher = {American Physical Society},
  doi = {10.1103/z3vr-w5c5},
  url = {https://link.aps.org/doi/10.1103/z3vr-w5c5}
}

@article{p7xt-s9nz,
  title = {Magic Resources of the Heisenberg Picture},
  author = {Dowling, Neil and Kos, Pavel and Turkeshi, Xhek},
  journal = {Phys. Rev. Lett.},
  volume = {135},
  issue = {5},
  pages = {050401},
  numpages = {10},
  year = {2025},
  month = {Jul},
  publisher = {American Physical Society},
  doi = {10.1103/p7xt-s9nz},
  url = {https://link.aps.org/doi/10.1103/p7xt-s9nz}
}

@article{c7k1-xcwy,
  title = {Bridging Entanglement and Magic Resources within Operator Space},
  author = {Dowling, Neil and Modi, Kavan and White, Gregory A. L.},
  journal = {Phys. Rev. Lett.},
  volume = {135},
  issue = {16},
  pages = {160201},
  numpages = {9},
  year = {2025},
  month = {Oct},
  publisher = {American Physical Society},
  doi = {10.1103/c7k1-xcwy},
  url = {https://link.aps.org/doi/10.1103/c7k1-xcwy}
}

@article{rudolph2025pauli,
  title={Pauli propagation: A computational framework for simulating quantum systems},
  author={Rudolph, Manuel S and Jones, Tyson and Teng, Yanting and Angrisani, Armando and Holmes, Zo{\"e}},
  journal={arXiv preprint arXiv:2505.21606},
  year={2025}
}

@article{aditya2025growth,
  title={Growth and spreading of quantum resources under random circuit dynamics},
  author={Aditya, Sreemayee and Turkeshi, Xhek and Sierant, Piotr},
  journal={arXiv preprint arXiv:2512.14827},
  year={2025}
}

@article{deutsch1985quantum,
  title={Quantum theory, the Church--Turing principle and the universal quantum computer},
  author={Deutsch, David},
  journal={Proceedings of the Royal Society of London. A. Mathematical and Physical Sciences},
  volume={400},
  number={1818},
  pages={97--117},
  year={1985},
  publisher={The Royal Society London}
}

@article{hopfield1982neural,
  title={Neural networks and physical systems with emergent collective computational abilities.},
  author={Hopfield, John J},
  journal={Proceedings of the national academy of sciences},
  volume={79},
  number={8},
  pages={2554--2558},
  year={1982}
}

@article{nakajima2020physical,
  title={Physical reservoir computing—an introductory perspective},
  author={Nakajima, Kohei},
  journal={Japanese Journal of Applied Physics},
  volume={59},
  number={6},
  pages={060501},
  year={2020},
  publisher={IOP Publishing}
}

@article{roberts2017chaos,
  title={Chaos and complexity by design},
  author={Roberts, Daniel A and Yoshida, Beni},
  journal={Journal of High Energy Physics},
  volume={2017},
  number={4},
  pages={1--64},
  year={2017},
  publisher={Springer}
}

@article{cotler2017chaos,
  title={Chaos, complexity, and random matrices},
  author={Cotler, Jordan and Hunter-Jones, Nicholas and Liu, Junyu and Yoshida, Beni},
  journal={Journal of High Energy Physics},
  volume={2017},
  number={11},
  pages={1--60},
  year={2017},
  publisher={Springer}
}

@article{mi2021information,
  title={Information scrambling in quantum circuits},
  author={Mi, Xiao and Roushan, Pedram and Quintana, Chris and Mandra, Salvatore and Marshall, Jeffrey and Neill, Charles and Arute, Frank and Arya, Kunal and Atalaya, Juan and Babbush, Ryan and others},
  journal={Science},
  volume={374},
  number={6574},
  pages={1479--1483},
  year={2021},
  publisher={American Association for the Advancement of Science}
}

@article{RevModPhys.91.025001,
  title = {Quantum resource theories},
  author = {Chitambar, Eric and Gour, Gilad},
  journal = {Rev. Mod. Phys.},
  volume = {91},
  issue = {2},
  pages = {025001},
  numpages = {48},
  year = {2019},
  month = {Apr},
  publisher = {American Physical Society},
  doi = {10.1103/RevModPhys.91.025001},
  url = {https://link.aps.org/doi/10.1103/RevModPhys.91.025001}
}

@article{gottesman1998heisenberg,
  title={The Heisenberg representation of quantum computers},
  author={Gottesman, Daniel},
  journal={arXiv preprint quant-ph/9807006},
  year={1998}
}

@article{PhysRevA.70.052328,
  title = {Improved simulation of stabilizer circuits},
  author = {Aaronson, Scott and Gottesman, Daniel},
  journal = {Phys. Rev. A},
  volume = {70},
  issue = {5},
  pages = {052328},
  numpages = {14},
  year = {2004},
  month = {Nov},
  publisher = {American Physical Society},
  doi = {10.1103/PhysRevA.70.052328},
  url = {https://link.aps.org/doi/10.1103/PhysRevA.70.052328}
}

@article{leone2021quantum,
  title={Quantum chaos is quantum},
  author={Leone, Lorenzo and Oliviero, Salvatore FE and Zhou, You and Hamma, Alioscia},
  journal={Quantum},
  volume={5},
  pages={453},
  year={2021},
  publisher={Verein zur F{\"o}rderung des Open Access Publizierens in den Quantenwissenschaften}
}

@article{haug2025probing,
  title={Probing quantum complexity via universal saturation of stabilizer entropies},
  author={Haug, Tobias and Aolita, Leandro and Kim, MS},
  journal={Quantum},
  volume={9},
  pages={1801},
  year={2025},
  publisher={Verein zur F{\"o}rderung des Open Access Publizierens in den Quantenwissenschaften}
}

@article{PhysRevA.110.L040403,
  title = {Stabilizer entropies are monotones for magic-state resource theory},
  author = {Leone, Lorenzo and Bittel, Lennart},
  journal = {Phys. Rev. A},
  volume = {110},
  issue = {4},
  pages = {L040403},
  numpages = {6},
  year = {2024},
  month = {Oct},
  publisher = {American Physical Society},
  doi = {10.1103/PhysRevA.110.L040403},
  url = {https://link.aps.org/doi/10.1103/PhysRevA.110.L040403}
}

@article{viscardi2026interplay,
  title={Interplay of entanglement structures and stabilizer entropy in spin models},
  author={Viscardi, Michele and Dalmonte, Marcello and Hamma, Alioscia and Tirrito, Emanuele},
  journal={SciPost Physics Core},
  volume={9},
  number={1},
  pages={012},
  year={2026}
}

@article{PhysRevLett.105.080501,
  title = {Entanglement Entropy and Entanglement Spectrum of the Kitaev Model},
  author = {Yao, Hong and Qi, Xiao-Liang},
  journal = {Phys. Rev. Lett.},
  volume = {105},
  issue = {8},
  pages = {080501},
  numpages = {4},
  year = {2010},
  month = {Aug},
  publisher = {American Physical Society},
  doi = {10.1103/PhysRevLett.105.080501},
  url = {https://link.aps.org/doi/10.1103/PhysRevLett.105.080501}
}

@article{kawabata2021probing,
  title={Probing Hawking radiation through capacity of entanglement},
  author={Kawabata, Kohki and Nishioka, Tatsuma and Okuyama, Yoshitaka and Watanabe, Kento},
  journal={Journal of High Energy Physics},
  volume={2021},
  number={5},
  pages={1--27},
  year={2021},
  publisher={Springer}
}

@article{geraedts2016many,
  title={Many-body localization and thermalization: Insights from the entanglement spectrum},
  author={Geraedts, Scott D and Nandkishore, Rahul and Regnault, Nicolas},
  journal={Physical Review B},
  volume={93},
  number={17},
  pages={174202},
  year={2016},
  publisher={APS}
}

@article{marconi2008fluctuation,
  title={Fluctuation--dissipation: response theory in statistical physics},
  author={Marconi, Umberto Marini Bettolo and Puglisi, Andrea and Rondoni, Lamberto and Vulpiani, Angelo},
  journal={Physics reports},
  volume={461},
  number={4-6},
  pages={111--195},
  year={2008},
  publisher={Elsevier}
}

@article{PhysRevB.83.115322,
  title = {Entanglement spectrum and entanglement thermodynamics of quantum Hall bilayers at \(\ensuremath{\nu}=1\)},
  author = {Schliemann, John},
  journal = {Phys. Rev. B},
  volume = {83},
  issue = {11},
  pages = {115322},
  numpages = {5},
  year = {2011},
  month = {Mar},
  publisher = {American Physical Society},
  doi = {10.1103/PhysRevB.83.115322},
  url = {https://link.aps.org/doi/10.1103/PhysRevB.83.115322}
}

@article{PhysRevLett.71.1291,
  title = {Average entropy of a subsystem},
  author = {Page, Don N.},
  journal = {Phys. Rev. Lett.},
  volume = {71},
  issue = {9},
  pages = {1291--1294},
  numpages = {0},
  year = {1993},
  month = {Aug},
  publisher = {American Physical Society},
  doi = {10.1103/PhysRevLett.71.1291},
  url = {https://link.aps.org/doi/10.1103/PhysRevLett.71.1291}
}

@article{baiguera2026quantum,
  title={Quantum complexity in gravity, quantum field theory, and quantum information science},
  author={Baiguera, Stefano and Balasubramanian, Vijay and Caputa, Pawel and Chapman, Shira and Haferkamp, Jonas and Heller, Michal P and Halpern, Nicole Yunger},
  journal={Physics Reports},
  volume={1159},
  pages={1--77},
  year={2026},
  publisher={Elsevier}
}

@article{PRXQuantum.2.030316,
  title = {Models of Quantum Complexity Growth},
  author = {Brand\~ao, Fernando G.S.L. and Chemissany, Wissam and Hunter-Jones, Nicholas and Kueng, Richard and Preskill, John},
  journal = {PRX Quantum},
  volume = {2},
  issue = {3},
  pages = {030316},
  numpages = {40},
  year = {2021},
  month = {Jul},
  publisher = {American Physical Society},
  doi = {10.1103/PRXQuantum.2.030316},
  url = {https://link.aps.org/doi/10.1103/PRXQuantum.2.030316}
}

@article{thanasilp2024exponential,
  title={Exponential concentration in quantum kernel methods},
  author={Thanasilp, Supanut and Wang, Samson and Cerezo, Marco and Holmes, Zo{\"e}},
  journal={Nature communications},
  volume={15},
  number={1},
  pages={5200},
  year={2024},
  publisher={Nature Publishing Group UK London}
}

@article{larocca2025barren,
  title={Barren plateaus in variational quantum computing},
  author={Larocca, Martin and Thanasilp, Supanut and Wang, Samson and Sharma, Kunal and Biamonte, Jacob and Coles, Patrick J and Cincio, Lukasz and McClean, Jarrod R and Holmes, Zo{\"e} and Cerezo, Marco},
  journal={Nature Reviews Physics},
  volume={7},
  number={4},
  pages={174--189},
  year={2025},
  publisher={Nature Publishing Group UK London}
}

@article{PhysRevA.106.042419,
  title = {Capacity of entanglement for a nonlocal Hamiltonian},
  author = {Shrimali, Divyansh and Bhowmick, Swapnil and Pandey, Vivek and Pati, Arun Kumar},
  journal = {Phys. Rev. A},
  volume = {106},
  issue = {4},
  pages = {042419},
  numpages = {12},
  year = {2022},
  month = {Oct},
  publisher = {American Physical Society},
  doi = {10.1103/PhysRevA.106.042419},
  url = {https://link.aps.org/doi/10.1103/PhysRevA.106.042419}
}

@article{j2qj-vwcl,
  title = {Edge of Many-Body Quantum Chaos in Quantum Reservoir Computing},
  author = {Kobayashi, Kaito and Motome, Yukitoshi},
  journal = {Phys. Rev. Lett.},
  volume = {136},
  issue = {4},
  pages = {040602},
  numpages = {8},
  year = {2026},
  month = {Jan},
  publisher = {American Physical Society},
  doi = {10.1103/j2qj-vwcl},
  url = {https://link.aps.org/doi/10.1103/j2qj-vwcl}
}

@article{PhysRevLett.127.100502,
  title = {Dynamical Phase Transitions in Quantum Reservoir Computing},
  author = {Mart\'{\i}nez-Pe\~na, Rodrigo and Giorgi, Gian Luca and Nokkala, Johannes and Soriano, Miguel C. and Zambrini, Roberta},
  journal = {Phys. Rev. Lett.},
  volume = {127},
  issue = {10},
  pages = {100502},
  numpages = {7},
  year = {2021},
  month = {Aug},
  publisher = {American Physical Society},
  doi = {10.1103/PhysRevLett.127.100502},
  url = {https://link.aps.org/doi/10.1103/PhysRevLett.127.100502}
}

@article{gq9r-d5q8,
  title = {Quantum reservoir computing on random regular graphs},
  author = {Ivaki, Moein N. and Lazarides, Achilleas and Ala-Nissila, Tapio},
  journal = {Phys. Rev. A},
  volume = {112},
  issue = {1},
  pages = {012622},
  numpages = {9},
  year = {2025},
  month = {Jul},
  publisher = {American Physical Society},
  doi = {10.1103/gq9r-d5q8},
  url = {https://link.aps.org/doi/10.1103/gq9r-d5q8}
}

@article{xia2022reservoir,
  title={The reservoir learning power across quantum many-body localization transition},
  author={Xia, Wei and Zou, Jie and Qiu, Xingze and Li, Xiaopeng},
  journal={Frontiers of Physics},
  volume={17},
  number={3},
  pages={33506},
  year={2022},
  publisher={Springer}
}

@article{vcindrak2026memory,
  title={Memory-Nonlinearity Trade-off across Quantum Reservoir Computing Frameworks},
  author={{\v{C}}indrak, Saud and Giebeler, Lara and G{\"o}tting, Niclas and Gies, Christopher and L{\"u}dge, Kathy},
  journal={arXiv preprint arXiv:2603.21371},
  year={2026}
}

@article{PhysRevA.67.042313,
  title = {Geometric theory of nonlocal two-qubit operations},
  author = {Zhang, Jun and Vala, Jiri and Sastry, Shankar and Whaley, K. Birgitta},
  journal = {Phys. Rev. A},
  volume = {67},
  issue = {4},
  pages = {042313},
  numpages = {18},
  year = {2003},
  month = {Apr},
  publisher = {American Physical Society},
  doi = {10.1103/PhysRevA.67.042313},
  url = {https://link.aps.org/doi/10.1103/PhysRevA.67.042313}
}

@article{PhysRevResearch.2.043126,
  title = {Entanglement measures of bipartite quantum gates and their thermalization under arbitrary interaction strength},
  author = {Jonnadula, Bhargavi and Mandayam, Prabha and \ifmmode \dot{Z}\else \.{Z}\fi{}yczkowski, Karol and Lakshminarayan, Arul},
  journal = {Phys. Rev. Res.},
  volume = {2},
  issue = {4},
  pages = {043126},
  numpages = {19},
  year = {2020},
  month = {Oct},
  publisher = {American Physical Society},
  doi = {10.1103/PhysRevResearch.2.043126},
  url = {https://link.aps.org/doi/10.1103/PhysRevResearch.2.043126}
}

@article{PhysRevA.95.040302,
  title = {Impact of local dynamics on entangling power},
  author = {Jonnadula, Bhargavi and Mandayam, Prabha and \ifmmode \dot{Z}\else \.{Z}\fi{}yczkowski, Karol and Lakshminarayan, Arul},
  journal = {Phys. Rev. A},
  volume = {95},
  issue = {4},
  pages = {040302},
  numpages = {5},
  year = {2017},
  month = {Apr},
  publisher = {American Physical Society},
  doi = {10.1103/PhysRevA.95.040302},
  url = {https://link.aps.org/doi/10.1103/PhysRevA.95.040302}
}

@article{PhysRevApplied.8.024030,
  title = {Harnessing Disordered-Ensemble Quantum Dynamics for Machine Learning},
  author = {Fujii, Keisuke and Nakajima, Kohei},
  journal = {Phys. Rev. Appl.},
  volume = {8},
  issue = {2},
  pages = {024030},
  numpages = {20},
  year = {2017},
  month = {Aug},
  publisher = {American Physical Society},
  doi = {10.1103/PhysRevApplied.8.024030},
  url = {https://link.aps.org/doi/10.1103/PhysRevApplied.8.024030}
}

@article{mujal2021opportunities,
  title={Opportunities in quantum reservoir computing and extreme learning machines},
  author={Mujal, Pere and Mart{\'\i}nez-Pe{\~n}a, Rodrigo and Nokkala, Johannes and Garc{\'\i}a-Beni, Jorge and Giorgi, Gian Luca and Soriano, Miguel C and Zambrini, Roberta},
  journal={Advanced Quantum Technologies},
  volume={4},
  number={8},
  pages={2100027},
  year={2021},
  publisher={Wiley Online Library}
}

@article{cerezo2021variational,
  title={Variational quantum algorithms},
  author={Cerezo, Marco and Arrasmith, Andrew and Babbush, Ryan and Benjamin, Simon C and Endo, Suguru and Fujii, Keisuke and McClean, Jarrod R and Mitarai, Kosuke and Yuan, Xiao and Cincio, Lukasz and others},
  journal={Nature Reviews Physics},
  volume={3},
  number={9},
  pages={625--644},
  year={2021},
  publisher={Nature Publishing Group UK London}
}

@article{gottesman2024surviving,
  title={Surviving as a quantum computer in a classical world},
  author={Gottesman, Daniel},
  journal={Textbook manuscript preprint},
  volume={8},
  number={8.1},
  pages={8--2},
  year={2024}
}

@article{tarabunga2025efficient,
  title={Efficient mutual magic and magic capacity with matrix product states},
  author={Tarabunga, Poetri Sonya and Haug, Tobias},
  journal={SciPost Physics},
  volume={19},
  number={4},
  pages={085},
  year={2025}
}

@article{PhysRevX.12.011038,
  title = {Unifying Quantum and Classical Speed Limits on Observables},
  author = {Garc\'{\i}a-Pintos, Luis Pedro and Nicholson, Schuyler B. and Green, Jason R. and del Campo, Adolfo and Gorshkov, Alexey V.},
  journal = {Phys. Rev. X},
  volume = {12},
  issue = {1},
  pages = {011038},
  numpages = {22},
  year = {2022},
  month = {Feb},
  publisher = {American Physical Society},
  doi = {10.1103/PhysRevX.12.011038},
  url = {https://link.aps.org/doi/10.1103/PhysRevX.12.011038}
}

@article{PhysRevLett.110.050403,
  title = {Quantum Speed Limits in Open System Dynamics},
  author = {del Campo, A. and Egusquiza, I. L. and Plenio, M. B. and Huelga, S. F.},
  journal = {Phys. Rev. Lett.},
  volume = {110},
  issue = {5},
  pages = {050403},
  numpages = {5},
  year = {2013},
  month = {Jan},
  publisher = {American Physical Society},
  doi = {10.1103/PhysRevLett.110.050403},
  url = {https://link.aps.org/doi/10.1103/PhysRevLett.110.050403}
}

@article{PhysRevLett.110.050402,
  title = {Quantum Speed Limit for Physical Processes},
  author = {Taddei, M. M. and Escher, B. M. and Davidovich, L. and de Matos Filho, R. L.},
  journal = {Phys. Rev. Lett.},
  volume = {110},
  issue = {5},
  pages = {050402},
  numpages = {5},
  year = {2013},
  month = {Jan},
  publisher = {American Physical Society},
  doi = {10.1103/PhysRevLett.110.050402},
  url = {https://link.aps.org/doi/10.1103/PhysRevLett.110.050402}
}

@article{RevModPhys.89.035002,
  title = {Quantum sensing},
  author = {Degen, C. L. and Reinhard, F. and Cappellaro, P.},
  journal = {Rev. Mod. Phys.},
  volume = {89},
  issue = {3},
  pages = {035002},
  numpages = {39},
  year = {2017},
  month = {Jul},
  publisher = {American Physical Society},
  doi = {10.1103/RevModPhys.89.035002},
  url = {https://link.aps.org/doi/10.1103/RevModPhys.89.035002}
}

@article{nicholson2020time,
  title={Time--information uncertainty relations in thermodynamics},
  author={Nicholson, Schuyler B and Garc{\'\i}a-Pintos, Luis Pedro and del Campo, Adolfo and Green, Jason R},
  journal={Nature Physics},
  volume={16},
  number={12},
  pages={1211--1215},
  year={2020},
  publisher={Nature Publishing Group UK London}
}

@article{xiong2025role,
  title={Role of scrambling and noise in temporal information processing with quantum systems},
  author={Xiong, Weijie and Holmes, Zo{\"e} and Angrisani, Armando and Suzuki, Yudai and Chotibut, Thiparat and Thanasilp, Supanut},
  journal={arXiv preprint arXiv:2505.10080},
  year={2025}
}

@article{PhysRevLett.127.200402,
  title = {Fisher Information Universally Identifies Quantum Resources},
  author = {Tan, Kok Chuan and Narasimhachar, Varun and Regula, Bartosz},
  journal = {Phys. Rev. Lett.},
  volume = {127},
  issue = {20},
  pages = {200402},
  numpages = {7},
  year = {2021},
  month = {Nov},
  publisher = {American Physical Society},
  doi = {10.1103/PhysRevLett.127.200402},
  url = {https://link.aps.org/doi/10.1103/PhysRevLett.127.200402}
}

@article{montenegro2025quantum,
  title={Quantum metrology and sensing with many-body systems},
  author={Montenegro, Victor and Mukhopadhyay, Chiranjib and Yousefjani, Rozhin and Sarkar, Saubhik and Mishra, Utkarsh and Paris, Matteo GA and Bayat, Abolfazl},
  journal={Physics Reports},
  volume={1134},
  pages={1--62},
  year={2025},
  publisher={Elsevier}
}

@article{PRXQuantum.3.010325,
  title = {Variance of Relative Surprisal as Single-Shot Quantifier},
  author = {Boes, Paul and Ng, Nelly H.Y. and Wilming, Henrik},
  journal = {PRX Quantum},
  volume = {3},
  issue = {1},
  pages = {010325},
  numpages = {31},
  year = {2022},
  month = {Feb},
  publisher = {American Physical Society},
  doi = {10.1103/PRXQuantum.3.010325},
  url = {https://link.aps.org/doi/10.1103/PRXQuantum.3.010325}
}

@article{RevModPhys.81.865,
  title = {Quantum entanglement},
  author = {Horodecki, Ryszard and Horodecki, Pawe\l{} and Horodecki, Micha\l{} and Horodecki, Karol},
  journal = {Rev. Mod. Phys.},
  volume = {81},
  issue = {2},
  pages = {865--942},
  numpages = {0},
  year = {2009},
  month = {Jun},
  publisher = {American Physical Society},
  doi = {10.1103/RevModPhys.81.865},
  url = {https://link.aps.org/doi/10.1103/RevModPhys.81.865}
}

@article{hayden2007black,
  title={Black holes as mirrors: quantum information in random subsystems},
  author={Hayden, Patrick and Preskill, John},
  journal={Journal of high energy physics},
  volume={2007},
  number={09},
  pages={120--120},
  year={2007}
}

@article{yoshida2017efficient,
  title={Efficient decoding for the Hayden-Preskill protocol},
  author={Yoshida, Beni and Kitaev, Alexei},
  journal={arXiv preprint arXiv:1710.03363},
  year={2017}
}

@article{gu2010fidelity,
  title={Fidelity approach to quantum phase transitions},
  author={Gu, Shi-Jian},
  journal={International Journal of Modern Physics B},
  volume={24},
  number={23},
  pages={4371--4458},
  year={2010},
  publisher={World Scientific}
}

@article{tang2025estimating,
  title={Estimating time in quantum chaotic systems and black holes},
  author={Tang, Haifeng and Vardhan, Shreya and Wang, Jinzhao},
  journal={SciPost Physics},
  volume={19},
  number={4},
  pages={095},
  year={2025}
}

@article{chapman2022quantum,
  title={Quantum computational complexity from quantum information to black holes and back},
  author={Chapman, Shira and Policastro, Giuseppe},
  journal={The European Physical Journal C},
  volume={82},
  number={2},
  pages={128},
  year={2022},
  publisher={Springer}
}

@article{PRXQuantum.5.040325,
  title = {Feedback-Driven Quantum Reservoir Computing for Time-Series Analysis},
  author = {Kobayashi, Kaito and Fujii, Keisuke and Yamamoto, Naoki},
  journal = {PRX Quantum},
  volume = {5},
  issue = {4},
  pages = {040325},
  numpages = {16},
  year = {2024},
  month = {Nov},
  publisher = {American Physical Society},
  doi = {10.1103/PRXQuantum.5.040325},
  url = {https://link.aps.org/doi/10.1103/PRXQuantum.5.040325}
}

@article{jasser2026journey,
  title={A journey through Flatland: What does the antiflatness of a spectrum teach us?},
  author={Jasser, Barbara and Iannotti, Daniele and Hamma, Alioscia},
  journal={arXiv preprint arXiv:2605.21664},
  year={2026}
}

@article{RevModPhys.80.517,
  title = {Entanglement in many-body systems},
  author = {Amico, Luigi and Fazio, Rosario and Osterloh, Andreas and Vedral, Vlatko},
  journal = {Rev. Mod. Phys.},
  volume = {80},
  issue = {2},
  pages = {517--576},
  numpages = {0},
  year = {2008},
  month = {May},
  publisher = {American Physical Society},
  doi = {10.1103/RevModPhys.80.517},
  url = {https://link.aps.org/doi/10.1103/RevModPhys.80.517}
}

@article{PhysRevLett.133.247101,
  title = {Thermodynamic Concentration Inequalities and Trade-Off Relations},
  author = {Hasegawa, Yoshihiko and Nishiyama, Tomohiro},
  journal = {Phys. Rev. Lett.},
  volume = {133},
  issue = {24},
  pages = {247101},
  numpages = {7},
  year = {2024},
  month = {Dec},
  publisher = {American Physical Society},
  doi = {10.1103/PhysRevLett.133.247101},
  url = {https://link.aps.org/doi/10.1103/PhysRevLett.133.247101}
}

@article{paris2009quantum,
  title={Quantum estimation for quantum technology},
  author={Paris, Matteo GA},
  journal={International Journal of Quantum Information},
  volume={7},
  number={supp01},
  pages={125--137},
  year={2009},
  publisher={World Scientific}
}

@article{PhysRevLett.72.3439,
  title = {Statistical distance and the geometry of quantum states},
  author = {Braunstein, Samuel L. and Caves, Carlton M.},
  journal = {Phys. Rev. Lett.},
  volume = {72},
  issue = {22},
  pages = {3439--3443},
  numpages = {0},
  year = {1994},
  month = {May},
  publisher = {American Physical Society},
  doi = {10.1103/PhysRevLett.72.3439},
  url = {https://link.aps.org/doi/10.1103/PhysRevLett.72.3439}
}

@article{deffner2017quantum,
  title={Quantum speed limits: from Heisenberg’s uncertainty principle to optimal quantum control},
  author={Deffner, Sebastian and Campbell, Steve},
  journal={Journal of Physics A: Mathematical and Theoretical},
  volume={50},
  number={45},
  pages={453001},
  year={2017},
  publisher={IOP Publishing}
}

@article{RevModPhys.94.015004,
  title = {Noisy intermediate-scale quantum algorithms},
  author = {Bharti, Kishor and Cervera-Lierta, Alba and Kyaw, Thi Ha and Haug, Tobias and Alperin-Lea, Sumner and Anand, Abhinav and Degroote, Matthias and Heimonen, Hermanni and Kottmann, Jakob S. and Menke, Tim and Mok, Wai-Keong and Sim, Sukin and Kwek, Leong-Chuan and Aspuru-Guzik, Al\'an},
  journal = {Rev. Mod. Phys.},
  volume = {94},
  issue = {1},
  pages = {015004},
  numpages = {69},
  year = {2022},
  month = {Feb},
  publisher = {American Physical Society},
  doi = {10.1103/RevModPhys.94.015004},
  url = {https://link.aps.org/doi/10.1103/RevModPhys.94.015004}
}

@article{wysocki2026volume,
  title={Volume-law protection of metrological advantage},
  author={Wysocki, Piotr and others},
  journal={arXiv preprint arXiv:2602.09086},
  year={2026}
}

@article{PhysRevLett.126.010602,
  title = {Thermodynamic Uncertainty Relation for General Open Quantum Systems},
  author = {Hasegawa, Yoshihiko},
  journal = {Phys. Rev. Lett.},
  volume = {126},
  issue = {1},
  pages = {010602},
  numpages = {7},
  year = {2021},
  month = {Jan},
  publisher = {American Physical Society},
  doi = {10.1103/PhysRevLett.126.010602},
  url = {https://link.aps.org/doi/10.1103/PhysRevLett.126.010602}
}

@article{escher2011general,
  title={General framework for estimating the ultimate precision limit in noisy quantum-enhanced metrology},
  author={Escher, BM and de Matos Filho, Ruynet Lima and Davidovich, Luiz},
  journal={Nature Physics},
  volume={7},
  number={5},
  pages={406--411},
  year={2011},
  publisher={Nature Publishing Group UK London}
}

@article{tnfv-lzfx,
  title = {Optimal quantum reservoir learning in proximity to universality},
  author = {Ivaki, Moein N. and Karjula, Matias and Ala-Nissila, Tapio},
  journal = {Phys. Rev. A},
  volume = {113},
  issue = {6},
  pages = {L060401},
  numpages = {8},
  year = {2026},
  month = {Jun},
  publisher = {American Physical Society},
  doi = {10.1103/tnfv-lzfx},
  url = {https://link.aps.org/doi/10.1103/tnfv-lzfx}
}

@article{krakauer2011darwinian,
  title={Darwinian demons, evolutionary complexity, and information maximization},
  author={Krakauer, David C},
  journal={Chaos: An Interdisciplinary Journal of Nonlinear Science},
  volume={21},
  number={3},
  year={2011},
  publisher={AIP Publishing}
}

@article{mora2011biological,
  title={Are biological systems poised at criticality?},
  author={Mora, Thierry and Bialek, William},
  journal={Journal of Statistical Physics},
  volume={144},
  number={2},
  pages={268--302},
  year={2011},
  publisher={Springer}
}

@article{PhysRevLett.113.068102,
  title = {Zipf's Law and Criticality in Multivariate Data without Fine-Tuning},
  author = {Schwab, David J. and Nemenman, Ilya and Mehta, Pankaj},
  journal = {Phys. Rev. Lett.},
  volume = {113},
  issue = {6},
  pages = {068102},
  numpages = {5},
  year = {2014},
  month = {Aug},
  publisher = {American Physical Society},
  doi = {10.1103/PhysRevLett.113.068102},
  url = {https://link.aps.org/doi/10.1103/PhysRevLett.113.068102}
}

@article{langton1990computation,
  title={Computation at the edge of chaos: Phase transitions and emergent computation},
  author={Langton, Chris G},
  journal={Physica D: nonlinear phenomena},
  volume={42},
  number={1-3},
  pages={12--37},
  year={1990},
  publisher={Elsevier}
}

@article{mitchell1993revisiting,
  title={Revisiting the edge of chaos: Evolving cellular automata to perform computations},
  author={Mitchell, Melanie and Hraber, Peter and Crutchfield, James P},
  journal={arXiv preprint adap-org/9303003},
  year={1993}
}

@article{PhysRevLett.109.237208,
  title = {Entanglement Spectrum, Critical Exponents, and Order Parameters in Quantum Spin Chains},
  author = {De Chiara, G. and Lepori, L. and Lewenstein, M. and Sanpera, A.},
  journal = {Phys. Rev. Lett.},
  volume = {109},
  issue = {23},
  pages = {237208},
  numpages = {5},
  year = {2012},
  month = {Dec},
  publisher = {American Physical Society},
  doi = {10.1103/PhysRevLett.109.237208},
  url = {https://link.aps.org/doi/10.1103/PhysRevLett.109.237208}
}

@article{shaffer2014irreversibility,
  title={Irreversibility and entanglement spectrum statistics in quantum circuits},
  author={Shaffer, Daniel and Chamon, Claudio and Hamma, Alioscia and Mucciolo, Eduardo R},
  journal={Journal of Statistical Mechanics: Theory and Experiment},
  volume={2014},
  number={12},
  pages={P12007},
  year={2014},
  publisher={IOP Publishing and SISSA}
}

@article{PhysRevLett.96.010401,
  title = {Quantum Metrology},
  author = {Giovannetti, Vittorio and Lloyd, Seth and Maccone, Lorenzo},
  journal = {Phys. Rev. Lett.},
  volume = {96},
  issue = {1},
  pages = {010401},
  numpages = {4},
  year = {2006},
  month = {Jan},
  publisher = {American Physical Society},
  doi = {10.1103/PhysRevLett.96.010401},
  url = {https://link.aps.org/doi/10.1103/PhysRevLett.96.010401}
}

@article{margolus1998maximum,
  title={The maximum speed of dynamical evolution},
  author={Margolus, Norman and Levitin, Lev B},
  journal={Physica D: Nonlinear Phenomena},
  volume={120},
  number={1-2},
  pages={188--195},
  year={1998},
  publisher={Elsevier}
}

@incollection{mandelstam1991uncertainty,
  title={The uncertainty relation between energy and time in non-relativistic quantum mechanics},
  author={Mandelstam, Leonid and Tamm, IG},
  booktitle={Selected papers},
  pages={115--123},
  year={1991},
  publisher={Springer}
}

@article{peruzzo2014variational,
  title={A variational eigenvalue solver on a photonic quantum processor},
  author={Peruzzo, Alberto and McClean, Jarrod and Shadbolt, Peter and Yung, Man-Hong and Zhou, Xiao-Qi and Love, Peter J and Aspuru-Guzik, Al{\'a}n and O’brien, Jeremy L},
  journal={Nature communications},
  volume={5},
  number={1},
  pages={4213},
  year={2014},
  publisher={Nature Publishing Group UK London}
}

@article{farhi2014quantum,
  title={A quantum approximate optimization algorithm},
  author={Farhi, Edward and Goldstone, Jeffrey and Gutmann, Sam},
  journal={arXiv preprint arXiv:1411.4028},
  year={2014}
}

@article{harrow2009random,
  title={Random quantum circuits are approximate 2-designs},
  author={Harrow, Aram W and Low, Richard A},
  journal={Communications in Mathematical Physics},
  volume={291},
  number={1},
  pages={257--302},
  year={2009},
  publisher={Springer}
}

@article{PhysRevA.80.012304,
  title = {Exact and approximate unitary 2-designs and their application to fidelity estimation},
  author = {Dankert, Christoph and Cleve, Richard and Emerson, Joseph and Livine, Etera},
  journal = {Phys. Rev. A},
  volume = {80},
  issue = {1},
  pages = {012304},
  numpages = {6},
  year = {2009},
  month = {Jul},
  publisher = {American Physical Society},
  doi = {10.1103/PhysRevA.80.012304},
  url = {https://link.aps.org/doi/10.1103/PhysRevA.80.012304}
}

@article{mcclean2018barren,
  title={Barren plateaus in quantum neural network training landscapes},
  author={McClean, Jarrod R and Boixo, Sergio and Smelyanskiy, Vadim N and Babbush, Ryan and Neven, Hartmut},
  journal={Nature communications},
  volume={9},
  number={1},
  pages={4812},
  year={2018},
  publisher={Nature Publishing Group UK London}
}

@article{PhysRevLett.101.010504,
  title = {Entanglement Spectrum as a Generalization of Entanglement Entropy: Identification of Topological Order in Non-Abelian Fractional Quantum Hall Effect States},
  author = {Li, Hui and Haldane, F. D. M.},
  journal = {Phys. Rev. Lett.},
  volume = {101},
  issue = {1},
  pages = {010504},
  numpages = {4},
  year = {2008},
  month = {Jul},
  publisher = {American Physical Society},
  doi = {10.1103/PhysRevLett.101.010504},
  url = {https://link.aps.org/doi/10.1103/PhysRevLett.101.010504}
}

@article{martinez2023information,
  title={Information processing capacity of spin-based quantum reservoir computing systems},
  author={Mart{\'\i}nez-Pe{\~n}a, Rodrigo and Nokkala, Johannes and Giorgi, Gian Luca and Zambrini, Roberta and Soriano, Miguel C},
  journal={Cognitive Computation},
  volume={15},
  number={5},
  pages={1440--1451},
  year={2023},
  publisher={Springer}
}

@article{dambre2012information,
  title={Information processing capacity of dynamical systems},
  author={Dambre, Joni and Verstraeten, David and Schrauwen, Benjamin and Massar, Serge},
  journal={Scientific reports},
  volume={2},
  number={1},
  pages={514},
  year={2012},
  publisher={Nature Publishing Group UK London}
}

@article{84f3-63mz,
  title = {Engineering quantum reservoirs through Krylov complexity, expressivity, and observability},
  author = {\ifmmode \check{C}\else \v{C}\fi{}indrak, Saud and Jaurigue, Lina and L\"udge, Kathy},
  journal = {Phys. Rev. Res.},
  volume = {7},
  issue = {4},
  pages = {043190},
  numpages = {21},
  year = {2025},
  month = {Nov},
  publisher = {American Physical Society},
  doi = {10.1103/84f3-63mz},
  url = {https://link.aps.org/doi/10.1103/84f3-63mz}
}

@article{ProvostVallee1980,
  author  = {Provost, J. P. and Vall{\'e}e, G.},
  title   = {Riemannian structure on manifolds of quantum states},
  journal = {Communications in Mathematical Physics},
  volume  = {76},
  pages   = {289--301},
  year    = {1980},
  doi     = {10.1007/BF02193559}
}

@article{liu2020quantum,
  title={Quantum Fisher information matrix and multiparameter estimation},
  author={Liu, Jing and Yuan, Haidong and Lu, Xiao-Ming and Wang, Xiaoguang},
  journal={Journal of Physics A: Mathematical and Theoretical},
  volume={53},
  number={2},
  pages={023001},
  year={2020},
  publisher={IOP Publishing}
}

@article{bennett1982thermodynamics,
  title={The thermodynamics of computation—a review},
  author={Bennett, Charles H},
  journal={International Journal of Theoretical Physics},
  volume={21},
  number={12},
  pages={905--940},
  year={1982},
  publisher={Springer}
}

@article{PhysRevLett.131.180403,
  title = {Scrambling Is Necessary but Not Sufficient for Chaos},
  author = {Dowling, Neil and Kos, Pavel and Modi, Kavan},
  journal = {Phys. Rev. Lett.},
  volume = {131},
  issue = {18},
  pages = {180403},
  numpages = {6},
  year = {2023},
  month = {Nov},
  publisher = {American Physical Society},
  doi = {10.1103/PhysRevLett.131.180403},
  url = {https://link.aps.org/doi/10.1103/PhysRevLett.131.180403}
}

@article{meyer2021fisher,
  title={Fisher information in noisy intermediate-scale quantum applications},
  author={Meyer, Johannes Jakob},
  journal={Quantum},
  volume={5},
  pages={539},
  year={2021},
  publisher={Verein zur F{\"o}rderung des Open Access Publizierens in den Quantenwissenschaften}
}

@article{v7mf-yh8n,
  title = {Critical Quantum Sensing: A Tutorial on Parameter Estimation Near Quantum Phase Transitions},
  author = {Mihailescu, George and Alushi, Uesli and Di Candia, Roberto and Felicetti, Simone and Gietka, Karol},
  journal = {PRX Quantum},
  volume = {7},
  issue = {2},
  pages = {020201},
  numpages = {68},
  year = {2026},
  month = {Jun},
  publisher = {American Physical Society},
  doi = {10.1103/v7mf-yh8n},
  url = {https://link.aps.org/doi/10.1103/v7mf-yh8n}
}

@article{robin2025anti,
  title={Anti-flatness and non-local magic in two-particle scattering processes},
  author={Robin, Caroline EP and Savage, Martin J},
  journal={arXiv preprint arXiv:2510.23426},
  year={2025}
}

@article{grieninger2026nonlocal,
  title={The nonlocal magic of a holographic Schwinger pair},
  author={Grieninger, Sebastian},
  journal={arXiv preprint arXiv:2605.04210},
  year={2026}
}

@article{ahmad2025experimental,
  title={Experimental demonstration of non-local magic in a superconducting quantum processor},
  author={Ahmad, Halima Giovanna and Esposito, Gianluca and Stasino, Viviana and Odavic, Jovan and Cosenza, Carlo and Sarno, Alessandro and Mastrovito, Pasquale and Viscardi, Michele and Cusumano, Stefano and Tafuri, Francesco and others},
  journal={arXiv preprint arXiv:2511.15576},
  year={2025}
}

@article{ivaki2025dynamical,
  title={Dynamical learning and quantum memory with non-Hermitian many-body systems},
  author={Ivaki, Moein N and Szuminsky, Austin J and Lazarides, Achilleas and Zagoskin, Alexandre and McCaul, Gerard and Ala-Nissila, Tapio},
  journal={arXiv preprint arXiv:2506.07676},
  year={2025}
}

@article{ding2026thermodynamics,
  title={Thermodynamics of Quantum Reservoir Computing},
  author={Ding, Lixiang and Qiu, Xingze},
  journal={arXiv preprint arXiv:2607.02157},
  year={2026}
}

@article{torre2026non,
  title={Non-Local Magic from the Entanglement Spectrum},
  author={Torre, Gianpaolo and Franchini, Fabio and Giampaolo, Salvatore Marco},
  journal={arXiv preprint arXiv:2607.07808},
  year={2026}
}

@article{yasuda2023quantum,
  title={Quantum reservoir computing with repeated measurements on superconducting devices},
  author={Yasuda, Toshiki and Suzuki, Yudai and Kubota, Tomoyuki and Nakajima, Kohei and Gao, Qi and Zhang, Wenlong and Shimono, Satoshi and Nurdin, Hendra I and Yamamoto, Naoki},
  journal={arXiv preprint arXiv:2310.06706},
  year={2023}
}

@article{wootters2001entanglement,
  title={Entanglement of formation and concurrence.},
  author={Wootters, William K},
  journal={Quantum Inf. Comput.},
  volume={1},
  number={1},
  pages={27--44},
  year={2001}
}

@article{sidhu2020geometric,
  title={Geometric perspective on quantum parameter estimation},
  author={Sidhu, Jasminder S and Kok, Pieter},
  journal={AVS Quantum Science},
  volume={2},
  number={1},
  year={2020},
  publisher={AIP Publishing}
}

@article{plodzien2026operator,
  title={Operator spreading and recoverability of local quantum Fisher information in a $ U (1) $-broken spin chain},
  author={P{\l}odzie{\'n}, Marcin and Chwede{\'n}czuk, Jan},
  journal={arXiv preprint arXiv:2605.02774},
  year={2026}
}

@article{schuld2021effect,
  title={Effect of data encoding on the expressive power of variational quantum-machine-learning models},
  author={Schuld, Maria and Sweke, Ryan and Meyer, Johannes Jakob},
  journal={Physical Review A},
  volume={103},
  number={3},
  pages={032430},
  year={2021},
  publisher={APS}
}

@article{schutte2025expressivity,
  title={Expressivity Limits of Quantum Reservoir Computing},
  author={Sch{\"u}tte, Nils-Erik and G{\"o}tting, Niclas and M{\"u}ntinga, Hauke and List, Meike and Brunner, Daniel and Gies, Christopher},
  journal={arXiv preprint arXiv:2501.15528},
  year={2025}
}

@article{mujal2021analytical,
  title={Analytical evidence of nonlinearity in qubits and continuous-variable quantum reservoir computing},
  author={Mujal, Pere and Nokkala, Johannes and Mart{\'\i}nez-Pe{\~n}a, Rodrigo and Giorgi, Gian Luca and Soriano, Miguel C and Zambrini, Roberta},
  journal={Journal of Physics: Complexity},
  volume={2},
  number={4},
  pages={045008},
  year={2021},
  publisher={IOP Publishing}
}

@article{mccaul2025minimal,
  title={Minimal quantum reservoirs with Hamiltonian encoding},
  author={McCaul, Gerard and Totero Gongora, Juan Sebastian and Otieno, Wendy and Savel’ev, Sergey and Zagoskin, Alexandre and Balanov, Alexander G},
  journal={Chaos: An Interdisciplinary Journal of Nonlinear Science},
  volume={35},
  number={9},
  year={2025},
  publisher={AIP Publishing}
}

@article{r8ww-qw7j,
  title = {High-Accuracy Temporal Prediction via Experimental Quantum Reservoir Computing in Correlated Spins},
  author = {Hou, Yanjun and Hua, Juncheng and Wu, Ze and Xia, Wei and Chen, Yuquan and Li, Xiaopeng and Li, Zhaokai and Peng, Xinhua and Du, Jiangfeng},
  journal = {Phys. Rev. Lett.},
  volume = {136},
  issue = {12},
  pages = {120602},
  numpages = {8},
  year = {2026},
  month = {Mar},
  publisher = {American Physical Society},
  doi = {10.1103/r8ww-qw7j},
  url = {https://link.aps.org/doi/10.1103/r8ww-qw7j}
}

@article{markovic2020quantum,
  title={Quantum neuromorphic computing},
  author={Markovi{\'c}, Danijela and Grollier, Julie},
  journal={Applied physics letters},
  volume={117},
  number={15},
  year={2020},
  publisher={AIP Publishing}
}

@article{ricci2026quantum,
  title={Quantum reservoir computing induced by controllable damping},
  author={Ricci, Emanuele and Monzani, Francesco and Nigro, Luca and Prati, Enrico},
  journal={npj Quantum Information},
  year={2026},
  publisher={Nature Publishing Group UK London}
}

@article{huang2026intrinsic,
  title={Intrinsic spectral structure of bipartite nonlocal magic resource},
  author={Huang, Xiao and Chen, Guanhua and Yao, Yao},
  journal={arXiv preprint arXiv:2606.24368},
  year={2026}
}

@article{xia2026quantum,
  title={Quantum magic and non-commutativity as computational resources in quantum reservoir computing},
  author={Xia, Wei and Cao, Shuaifan and Qiu, Xingze and Li, Xiaopeng},
  journal={arXiv preprint arXiv:2607.12035},
  year={2026}
}

@article{PhysRevLett.109.120604,
  title = {Thermodynamics of Prediction},
  author = {Still, Susanne and Sivak, David A. and Bell, Anthony J. and Crooks, Gavin E.},
  journal = {Phys. Rev. Lett.},
  volume = {109},
  issue = {12},
  pages = {120604},
  numpages = {5},
  year = {2012},
  month = {Sep},
  publisher = {American Physical Society},
  doi = {10.1103/PhysRevLett.109.120604},
  url = {https://link.aps.org/doi/10.1103/PhysRevLett.109.120604}
}

@article{ivaki2025noise,
  title={Noise resilience in adaptive and symmetric monitored quantum circuits},
  author={Ivaki, Moein N and Ojanen, Teemu and Moghaddam, Ali G},
  journal={npj Quantum Information},
  volume={11},
  number={1},
  pages={111},
  year={2025},
  publisher={Nature Publishing Group UK London}
}

@article{veitch2014resource,
  title={The resource theory of stabilizer quantum computation},
  author={Veitch, Victor and Hamed Mousavian, SA and Gottesman, Daniel and Emerson, Joseph},
  journal={New Journal of Physics},
  volume={16},
  number={1},
  pages={013009},
  year={2014},
  publisher={IOP Publishing}
}

\end{document}